\documentclass[twocolumn,showpacs,preprintnumbers,amsmath,amssymb,prb,aps]{revtex4-1}
\usepackage{epsfig}
\usepackage{epstopdf}
\usepackage{multirow}
\usepackage{graphicx}
\usepackage{dcolumn}
\usepackage{bm}
\usepackage{textcomp}
\usepackage{array}
\usepackage{caption}
\usepackage{xfrac}
\usepackage{mathtools}
\usepackage{amsmath}
\usepackage{advdate}
\usepackage{subfigure}
\usepackage{hyperref}
\hypersetup{
    colorlinks=true,
    linkcolor=blue,
    filecolor=magenta,      
    urlcolor=blue,
    pdftitle={},
    }

\newcommand{\beginsupplement}{%
        \setcounter{table}{0}
        \renewcommand{\thetable}{S\arabic{table}}%
        \setcounter{figure}{0}
        \renewcommand{\thefigure}{S\arabic{figure}}%
     }

\begin{document}

\title{An \textit{ab-initio} study of \textit{nodal-arcs}, axial strain\textquoteright s effect on nodal-lines \& Weyl nodes and Weyl-contributed Seebeck coefficient in TaAs class of Weyl semimetals.}

\author{Vivek Pandey$^{1}$ }
\altaffiliation{vivek6422763@gmail.com}
\author{Sudhir K. Pandey$^{2}$}
\altaffiliation{sudhir@iitmandi.ac.in}
\affiliation{$^{1}$School of Physical Sciences, Indian Institute of Technology Mandi, Kamand - 175075, India\\
$^{2}$School of Mechanical and Materials Engineering, Indian Institute of Technology Mandi, Kamand - 175075, India}

\date{\today}

\begin{abstract}

This work verifies the existence of dispersive \textit{nodal-arcs} and their evolution into Weyl nodes under the effect of spin-orbit coupling (SOC) in NbAs \& NbP. The obtained features mimic the observations as reported for TaAs \& TaP in our previous work\cite{pandey2023existence}. In addition, this work reports that the number of nodes in TaAs class of Weyl semimetals (WSMs) can be altered via creating strain along $a$ or $c$ direction of the crystal. For instance, the number of nodes in NbAs under SOC-effect along with 2\% (3\%) tensile-strain in $a$ direction is found to be 40 (56) in its full Brillouin zone (BZ). Besides the nodes, such strain are found to have considerable impact on the nodal-lines of these WSMs when effect of SOC is ignored. A 3\% tensile (compressive) strain along the $a$ ($c$) direction leads to the partially merging of nodal-lines (without SOC) in the extended BZ of NbAs \& NbP, which is not observed in TaAs \& TaP within the range of -3\% to 3\% strain. Apart from this, the work discusses the role of Weyl physics in affecting the Seebeck coefficient ($S$) of any WSM. In this direction, it is discussed that how a symmetric Weyl cone, even if tilted, will have no contribution to the $S$ of WSMs. Furthermore, the work highlights the conditions under which a Weyl cone can contribute to the $S$ of a given WSM. Lastly, the discussion of Weyl contribution to $S$ is validated over TaAs class of WSMs via investigating the features of their Weyl cones and calculating the contributions of such cones to the $S$ of these semimetals. The value of $S$ contributed from Weyl cone is found to be as large as $\sim$65 $\mu$\textit{V}/\textit{K} below 25 K in case of TaAs. The findings of this work present a possibility of engineering the topological properties of TaAs class of WSMs via creating strain in their crystal. It also makes the picture of Weyl physics\textquoteright\hspace{0.1cm}  impact on the $S$ of WSMs a more clear.

\end{abstract}

\maketitle

\section{Introduction} 
\setlength{\parindent}{3em}

In the realm of condensed matter physics, researchers are increasingly attracted by the exploration of topological materials\cite{bernevig2022progress,vergniory2019complete,kawabata2022nonlinear,tokura2019magnetic,gao2019topological,hu2020realization}. Among them, Weyl semimetals (WSMs) stand out for their fascinating electronic characteristics that challenge conventional understanding\cite{das2019electronic,bodo2022spin,armitage2018weyl,shekhar2015extremely,chen2016superconductivity}. At the heart of WSMs are Weyl nodes, mysterious points in momentum space where electronic bands intersect linearly, forming cone-like structures\cite{lv2015observation,meng2020ternary,yang2016acoustic}. These nodes are sources or sinks of Berry curvature\cite{zeng2021nonlinear,imran2018berry}, leading to intriguing phenomena like the chiral anomaly\cite{zyuzin2012topological,ashby2014chiral} and Fermi arc surface states\cite{ojanen2013helical,sun2015topological}. Weyl semimetals offer not only fundamental insights but also potential for transformative technologies, from ultrafast electronics\cite{zhuo2021dynamical,gao2020chiral} to quantum computing\cite{sie2019ultrafast}.

Tantalum arsenide (TaAs) and tantalum phosphide (TaP) serve as prime examples of these materials\cite{caglieris2018anomalous,sun2016strong,grassano2020influence,du2016large}. Studies have revealed the presence of nodal-lines (also referred as nodal-rings) in their electronic band structures, when the effect of spin-orbit coupling (SOC) is ignored\cite{armitage2018weyl}. These nodal-lines are emerging from special kind of band inversion phenomena known as \textit{intraband dp inversion (IDPI)}\cite{pandey2023existence}. Notably, these nodal-lines are getting generated from \textit{nodal-arcs} under the effect of crystal symmetries including the time-reversal symmetry\cite{pandey2023existence}. However, with SOC, these arcs transform into Weyl nodes, giving the material unique topological properties and chiral characteristics.

Besides TaAs \& TaP, the TaAs class of WSMs include NbAs \& NbP\cite{huang2015inversion,liu2016evolution}. These WSMs share similar electronic structures and semimetalic features as that of TaAs \& TaP\cite{grassano2020influence}. Particularly, NbAs \& NbP also possess four nodal-lines on their mirror planes as possessed by the TaAs \& TaP. Moreover, previous works reveal that the evolution of Weyl nodes under the effect of SOC in NbAs \& NbP is also similar to TaAs \& TaP, in terms of their number and coordinates\cite{weng2015weyl}. These features indicate that there is a possibility of the existence of \textit{nodal-arcs} in NbAs \& NbP with similar properties as that were obtained in the case of TaAs \& TaP. Such an analysis of NbAs \& NbP in order to search the \textit{nodal-arcs} and explore their properties is still missing. Moreover, the comparative analysis of features of accidental-degeneracy in the four members of TaAs class of WSMs will highlight the role of hybridization-strength and SOC in formation of such degenerate points in a more clear manner. This is because different members of this family possess different hybridization-strength and the magnitude of SOC.

\begin{figure*}
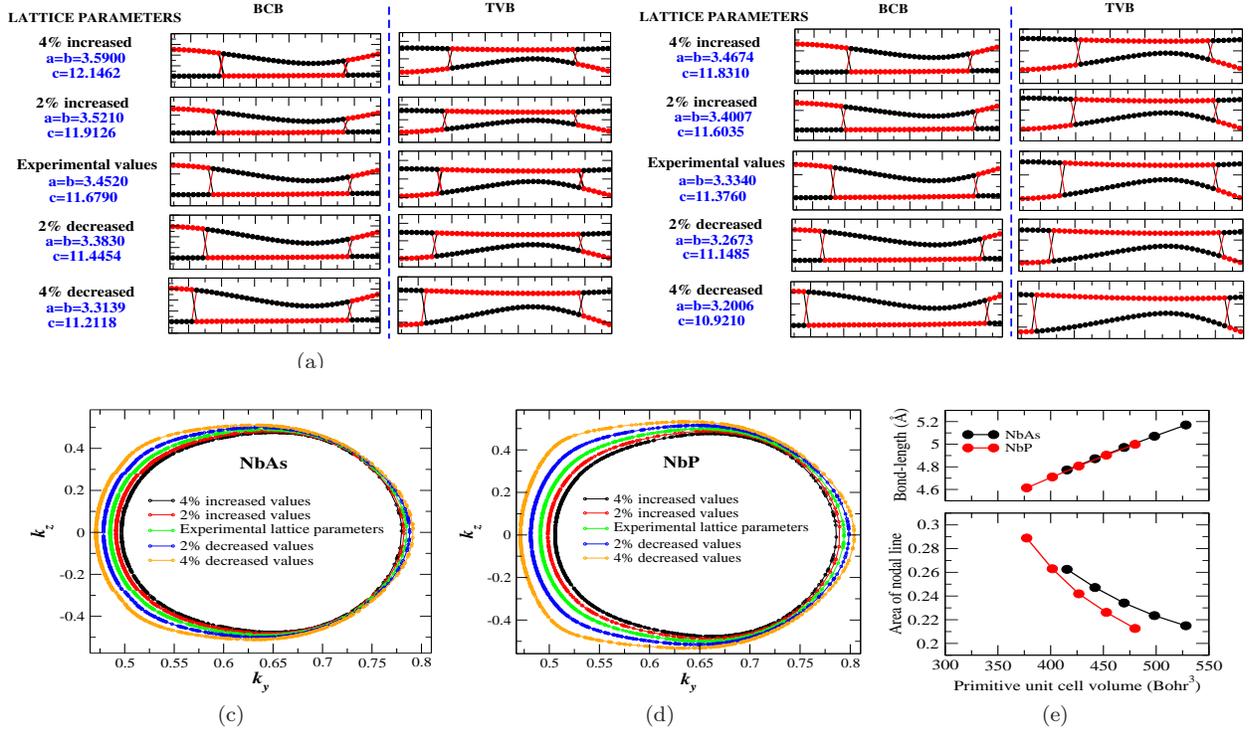

    \centering
    \subfigure[]
    {
        \includegraphics[width=0.45\linewidth, height=4.5cm]{fig1a.eps}
    }
    \subfigure[]
    {
        \includegraphics[width=0.45\linewidth, height=4.5cm]{fig1b.eps}
    }\vspace*{0.1in}
    \subfigure[]
    {
        \includegraphics[width=0.30\linewidth, height=3.8cm]{fig1c.eps}
    }
    \subfigure[]
    {
        \includegraphics[width=0.30\linewidth, height=3.8cm]{fig1d.eps}
    }
    \subfigure[]
    {
        \includegraphics[width=0.25\linewidth, height=3.8cm]{fig1e.eps}
    }
    \caption
    { {{\footnotesize (Color online) Plot (a) ((b)) shows the effect of change in lattice parameters on \textit{IDPI} in NbAs (NbP). Plots (c) \& (d) shows the change in nodal-line in $k_x$=0
plane obtained corresponding to change in lattice parameters of NbAs \& NbP, respectively. Plot (e) shows the area of nodal-line versus primitive unit cell volume for NbAs \& NbP.}}
    }
    \label{fig1}
\end{figure*} 

Moving further, the Weyl nodes (nodal-lines) in any material generally arises due to the bands-inversion phenomena resulting from the hybridization of various atomic orbitals\cite{du2017emergence,markovic2019weyl}. Such hybridization of orbitals are extremely sesitive to the the inter-atomic distances. Thus, the hybridization-strength can be altered through the application of strain along certain directions of materials hosting non-trivial bands-degeneracy. In this direction, several analysis have been carried out over various WSMs to highlight the ability of the strain-induced engineering of Weyl nature of WSMs. It has been shown that in HfCuP, which hosts both the type I and type II Weyl nodes at its experiamntal lattice parameters, application of 1\% biaxial compressive-strain in the \textit{a}-\textit{b} plane leads to the annihilation of type I Weyl nodes leading the material to become purely type II WSM\cite{meng2020ternary}. In another study on HgTe, it was found that under a large enough tensile strain, the eight Weyl points annihilate with each other in the $k_x$=$\pm$$k_y$ planes, leading to a strong
topological insulator phase\cite{ruan2016symmetry}. These study clearly shows that Weyl semimetallic features can be easily altered through the application of strain. Now, we shall discuss the possible impact of strain along the $a$ or $c$ directions of body-centered tetragonal crystal-structure of TaAs class of WSMs.

Strain can be of two type: (i) compressive strain \& tensile strain which reduces or increases the size of crystal along the direction their application, respectively. Particularly for TaAs class of WSMs, tensile strain along the $a$-axis will stretch the lattice, affecting hopping parameters between atoms. This may influence the band inversion, leading to the emergence of new Weyl nodes. Conversely, compressive strain along the $a$-axis will compress the lattice and may potentially cause existing Weyl nodes to merge. Along the $c$-axis, strain will influence interlayer coupling, which may affect hopping and Weyl node distribution. Additionally, in the absence of SOC, these strain will affect the nodal-lines degeneracy due to the above mentioned mechanism. To the best of our knowledge, such aspects are still remaining untouched corresponding to the nodal-lines and Weyl nodes in TaAs class of WSMs. Understanding these strain-induced effects on nodal-lines and Weyl nodes will be crucial for tailoring the electronic and topological properties of TaAs class of WSMs. 

Another necessary topic of discussion is the extent to which Weyl physics will affect the transport properties of TaAs class of WSMs. Weyl physics is associated with the states around the linear region of the Weyl cone of a given WSM\cite{armitage2018weyl}. Thus, the extent to which these Weyl-dominated states affect the total transport of a given WSM measures the significance of Weyl physics in influencing the properties of the semimetal. Generally, the Weyl nodes are situated close to the Fermi energy of WSMs. Thus, performing experiments at low temperature may provide some insight to Weyl physics contributions to the transport of WSMs. However, particularly in the TaAs class of WSMs, along with the Weyl cones there are also trival states present close to their Fermi energy. This makes it experimentaly a challenging task to extract the Weyl-contributed transport of these WSMs. However, such an analysis, which is still found to be missing in our literature survey, can be carried out through computational methods. 

In the present work, we have established that each nodal-line in NbAs \& NbP are formed from a pair of \textit{nodal-arcs}. These arcs are related to each-other via crystal symmetry operations. It also highlights the role of SOC in evolving each \textit{nodal-arc} into two pairs of Weyl nodes out of which one pair is coming from the combined contribution of the two arcs forming a ring. Apart from this, the work presents a detailed analysis of the effect of compressive and tensile strain along $a$ or $c$ direction of crystals of TaAs class of WSMs on the nodal-lines and the node points (when SOC is considered). Beyond characterization, the work also explores the role of Weyl physics in affecting the thermoelectric properties of WSMs, with a specific focus on the Seebeck coefficient ($S$) contribution from Weyl cones. For this study, TaAs class of WSMs are used as prototype.


\section{Computational details}

In our study, we utilized the WIEN2k package for density functional theory (DFT) calculations, employing the PBESol as exchange-correlation functional. TaAs class of WSMs crystallize in body-centered tetragonal crystal structure with the space group $I4_1md$\cite{grassano2018validity}. The lattice constants and Wyckoff positions of atoms for TaAs \& TaP are taken same as mentioned in reference\cite{pandey2023existence}. The lattice constants for NbAs (NbP) are taken as \textit{a}=3.4520 \AA\hspace{0.03in} \& \textit{c}=11.6790 \AA \hspace{0.03in}\cite{lee2015fermi}(\textit{a}=3.3340 \AA\hspace{0.03in} \& \textit{c}= 11.3760 \AA\cite{lee2015fermi}). The Wyckoff position of Nb is (0,0,0) while for As \& P, it is (0,0,0.416). For these semimetals, self-consistent ground state energy calculations were conducted with a 10$\times$10$\times$10 \textit{k}-mesh size and a charge convergence limit of $10^{-10}$ electronic charge. The nodal-lines and Weyl nodes analysis are performed using the PY-Nodes code\cite{pandey2023py} that employs the \textit{Nelder-Mead's function minimization} method to identify bands-touching features in the electronic band structures. Additionally, the Weyl-cone-contributed $S$ were computed using the \textit{TRACK} code, which utilizes the Kubo linear-response formalism\cite{sihi2023track}.

\begin{figure}
    \centering
    \subfigure[]
    {
        \includegraphics[width=0.55\linewidth, height=3.5cm]{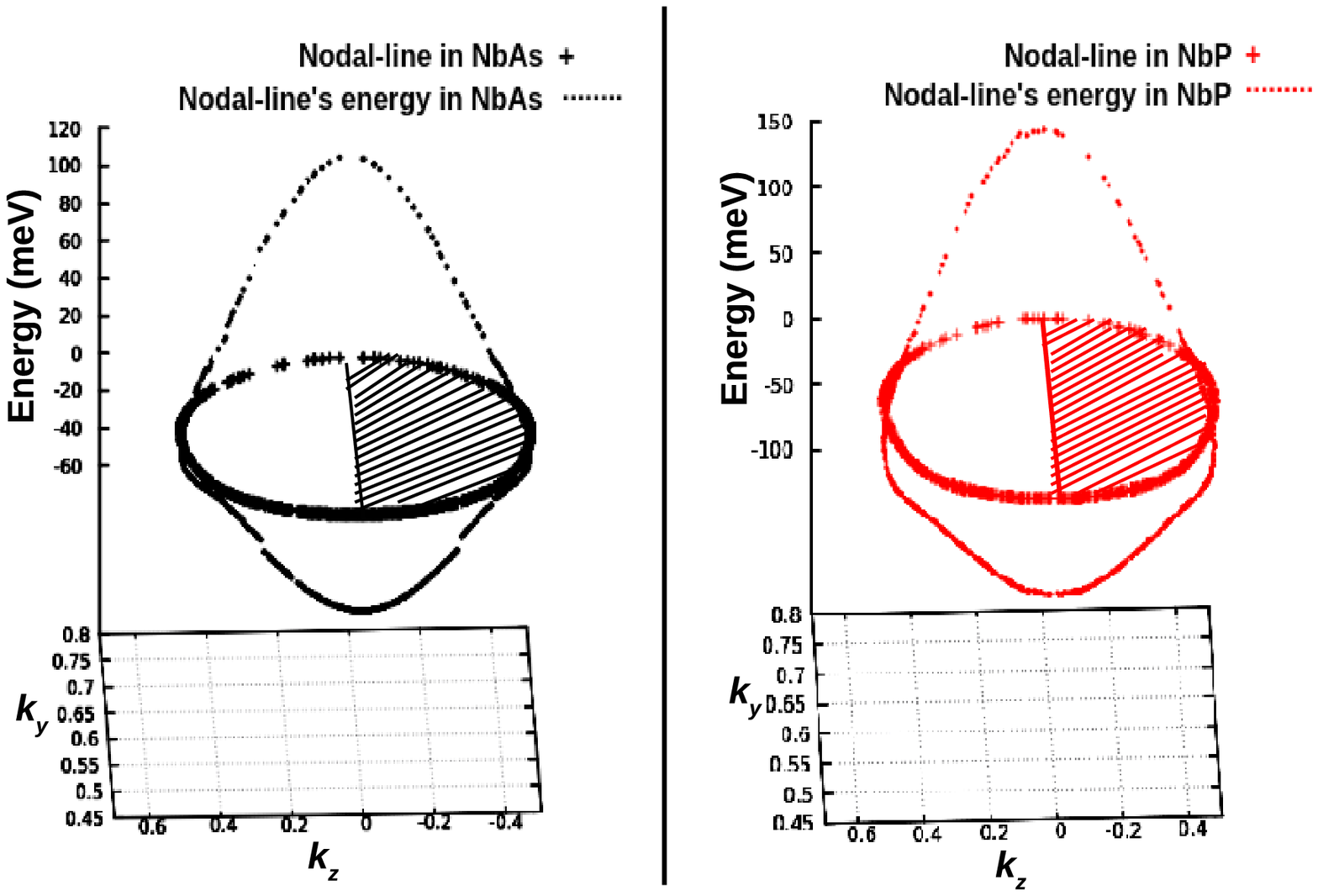}
    }
    \subfigure[]
    {
        \includegraphics[width=0.35\linewidth, height=3.5cm]{fig2b.eps}
    }\vspace*{0.1in}
    
    \caption
    { {{\footnotesize (Color online) Plot (a) shows the energy of nodal-line in NbAs \& NbP, respectively. The obtained energy is symmetric across $k_z$=0 plane. Plot (b) shows the quantitative comparison of energy of \textit{nodal-arc} along $k_y$ axis for NbAs \& NbP.}}
    }
    \label{fig2}
\end{figure}

\section{Results and Discussion} 

\subsection{Formation of \textit{nodal-arc}}
Inspired by our previous studies on TaAs \& TaP\cite{pandey2023existence}, there is a strong interest in investigating the presence of \textit{nodal-arcs} in NbAs \& NbP. Our focus is on analyzing the band characteristics of the topmost valence band (TVB) and bottommost conduction band (BCB) of NbAs \& NbP along the directions \textit{$\Gamma$-N-Z-P-X-$\Gamma$-P-N} in their electronic structure. Our analysis reveals that these states are primarily influenced by Nb-4\textit{d} and As-4\textit{p}/P-3\textit{p} orbitals. At the atomic level, Nb-4\textit{d} orbital possess higher energy when compared with the energy of As-4\textit{p}/P-3\textit{p}. Thus, it is typically expected that the BCB (TVB) must have higher (lower) contributions from Nb-4\textit{d} than that from As-4\textit{p}/P-3\textit{p} across the BZ. However, deviations from this behaviour are observed in certain regions of the BZ, where BCB (TVB) is found to have higher (lower) contributions from As-4\textit{p}/P-3\textit{p} than that from Nb-4\textit{d} orbital. This leads to what is termed as \textit{intraband dp inversion (IDPI)}\cite{pandey2023existence}. The \textit{IDPI} observed along the $N-Z$ high symmetric directions of NbAs \& NbP is shown in Fig. S1 of supplementary materials. It is well known that such bands-inversion generally leads to non-trivial bands-touchings. Further analysis in this direction reveal that NbAs \& NbP possess a pair of nodal-lines on each of its mirror planes ($k_x$=0 \& $k_y$=0 planes). Moreover, examination shows that \textit{IDPI} predominantly occurs within these nodal-lines when compared with the regions outside these rings, highlighting its crucial role in their formation.

Our previous work\cite{pandey2023existence} reveal that the four nodal-lines in TaAs \& TaP can be generated from a single \textit{nodal-arc} by applying on it the crystal symmetry operations associated with these materials. Based on this result, the \textit{nodal-arc} degeneracy was reported to be associated with the irreducible parts of the BZ of these WSMs and thus, considered as fundamental degeneracy. As NbAs \& NbP share similarities with TaAs \& TaP in terms of crystal symmetries\cite{armitage2018weyl}, it is anticipated that each nodal-line of NbAs \& NbP can be further devided into \textit{nodal-arcs}. These arcs are supposed to be related to each other through crystal symmetries including the time-reversal symmetry. In that case, one of the arcs can be considered as arising from fundamental degeneracy phenomena, while the others can be termed as derived arcs. This expectation is supported by energy analysis of nodal-line in NbAs \& NbP as shown in Fig. \ref{fig2}(a), revealing symmetric energy distribution about $k_z$=0 plane. Thus, the periphery of the shaded region can be considered as \textit{nodal-arc}. This arc can generate all the four nodal-lines upon the application of crystal symmetries as discussed in our previous work\cite{pandey2023py}. Moving further, the comparison of energy spread between NbAs \& NbP indicates a higher spread in NbP ($\sim$215 meV) as compared to NbAs ($\sim$157 meV). The relevant graphs are shown in Fig. \ref{fig2}(b). The \textit{nodal-arcs} in TaAs \& TaP were reported to be extended beyond the boundaries of first BZ\cite{pandey2023existence}. In this direction, investigation in NbAs \& NbP reveals a similar results. The extended \textit{nodal-arcs} of NbAs \& NbP beyond the boundaries of the first BZ is shown in Fig. S2 of supplementary materials. The figure also shows the degeneracy created in the first BZ when the extended regions of the \textit{nodal-arcs} are mapped inside the first BZ.

Another well-established aspect about TaAs \& TaP is that the strength of hybridization among the atomic orbitals play an important role in the occurrence of \textit{IDPI} and the formation of nodal-lines\cite{pandey2023py}. As the hybridization-strength depends on the inter-atomic distances, changing the lattice parameters was used to probe the hybridization-strength among these materials. In this direction, the study has been carried out to explore the sensitivity of \textit{IDPI} in TVB \& BCB of NbAs \& NbP to the variation in lattice parameters. The obtained results are shown in Fig. \ref{fig1}(a) (Fig. \ref{fig1}(b)) corresponding to NbAs (NbP). It is found that spatial-extent of \textit{IDPI} in the bands increases with the decrease in magnitude of lattice parameters. The size of nodal-lines also shows exactly similar variations with the change in lattice parameters of these WSMs. Corresponding plots are shown in Fig. \ref{fig1}(c) (Fig. \ref{fig1}(d)) for NbAs (NbP). A comparative analysis of change in the area of nodal-lines with the variation in primitive unit cell volume of NbAs \& NbP is presented in Fig. \ref{fig1}(e). These observations mimics to the one observed in TaAs \& TaP\cite{pandey2023py}. This establishes the crucial role of hybridization of atomic orbitals in the formation of nodal-lines in NbAs \& NbP. 

\begin{figure}
    \centering
    {
        \includegraphics[width=0.80\linewidth, height=4.0cm]{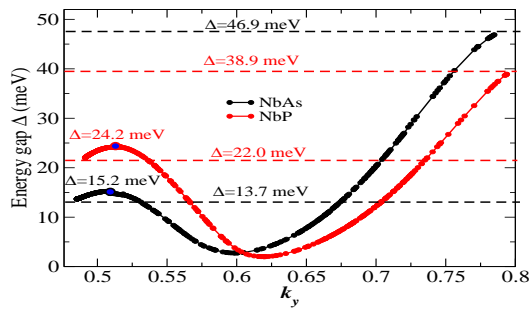}
    }  
    \caption
    { {{\footnotesize (Color online) Energy gap created at the \textit{nodal-arc} due to SOC in NbAs \& NbP.}}
    }
    \label{fig33}
\end{figure}

\begin{table*}\label{tabc}
\caption{\label{tabc}
{\footnotesize The coordinates and energy of the representative Weyl-nodes (W1 \& W2) in NbAs \& NbP obtained corresponding to different strength of SOC used in the calculations.}}
 \centering
 \begin{tabular}{|p{1.5cm}|c|c|c|c|c|c|c|}
   \hline
   \multirow{2}{1cm}{{\textbf{Material}}} &{\textbf{Value of \textit{c}}} & {\textbf{SOC-strength per Nb atom}}& \multicolumn{2}{c|}{\textbf{W1 point}}& \multicolumn{2}{c|}{\textbf{W2 point}}  \\
   \cline{4-7}
  & \textbf{(atomic unit)} & \textbf{(meV)} & \textbf{  Coordinates  } & \textbf{Energy (meV)} & \textbf{  Coordinates  } & \textbf{Energy (meV)}  \\
   \hline
          & 137.03     & 75.60      & (0.484,0.003,0.000)     & -39.98  & (0.281,0.007,0.581) &  12.86 \\
  NbAs    & 130.18     & 93.68     & (0.484,0.003,0.000)     & -38.50   & (0.281,0.007,0.581) & 12.71 \\
          & 123.32     & 117.63     & (0.484,0.003,0.000)     & -36.69   & (0.281,0.008,0.580) & 12.42 \\
          &            &             &                      &         &                  &        \\
          & 137.03     & 60.84      & (0.491,0.003,0.000)     & -68.40  & (0.271,0.006,0.578) &  8.08   \\
   NbP    & 130.18     & 75.49     & (0.491,0.003,0.000)     & -68.19  & (0.271,0.006,0.578) & 7.24   \\
          & 123.32     & 94.95     & (0.491,0.003,0.000)     & -67.98  & (0.271,0.006,0.578) & 6.08  \\
   \hline
 \end{tabular}
 \label{table1}
\end{table*}

\begin{figure*}
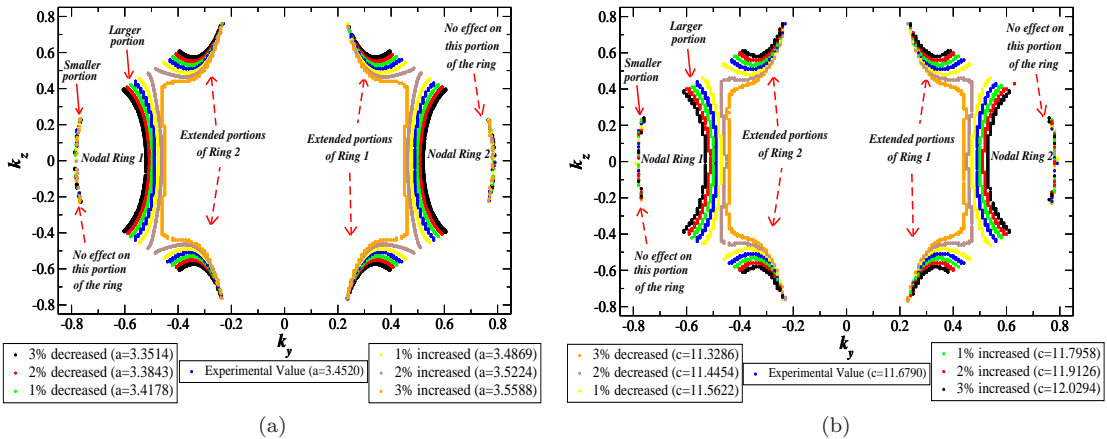

    \centering
    \subfigure[]
    {
        \includegraphics[width=0.40\linewidth, height=5.3cm]{fig4a.eps}
    }
    \subfigure[]
    {
        \includegraphics[width=0.40\linewidth, height=5.3cm]{fig4b.eps}
    }\vspace*{0.1in}
    
    \caption
    { {{\footnotesize (Color online) Plot (a) ((b)) shows the effect of axial strain, applied along the $a$ ($c$) direction, on the nodal-lines situated in $k_x$=0 plane of NbAs.}}
    }
    \label{fig_cone1_taas}
\end{figure*}

\subsection{Evolution of \textit{nodal-arc} to Weyl-nodes}

In the presence of SOC, the \textit{nodal-arcs} degeneracy in NbAs \& NbP gets destroyed. The magnitude of energy-gap created along these arcs varies along the $k_y$ direction, showing a different trend when compared to the case in TaAs \& TaP. Unlike TaAs \& TaP\cite{pandey2023existence}, where the energy-gap initially decreases, reaching a minimum non-zero value before increasing again to a maximum value as $k_y$ increases\cite{pandey2023py}, in NbAs \& NbP, it first increases to a local-maximum value (NbAs: 15.2 meV \& NbP: 24.2 meV) and then decreases to a non-zero minimum. Subsequently, it rises to a maximum value (NbAs: 46.9 meV \& NbP: 38.9 meV) at the end of the \textit{nodal-arc} along the $k_y$ direction. Comparing these results with TaAs \& TaP, it is found that the value of maximum energy gap created due to SOC follows the trend: TaAs (162.0 meV)\cite{pandey2023existence}$>$TaP (143.7 meV)\cite{pandey2023existence}$>$NbAs (46.9 meV)$>$NbP (38.9 meV). It is important to note that this trend is in accordance to the order of SOC-strength associated with these WSMs. The quatitative comparision of this feature in NbAs \& NbP is shown in Fig. \ref{fig33}. Before moving further, it is important to highlight that the \textit{nodal-arcs} in NbAs \& NbP, as already mentioned, are extended beyond the boundaries of the first BZ. Thus, in further discussion, the term \textit{nodal-arc} will refer to the one obtained when the extended portion is also mapped into the first BZ. Moving further, although SOC leads to the vanishing of \textit{nodal-arc}, it causes the formation of two pairs of Weyl nodes close to the plane containig the \textit{nodal-arc}. The two nodes forming each Weyl pair are found to be situated on either sides of the plane containing the \textit{nodal-arc} and also perpendicular to it. Out of the two pairs, one is positioned on either side of the middle portion of the \textit{nodal-arc} (called as W2 points), while the other pair is situated at the junction point of the two arcs forming a nodal-line (referred as W1 points). This second pair shares contributions from both arcs, resulting in a total of three pairs of Weyl nodes per nodal-line and 12 pairs of Weyl nodes in the entire BZ. These findings closely parallel to those observed in TaAs \& TaP\cite{pandey2023existence}. Detailed figures illustrating these features for NbAs \& NbP can be found in supplementary materials (Fig. S3). Furthermore, the Fermi-level-scaled energy of representative W1 \& W2 points in NbAs \& NbP are mentioned in table \ref{table1}. \textit{Grassano} \textit{et.al.} have previously investigated the energy of these nodes in TaAs class of WSMs\cite{grassano2020influence}. For NbAs (NbP) they reported the energy of W1 point as -33 (-56) meV whereas for W2, it is 4 (26) meV with respect to the Fermi energy. However, in present analysis, the energy of W1 points is found to be $\sim$-40 ($\sim$-68) meV while for W2 it is obtained as $\sim$13 ($\sim$8) meV with respect to Fermi energy in NbAs (NbP). The differences in the results of present work with that of \textit{Grassano} \textit{et.al.} is probably because they have used $\textit{\textbf{k}}$-mesh gridding method to search the Weyl nodes while in the present analysis, the highly efficient \textit{PY-Nodes} code is employed for searching these bands-touching points. On the top of that, \textit{Grassano} \textit{et.al.} have used the pseudopotential methods as implemented in the Quantum expresso\cite{giannozzi2009quantum} package while in the present analysis, full-potential Linearized augmented-plane-wave method (FP-LAPW) method as implemented in WIEN2k package is used. This FP-LAPW method is considered to be the most accurate one in the research community. Thus, the results of the present analysis can be considered as more reliable and accurate.

The next point of discussion is whether the strength of SOC affects the number of Weyl-nodes into which these arcs evolve. As the SOC Hamiltonian ($H_{so}$) is inversely proportional to the $c^2$\cite{blaha2001wien2k}, value of $c$ is decreased in our calculations to increase the strength of SOC. The method is similar to the one as followed in our previous work\cite{pandey2023existence}. The SOC-strength per Nb atom is calculated by finding the difference between the ground state energy obtained before and after adding the SOC-effect in our calculations. The obtained results are shown in table \ref{table1}. It is seen that as the value of $c$ decreases from $\sim$137 to $\sim$123, the strength of SOC in NbAs (NbP) increases from $\sim$0.08 ($\sim$0.06) eV to $\sim$0.11 ($\sim$0.09) eV. This increase is found to be extremely small when compared with increase in SOC strength in TaAs ($\sim$0.7 eV) \& TaP ($\sim$3.0 eV) for the same change in value of $c$\cite{pandey2023existence}. Similar to the case of TaAs \& TaP, the evolution of \textit{nodal-arc} into two pairs of Weyl-nodes in NbAs \& NbP is found to be robust against the given increase in the magnitude of SOC-strength. That is, the nodes forming a Weyl pair are still found to be situated on either sides of the plane containing the \textit{nodal-arc} and also at equidistant from the plane. The coordinates and energy of one node from each pair is mentioned in table \ref{table1}. The coordinated of obtained nodes are also found to be robust to the given increase in SOC strength of NbAs \& NbP, which is similar to the case as observed in TaAs \& TaP\cite{pandey2023existence}. As the increase in the SOC strength in NbAs \& NbP is extremenly small, it is expected that the energy of Weyl nodes in these semimetals will be less sensitive when compared to that of TaAs \& TaP. The obtained results as mentioned in table \ref{table1} reflects these aspects.

\setlength{\parindent}{3em}
\setlength{\parskip}{0.2em}

\subsection{Effect of axial strain}

Understanding the effects of strain along different crystallographic directions is crucial for tailoring the electronic properties of TaAs class of WSMs. The study of impact of strain on nodal lines is carried-out under two different cases: (i) strain created along $a$ direction and (ii) strain created along the $c$ direction. Fig. \ref{fig_cone1_taas} illustrates the results for nodal lines situated in the $k_x$=0 plane of NbAs. For clarity, we divide a nodal line into three parts: a larger portion, a smaller portion, and an extended portion forming beyond the boundaries of the first BZ, as indicated in the figure. It is found that strain (either tensile or compressive) along the $a$ or $c$ direction has no effect on the smaller portion of the rings. However, it significantly impacts the larger and extended portions. Increasing strain magnitude along the $a$ direction enlarges these portions, while the opposite trend occurs with strain along the $c$ direction. Notably, a 3\% tensile strain along $a$ (or $\geq$2\% compressive strain along $c$) leads to the partial merging of nodal lines 1 and 2 in the extended BZ. This merging is schematically depicted in Fig. S4 of supplementary materials. Similar observations hold for nodal lines in the $k_y$=0 plane. Also, nodal-lines in the $k_x$=0 plane of NbP, TaAs \& TaP are also found to be sensitive to the creation of such strain (see Figs. S5, S6 \& S7 of supplementary materials). It is important to note that partial merging of rings at specific strain values (in between -3\% to 3\%) is observed only in NbAs \& NbP but not in TaAs \& TaP. This difference may stem from the larger area covered by nodal lines in NbAs \& NbP at experimental lattice parameters when compared with that of TaAs \& TaP. Consequently, further ring enlargement upon creating a strain in NbAs \& NbP brings the rings closer in the extended BZ, resulting in their partial merging. This discussion establishes the effect of lattice strain on nodal lines in TaAs class of WSMs (under no SOC). Further discussion will investigate the impact of SOC on these strained nodal lines.

Fig. \ref{fig_cone3_taas}(a) (\ref{fig_cone3_taas}(b)) depict the influence of SOC on nodal lines in the $k_x$=0 plane of NbAs under compressive (tensile) strain along the $a$ direction. In Fig. \ref{fig_cone3_taas}(c) (\ref{fig_cone3_taas}(d)), the effect of SOC on these rings is shown for compressive (tensile) strain along the $c$ direction. It is seen in these figures that some of the nodes (clustered at discrete places of the BZ) are encircled in blue dashed-circles. These nodes are forming at the points in the $k_x$=0 plane, where end portions of \textit{nodal-arcs} existed in the absence of SOC. One must not confuse these nodes (encircled in blue-dashed circles) with the nodal-lines, because nodal-lines generally form due to the crossing of two bands to form a closed loop-like structure. However, these nodes (encircled in blue-dashed circles) do not form a closed loop. Instead, they are discrete bands-degenerate points forming in the $k_x$=0 plane. Apart from these nodes, there are also other node-points that are created away from the $k_x$=0 plane as shown in Fig. \ref{fig_cone3_taas}. One must note that each red dots that are not encircled in a dashed-circle denotes two nodes, which are getting projected at the same point when shown in 2D plots. Next, we shall discuss how the number of nodes get affected on creating strain along $a$ direction. One must keep in mind that for the experimental lattice parameter, nodal-lines of NbAs in $k_x$=0 plane evolve into 6 pairs of nodes off to the $k_x$=0 plane, as discussed before.  

Fig. \ref{fig_cone3_taas} (a) shows that compressive strain of 1\% or 2\% or 3\% along $a$ direction has no effect on the number of nodes created in $k_x\neq$0 plane. However, with increasing compressive strain, new nodes are found to get created within the $k_x$=0 plane, particularly where the ends of \textit{nodal-arcs} existed under no-SOC case. Now moving to the effect of tensile strain along $a$ direction (shown in Fig. \ref{fig_cone3_taas} (b)), it is observed that the 4 pairs of nodes that used to be formed in the $k_z\neq$0 plane under no-strain, get vanished upon the application of 1\% tensile strain. However, the two pairs of nodes are still generated in the $k_z$=0 plane as used to be in the no-strain case. With further increase in the strain, the number of nodes in $k_z\neq$0 plane starts increasing along with maintaining the number of nodes in $k_z$=0 plane as was in no-strain case. The total number of nodes generated from the nodal-lines in the $k_x$=0 plane under SOC-effect along with 2\% (3\%) strain in $a$ direction is found to be 20 (28). Regarding the impact of compressive strain on node energy, it is observed in Fig. \ref{fig_cone3_taas} (a) that nodes in the $k_z$=0 plane become increasingly negative in energy relative to the Fermi energy as the magnitude of compressive strain along the $a$ direction increases. Similar behavior is observed for nodes situated within the $k_x$=0 plane (encircled in blue-dashed circles). Conversely, nodes not encircled and located in $k_z\neq$0 planes approach the Fermi energy from the positive energy direction with increasing compressive strain magnitude. Moreover, the energy of nodes with respect to the Fermi-energy, obtained after tensile strain are also found to be sensitive to the magnitude of strain. Nextly, the effect of strain along $c$ direction on the number and energy of nodes is discussed.

Fig. \ref{fig_cone3_taas}(c) illustrates the impact of compressive strain (1\%, 2\%, and 3\%) along the $c$ direction on the number of nodes in NbAs. The strain of 1\% have no effect on the number of nodes generated from nodal-lines in the $k_x$=0 plane. However, at 2\% strain, the number increases to 28. Notably, 2\% compressive strain along $c$ has no effect on the number of nodes in the $k_z$=0 plane. At 3\% strain, the extra nodes vanish, reverting to the 1\% compressive strain scenario. Additionally, the effect of tensile strain along the $c$ direction is discussed in Fig. \ref{fig_cone3_taas}(d). Strain percentages of 1\%, 2\%, and 3\% have no effect on the number of nodes generated off the $k_x$=0 plane. However, the number of nodes in the $k_x$=0 plane increases with higher tensile strain percentages, clustering at the regions where the ends of \textit{nodal-arcs} existed when SOC was ignored. Similar to strain along the $a$ direction, increasing compressive or tensile strain along $c$ direction leads to nodes in NbAs spreading further apart in energy and with respect to the Fermi level, as depicted in Figs. \ref{fig_cone3_taas}(c) and \ref{fig_cone3_taas}(d).

It is important to note here that as the nodal-lines situated in $k_y$=0 plane are related with crystal-symmetry to those situated in the $k_x$=0 plane, similar observations as discussed above will be obtained for the effect of SOC on the nodal-lines in $k_y$=0 plane of NbAs. Moreover, similar effect of lattice strain along $a$ and $c$ direction of unit-cell is obtained on the nodal-lines and Weyl nodes of other family of TaAs class of WSMs (\textit{i.e.,} for TaAs, TaP \& NbP). Corresponding results are shown in supplementary materials (Figs. S8, S9 \& S10). However, as already mentioned, for strain of -3\% to 3\% along $a$ or $c$ direction do not lead to the partial merging of two nodal-lines in a given plane for TaAs \& TaP.

\begin{figure*}
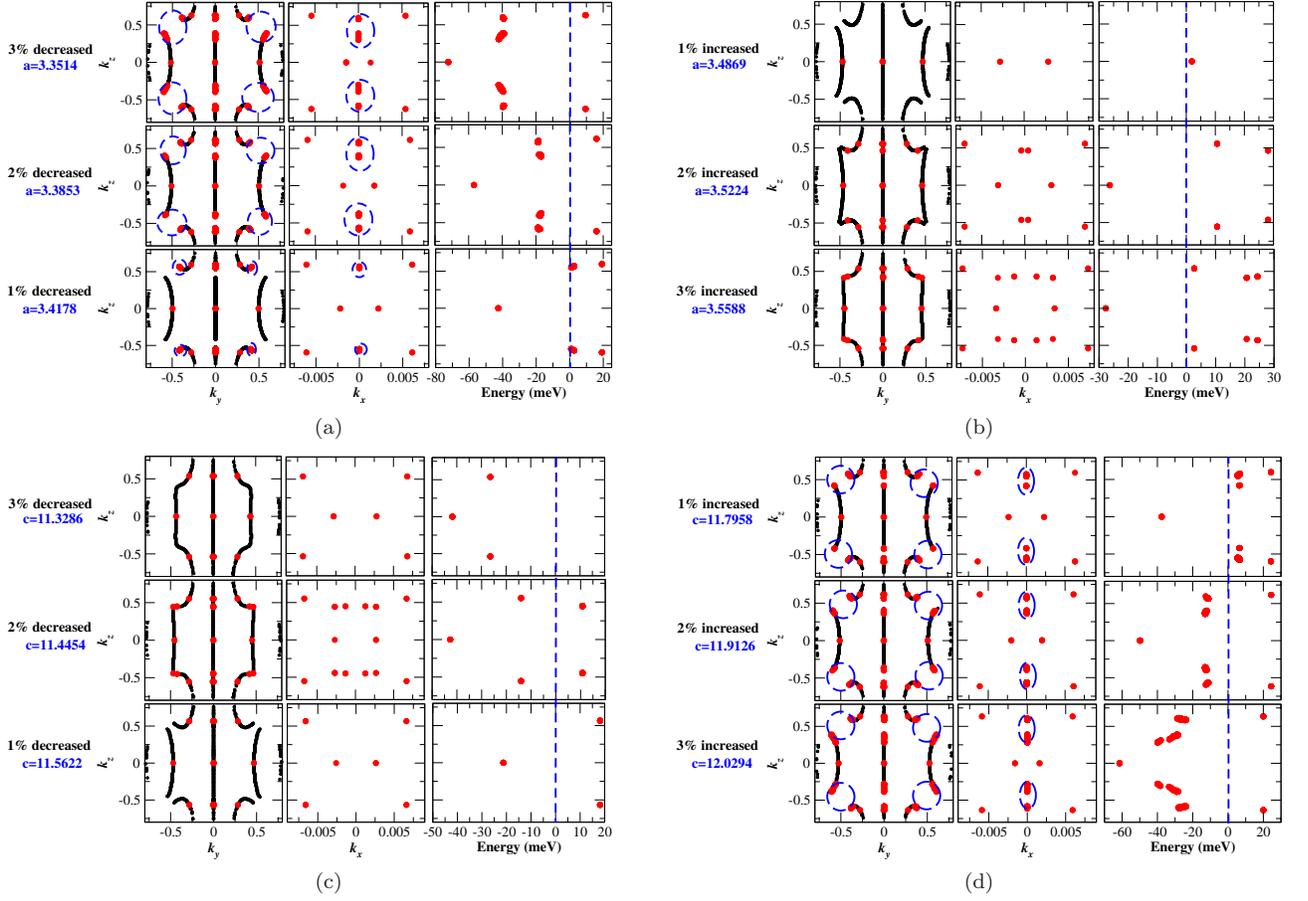

    \centering
    \subfigure[]
    {
        \includegraphics[width=0.45\linewidth, height=5.4cm]{fig5a.eps}\hspace*{0.2in}
    }
    \subfigure[]
    {
        \includegraphics[width=0.45\linewidth, height=5.4cm]{fig5b.eps}
    }
    \subfigure[]
    {
        \includegraphics[width=0.45\linewidth, height=5.4cm]{fig5c.eps}\hspace*{0.2in}
    }
    \subfigure[]
    {
        \includegraphics[width=0.45\linewidth, height=5.4cm]{fig5d.eps}
    }
    \caption
    { {{\footnotesize (Color online) Plots (a) \& (b) ((c) \& (d)) shows the position of node points and their energy with respect to Fermi energy (scalled to 0 meV shown in dashed blue line) when the strain is applied along $a$ ($c$) direction in NbAs.}}
    }
    \label{fig_cone3_taas}
\end{figure*}

\subsection{Weyl-contributed Seebeck Coefficient}

For studying the Seebeck coefficient ($S$) behaviour of any given material, it is important to know the physical parameters upon which it depends. Within the Kubo linear-response formalism, expression for $S$ is defined as\cite{tomczak2010thermopower,oudovenko2006electronic}-
\begin{equation}
S^{\alpha\alpha\textquoteright}=-\frac{k_B}{|e|}\frac{K_1^{\alpha\alpha\textquoteright}}{K_0^{\alpha\alpha\textquoteright}}
\label{E1}
\end{equation}
where $K_n^{\alpha\alpha\textquoteright}$ are called the kinetic coefficients. Their expression is given by\cite{sihi2023track}-
\begin{equation}
K_n^{\alpha\alpha\textquoteright}=N_{sp}\pi\hbar\int d\omega(\beta\omega)^nf(\omega)f(-\omega) \Gamma^{\alpha\alpha\textquoteright}(\omega,\omega)
\label{E2}
\end{equation}
where, $N_{sp}$ and $f(\omega)$ are the spin factor and Fermi function, respectively. $\hbar$ denotes the reduced Planck\textquoteright s constant\cite{sihi2023track}. Here, $\Gamma^{\alpha\alpha\textquoteright}(\omega_1,\omega_2)$ is known as transport distribution function and is defined as-
\begin{equation}
\Gamma^{\alpha\alpha\textquoteright}(\omega_1,\omega_2)=\frac{1}{V}\sum_{k}Tr\{ v_n^\alpha(\textbf{\textit{k}})A(\textbf{\textit{k}},\omega_1)v_n^{\alpha\textquoteright} (\textbf{\textit{k}})A(\textbf{\textit{k}},\omega_2)\} 
\label{E3}
\end{equation}
The term $A(\textbf{\textit{k}},\omega)$, known as spectral function in real $\omega$, is expressed as\cite{imada1998metal}-
\begin{equation}
A(\textbf{\textit{k}},\omega)=-\frac{1}{\pi}\frac{Im\Sigma(\omega)}{[\omega-\varepsilon_\textbf{\textit{k}}^0-Re\Sigma(\omega)]^2+[Im\Sigma(\omega)]^2} 
\label{E4}
\end{equation}
Here, $\omega$ is the energy with respect to the chemical potential ($\mu$) where the transport is supposed to be calculated. In the present work, the $S$ is calculated by the setting $\mu$ at the energy of Weyl node. Nextly, as the topological and transport analysis in the present work is carried out at DFT level, a slight modification in the above formalism is required to use it for studying the behaviour of $S$ obtained from Weyl cones of TaAs class of WSMs. In the Eq. \ref{E4}, the $Re\Sigma(\omega)$ must be taken as 0 while the value of $Im\Sigma(\omega)$ should be taken to be extremely small. As \textit{TRACK} code\cite{sihi2023track} provides such options, we have used it in our present study. The value of $Im\Sigma(\omega)$ is taken as 0.5 meV in our calculations. Before discussing the behaviour of $S$ obtained from Weyl cones, we are motivated in exploring that under what conditions will a Weyl cone contribute to a non-zero value of $S^{\alpha\alpha\textquoteright}$ to a given WSM. This aspect is discussed further.

\begin{figure*}
    \centering
    \subfigure[]
    {
        \includegraphics[width=0.25\linewidth, height=4.4cm]{fig6a.eps}
    }
    \subfigure[]
    {
        \includegraphics[width=0.35\linewidth, height=4.4cm]{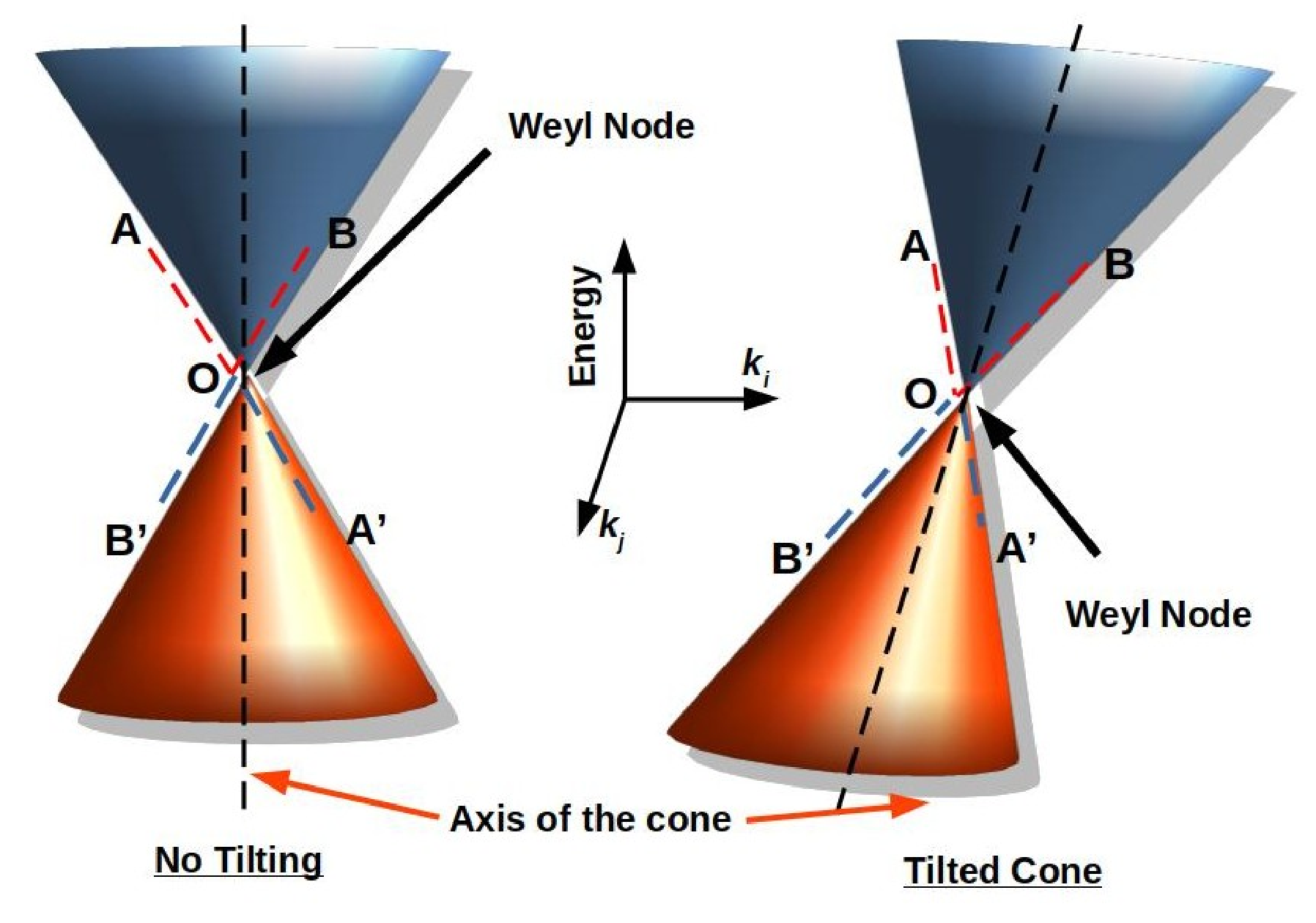}
    }
    \subfigure[]
    {
        \includegraphics[width=0.35\linewidth, height=4.4cm]{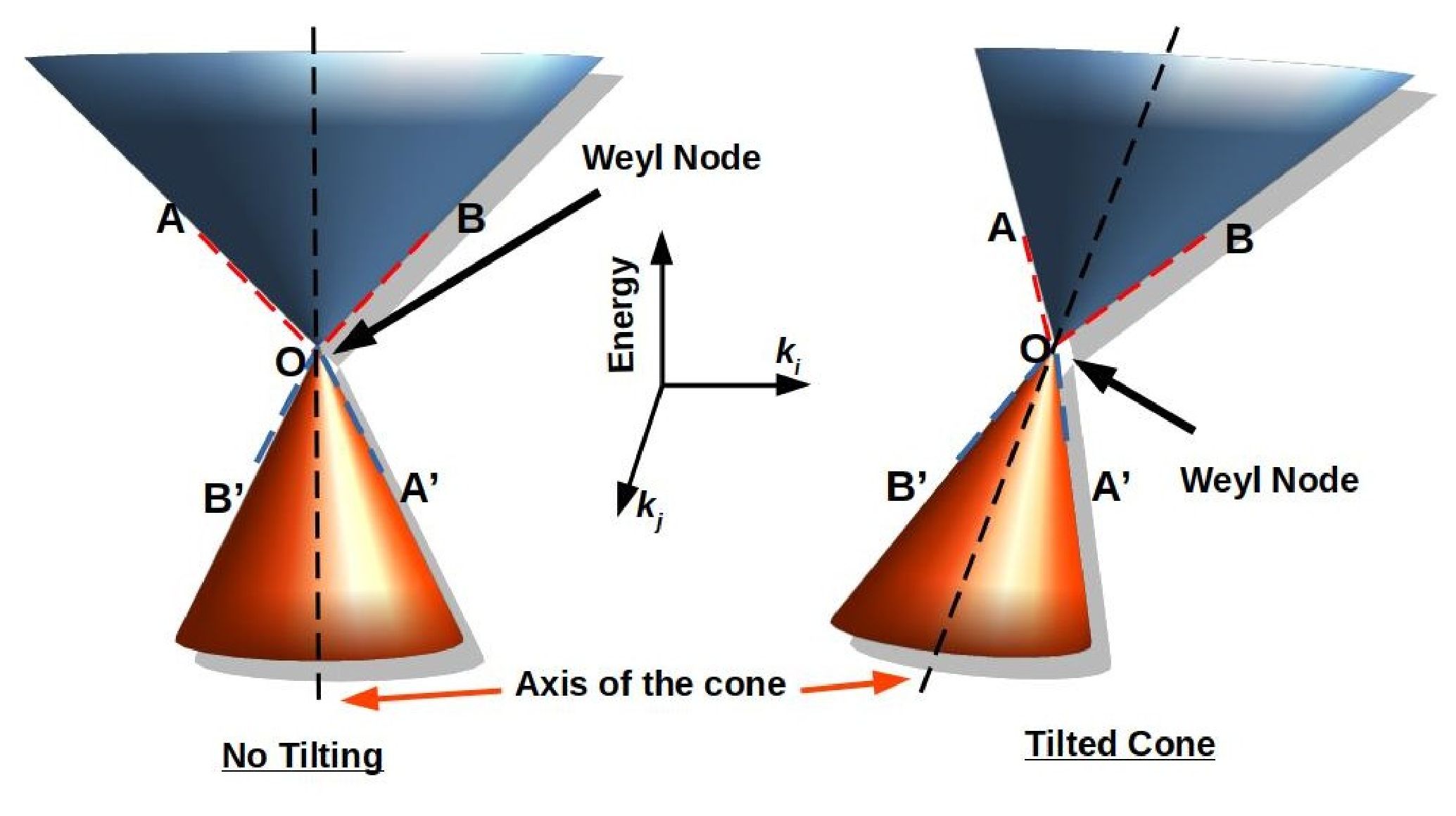}
    }\vspace*{0.1in}
    
    \caption
    { {{\footnotesize (Color online) Plot (a) shows the temperature-dependent curves of $\beta\omega f(\omega)f(-\omega)$ \& $f(\omega)f(-\omega)$ around a Weyl node. Plots (b) and (c) shows the symmetric and asymmetric Weyl cones, respectively.}}
    }
    \label{fig_cone}
\end{figure*}

To understand the contribution of Weyl cone to the $S$ of a given materials, it becomes extremely important to analyse the minute aspects of the Weyl cone. In this regard, the two most important aspects of the cone are (i) \textit{tilt of the Weyl cone} and (ii) \textit{asymmetry of the Weyl cone}. The term \textquoteleft \textit{tilt of the cone}\textquoteright\hspace*{0.1cm}  means that how much angle does the axis of the Weyl cone makes with the energy axis. Thus, if the axis of the Weyl cone is parallel with the energy axis, the Weyl cone is said to be not at all tilted (see Figs. \ref{fig_cone} (b) and (c)). Moving further, in this work the term  \textquoteleft \textit{symmetric cone}\textquoteright\hspace*{0.1cm} stands for the one which possess either the inversion symmetry with respect to the Weyl node or mirror symmetry with respect to plane perpendular to energy axis and passing through the Weyl node. To make it more clear, let us explain the meaning of \textit{asymmetry of the Weyl cone} using Fig. \ref{fig_cone}. If the cone is symmetric, then on applying inversion symmetry operation (with respect to Weyl node) or mirror symmetry operation (with respect to plane perpendular to energy axis and passing through Weyl node) on the red cone, full blue cone can be obtained. In the other case, the cone is said to be asymmetric. In addition to this, in the present work the term \textit{ideal cone} stands for the case in which the Weyl cone is symmetric and not tilted. In further discussions, the condition under which the Weyl cone will have non-zero contribution to $S$ is explained.

Firstly, consider the case of ideal cone as shown in Fig. \ref{fig_cone} (b), the non-tilted cone. It is well known that the velocity $v_n^\alpha(\textbf{\textit{k}})$ is defined as (1/$\hbar$)($\partial\varepsilon_n (\textbf{\textit{k}}) /\partial k_\alpha$)\cite{madsen2006boltztrap}. As the red \& blue cones are symmetric, corresponding to magnitude of velocity of the red cone at each $\textbf{\textit{k}}$-point, there will be a $\textbf{\textit{k}}$-point where the blue cone will have the same magnitude of velocity. If the Weyl cone is inversion symmetric about the Weyl node, these two $\textbf{\textit{k}}$-points will be different. However, if the Weyl cone possess mirror symmetry about the energy plane passing through the energy of Weyl nodes, the red \& blue cone will have same magnitude of velocity at the same $\textbf{\textit{k}}$-point.  Moreover, corresponding to these $\textbf{\textit{k}}$-points, the value of $A_n(\textbf{\textit{k}},\omega)$ for both the bands forming the Weyl cone will also be same. This is because, as the cone is symmetric at these $\textbf{\textit{k}}$-points, they will possess same magnitude of $[\omega-\varepsilon_\textbf{\textit{k}}^0]^2$. Moving further, due to such behaviour of $v_n^\alpha(\textbf{\textit{k}})$ \& $A_n(\textbf{\textit{k}},\omega)$, the quantity $\Gamma^{\alpha\alpha\textquoteright}(\omega,\omega)$ will also become symmetric with respect to the energy of the Weyl node. As can be seen in Fig. \ref{fig_cone} (a), at any temperature, the term $\beta\omega f(\omega)f(-\omega)$ is an odd function of $\omega$. Due to this, the value of $K_1^{\alpha\alpha\textquoteright}$, as mentioned in equation \ref{E1}, will vanish out at any given temperature ($T$). Hence, an ideal Weyl cone will have zero contribution to the $S$ of a material possessing it. Now moving to the case of symmetric Weyl cone having non-zero tilt as shown in Fig. \ref{fig_cone} (b). The discussion for such cone will be exactly similar to case of an ideal Weyl cone as discussed above. The only difference is that it will have different magnitude of velocity at a given $\textbf{\textit{k}}$-point when compared to case when the cone was not tilted. This is because the tilt in the cone with respect to the energy axis modifies the value of velocity at each $\textbf{\textit{k}}$-points. Thus, in this case also there will be a compensating effect from the red \& the blue cones resulting into zero contribution to the value of $S$ at any given temperature. Now we are left with the case of asymmetric Weyl cone which is discussed further.

\begin{figure}
    \centering
    \subfigure[]
    {
        \includegraphics[width=0.45\linewidth, height=3.4cm]{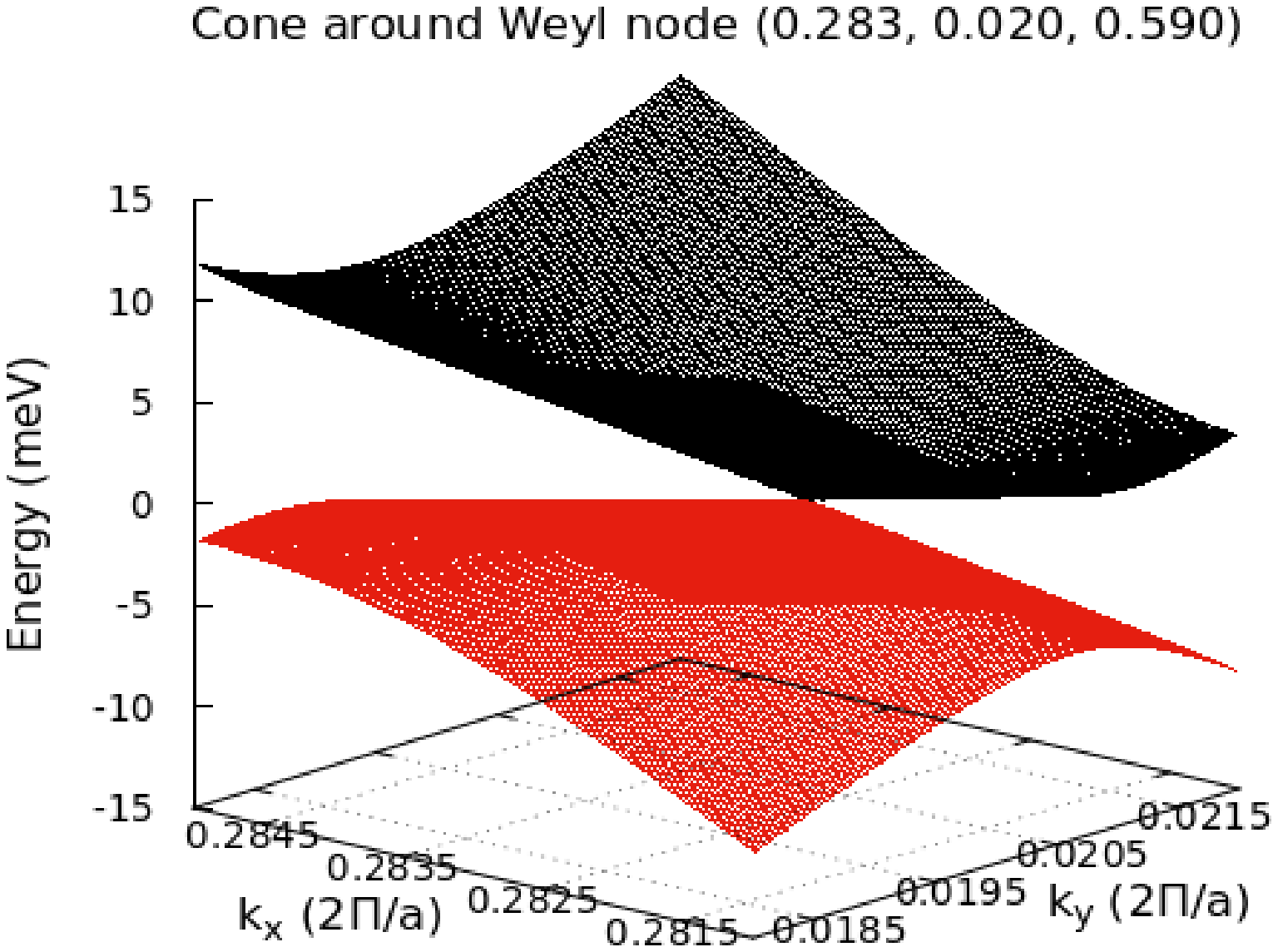}
    }
    \subfigure[]
    {
        \includegraphics[width=0.45\linewidth, height=3.4cm]{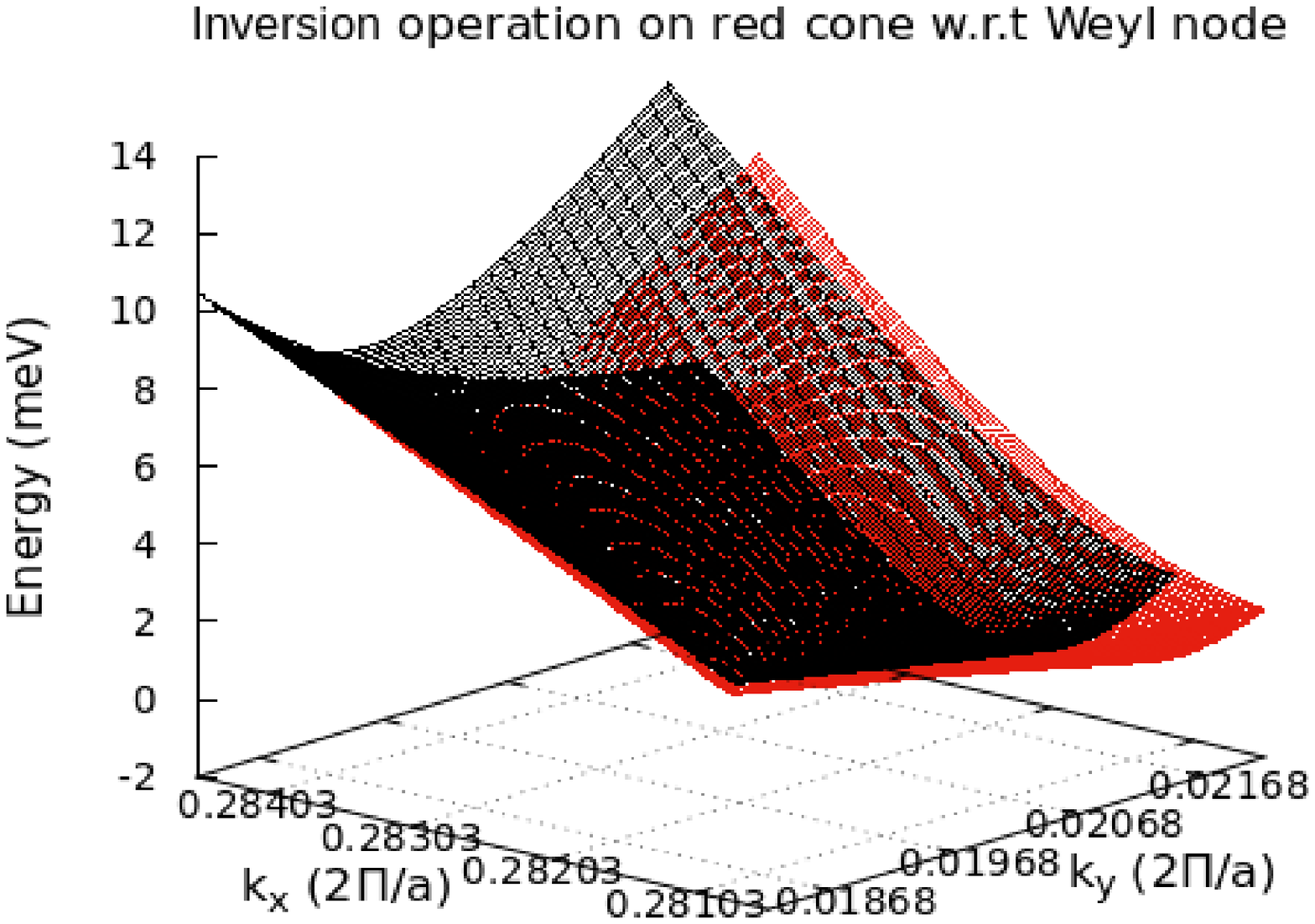}
    }\vspace*{0.1in}
    \subfigure[]
    {
        \includegraphics[width=0.45\linewidth, height=3.4cm]{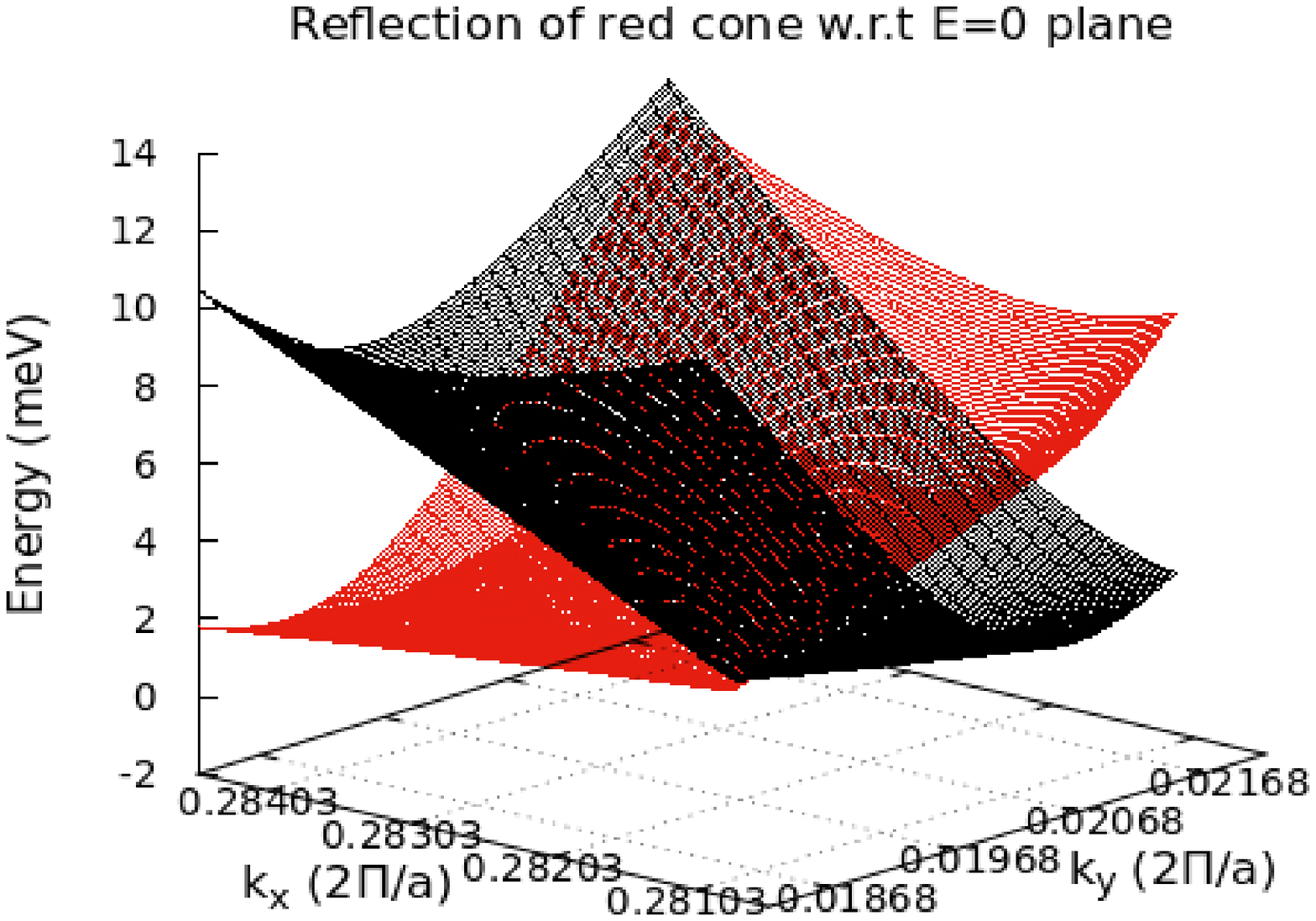}
    }
    \subfigure[]
    {
        \includegraphics[width=0.45\linewidth, height=3.4cm]{fig7d.eps}
    }
    \caption
    { {{\footnotesize (Color online) Plot (a) shows the Weyl cone around W2 point of TaAs. Plot (b) shows the extent to which red cone overlaps the balck one when inversion operation is applied on it with respect to the Weyl node. Plot (c) shows the extent to which red cone overlaps the balck one when mirror-reflection operation is applied on it with respect to the plane parallel to $k_x$-$k_y$ plane and passing through Weyl energy. Plot (d) shows the components of $S^{xx}$, $S^{yy}$ \& $S^{zz}$ contributed from this cone.}}
    }
    \label{fig7}
\end{figure}

For the asymmetric Weyl cone with no tilt, there will be some $\textbf{\textit{k}}$-points where corresponding to the magnitude of $v_n^\alpha(\textbf{\textit{k}})$ of red cone, there will not be a $\textbf{\textit{k}}$-point where the magnitude of $v_n^\alpha(\textbf{\textit{k}})$ of blue cone will be same with same magnitude of $A_n(\textbf{\textit{k}},\omega)$ for both the cones. Thus, for such a asymmetric Weyl cone there will be no compensating effect from the red \& the blue cone. This will result in a non-zero contribution to the value of $S$ at any given temperature. Moving further, for such an asymmetric Weyl cone, tilting of the cone will change the profile of temperature dependent value of $S^{\alpha\alpha}$. The reason behind this is that tilting of the cone will modify the value of $v_n^\alpha(\textbf{\textit{k}})$ and due to the non-compensating effect from red \& blue cones, the value of $S^{\alpha\alpha}$ will get affected. From these discussions, a general conclusion that can be drawn is that the basic requirements for a Weyl cone to have finite contribution to the $S$ of WSMs is its asymmetric nature. At the top of that, tilting of the cone may change the magnitude of contribution to the $S$ at any given T. In the further discussion, this conclusion is validated over the TaAs class of WSMs by analysing their Weyl cones and calculating the contribution of such cones to $S$ of these semimetals.

It is well-known that TaAs class of WSMs contain 24 Weyl nodes in their full BZ. These nodes are arranged in a particular fashion: 8 Weyl nodes in the $k_z$=0 plane and 8-8 nodes are also situated in a similar fashion on the $k_z\sim$0.58 \& $k_z\sim$-0.58 planes of these WSMs. In the present analysis of $S$, we have separately calculated $S$ for 3 different nodes for each semimetals, taking one from each plane. Also we have examined the features of each of these Weyl cones. The results corresponding to Weyl node (0.283, 0.020, 0.590) in case of TaAs is shown in Fig. \ref{fig7}. Let us first discuss the features of Weyl cone around this node (shown in Fig. \ref{fig7}(a)). The cone is drawn by taking the $\textbf{\textit{k}}$-points sampled in $k_x$-$k_y$ plane. In the preliminary analysis of the cone, it is seen that the axis of the cone is tilted with respect to the energy axis. Further analysis reveal that the cone is assymmetric for symmetry operations: (i) inversion symmetry with respect to Weyl node and (ii) mirror symmetry with respect to energy plane corresponding to Weyl energy. Corresponding plots are shown in Figs. \ref{fig7}(b) and (c), respectively. Thus, according to our previous discussion, the cone must have non-zero contribution to temperature dependent $S$. It is important to note here that such an asymmetry is found corresponding to all the three cones (in the above mentioned three planes) and also in such cones for other members of TaAs class of WSMs. The relevant plots are shown in supplementary materials (Figs. S11-S21). Now, we are interested in studying the magnitude of $S$ obtained from the above mentioned Weyl cone in TaAs, which is discussed below.

Generally, a Weyl cone is visualised as a 3-dimensional figure in the reciprocal space where the \textit{x}-\textit{y} plane consists of states and the \textit{z}-axis contains the energy of the two bands forming the cone. For studying the Weyl dominated transport, it is needed to consider the excitations only from the states close to the Weyl nodes. However, one must keep in mind that passing through a Weyl-node, infinite number of 2-dimensional plane can be constructed which will produce the Weyl cones (as all these planes will contain the Weyl node). Thus, for calculating the Weyl-contributed transport coefficients such as $S$, it is necessary to consider the states contained in all these planes. Hence, in the present work, a 3-dimensional \textit{\textbf{k}}-points sampling around the Weyl nodes is performed to calculate the $S$. The results obtained for Weyl-contributed $S$ to the TaAs form the Weyl node (0.283, 0.020, 0.590) is shown in Fig. \ref{fig7}(d). $S$ obtained corresponding to the Weyl cone is found to be anisotropic. That is, the $S^{xx}$, $S^{yy}$ and $S^{zz}$ components are all differently contributed from the Weyl cone. This occurs because in general, the $v_n^{x}$$\neq$$v_n^{y}$$\neq$$v_n^{z}$ for the states comprising the two bands forming Weyl cones in these WSMs. Another important feature that can be seen is that the magnitude of each components of $S$ (\textit{i.e.,} $S^{xx}$, $S^{yy}$ \& $S^{zz}$) changes in a non-monotonous fashion with the rise in the value of $T$. It is found that with the rise in temperature, each components attains a peak at some value of $T$. With the further increase in the magnitude of temperature, the value of each components of $S$ is found to decrease. The observation of such behavious of each components of $S$ is found to be common for the $S$ contributed from the Weyl cones corresponding to the nodes in $k_z$=-0.590 and $k_z$=0 planes in TaAs (see Figs. S11 \& S12). Furthermore, similar observations have been found for the Weyl cones in case of other members of TaAs class of WSMs, \textit{i.e.}, NbAs, NbP \& TaP. The relevant plots are shown in supplementary material (Figs. S13-S21). With this analysis, the fact has been successfully validated that the assymmetric nature of the Weyl cone is necessary for it to affect the $S$ of any WSM.

\section{Conclusions} 

This work confirms the presence of dispersive \textit{nodal-arcs} in NbAs and NbP Weyl semimetals (WSMs), when the effect of spin-orbit coupling (SOC) is ignored. These arcs result in the formation of four nodal-lines when crystal symmetries, including time-reversal symmetry, are applied. The \textit{nodal-arcs} are found to arise from band inversion involving Nb-4\textit{d} and As-4\textit{p} (P-3\textit{p}) orbitals and are highly sensitive to changes in hybridization-strength. Under the influence of SOC, each \textit{nodal-arc} are found to evolve into two pairs of Weyl nodes, out of which one pair is coming from the combined contribution of the two arcs forming a ring. The energy of Weyl nodes in these WSMs are found to be sensitive to the change in SOC-strength. An increase in SOC-strength of NbAs (NbP) from $\sim$75 ($\sim$61) meV to $\sim$117 ($\sim$95) meV decreases the energy gap of W1-W2 points from $\sim$52 meV to $\sim$49 meV ($\sim$76 meV to $\sim$74 meV). These features of \textit{nodal-arc} and its evolution into Weyl nodes are found to be similar to the case in TaAs \& TaP\cite{pandey2023existence}. Additionally, it is obtained that application of strain along specific directions can affect nodal-lines and the number of nodes in TaAs class of WSMs. It is important to note that a 3\% tensile (compressive) strain along $a$ ($c$) direction leads to the partial merging of nodal-lines (under no SOC) in the extended Brillouin zone of NbAs \& NbP. This feature is found to be missing in TaAs \& TaP for strain-percentage from -3\% to 3\%. In the presence of SOC, the number of nodes in NbAs under 2\% (3\%) tensile-strain in $a$ direction is found to increase from 24 to 40 (56). Similar effect of strain on number of nodes is also obtained in TaAs, TaP \& NbP. Lastly, it has been established that a symmetric Weyl cone (even if tilted) have negligible contribution to the Seebeck coefficient ($S$) of Weyl semimetals (WSMs). However, asymmetric ones will contribute significantly, and their tilt influences magnitude of $S$. This conclusion regarding $S$ is validated over the TaAs class of WSMs, which contains assymmetric cones. The value of $S$ contributed from Weyl cone is found to be as large as $\sim$65 $\mu$\textit{V}/\textit{K} below 25 K in case of TaAs. These findings clarify the impact of Weyl physics on $S$ and demonstrate the potential for engineering topological properties of these WSMs through crystal strain.

\section{References}
\bibliography{MS_fourth}

\begin{thebibliography}{44}%
\makeatletter
\providecommand \@ifxundefined [1]{%
 \@ifx{#1\undefined}
}%
\providecommand \@ifnum [1]{%
 \ifnum #1\expandafter \@firstoftwo
 \else \expandafter \@secondoftwo
 \fi
}%
\providecommand \@ifx [1]{%
 \ifx #1\expandafter \@firstoftwo
 \else \expandafter \@secondoftwo
 \fi
}%
\providecommand \natexlab [1]{#1}%
\providecommand \enquote  [1]{``#1''}%
\providecommand \bibnamefont  [1]{#1}%
\providecommand \bibfnamefont [1]{#1}%
\providecommand \citenamefont [1]{#1}%
\providecommand \href@noop [0]{\@secondoftwo}%
\providecommand \href [0]{\begingroup \@sanitize@url \@href}%
\providecommand \@href[1]{\@@startlink{#1}\@@href}%
\providecommand \@@href[1]{\endgroup#1\@@endlink}%
\providecommand \@sanitize@url [0]{\catcode `\\12\catcode `\$12\catcode
  `\&12\catcode `\#12\catcode `\^12\catcode `\_12\catcode `\%12\relax}%
\providecommand \@@startlink[1]{}%
\providecommand \@@endlink[0]{}%
\providecommand \url  [0]{\begingroup\@sanitize@url \@url }%
\providecommand \@url [1]{\endgroup\@href {#1}{\urlprefix }}%
\providecommand \urlprefix  [0]{URL }%
\providecommand \Eprint [0]{\href }%
\providecommand \doibase [0]{http://dx.doi.org/}%
\providecommand \selectlanguage [0]{\@gobble}%
\providecommand \bibinfo  [0]{\@secondoftwo}%
\providecommand \bibfield  [0]{\@secondoftwo}%
\providecommand \translation [1]{[#1]}%
\providecommand \BibitemOpen [0]{}%
\providecommand \bibitemStop [0]{}%
\providecommand \bibitemNoStop [0]{.\EOS\space}%
\providecommand \EOS [0]{\spacefactor3000\relax}%
\providecommand \BibitemShut  [1]{\csname bibitem#1\endcsname}%
\let\auto@bib@innerbib\@empty
\bibitem [{\citenamefont {Pandey}\ and\ \citenamefont
  {Pandey}(2023{\natexlab{a}})}]{pandey2023existence}%
  \BibitemOpen
  \bibfield  {author} {\bibinfo {author} {\bibfnamefont {V.}~\bibnamefont
  {Pandey}}\ and\ \bibinfo {author} {\bibfnamefont {S.~K.}\ \bibnamefont
  {Pandey}},\ }\href@noop {} {\bibfield  {journal} {\bibinfo  {journal}
  {Journal of Physics: Condensed Matter}\ }\textbf {\bibinfo {volume} {35}},\
  \bibinfo {pages} {455501} (\bibinfo {year} {2023}{\natexlab{a}})}\BibitemShut
  {NoStop}%
\bibitem [{\citenamefont {Bernevig}\ \emph {et~al.}(2022)\citenamefont
  {Bernevig}, \citenamefont {Felser},\ and\ \citenamefont
  {Beidenkopf}}]{bernevig2022progress}%
  \BibitemOpen
  \bibfield  {author} {\bibinfo {author} {\bibfnamefont {B.~A.}\ \bibnamefont
  {Bernevig}}, \bibinfo {author} {\bibfnamefont {C.}~\bibnamefont {Felser}}, \
  and\ \bibinfo {author} {\bibfnamefont {H.}~\bibnamefont {Beidenkopf}},\
  }\href@noop {} {\bibfield  {journal} {\bibinfo  {journal} {Nature}\ }\textbf
  {\bibinfo {volume} {603}},\ \bibinfo {pages} {41} (\bibinfo {year}
  {2022})}\BibitemShut {NoStop}%
\bibitem [{\citenamefont {Vergniory}\ \emph {et~al.}(2019)\citenamefont
  {Vergniory}, \citenamefont {Elcoro}, \citenamefont {Felser}, \citenamefont
  {Regnault}, \citenamefont {Bernevig},\ and\ \citenamefont
  {Wang}}]{vergniory2019complete}%
  \BibitemOpen
  \bibfield  {author} {\bibinfo {author} {\bibfnamefont {M.}~\bibnamefont
  {Vergniory}}, \bibinfo {author} {\bibfnamefont {L.}~\bibnamefont {Elcoro}},
  \bibinfo {author} {\bibfnamefont {C.}~\bibnamefont {Felser}}, \bibinfo
  {author} {\bibfnamefont {N.}~\bibnamefont {Regnault}}, \bibinfo {author}
  {\bibfnamefont {B.~A.}\ \bibnamefont {Bernevig}}, \ and\ \bibinfo {author}
  {\bibfnamefont {Z.}~\bibnamefont {Wang}},\ }\href@noop {} {\bibfield
  {journal} {\bibinfo  {journal} {Nature}\ }\textbf {\bibinfo {volume} {566}},\
  \bibinfo {pages} {480} (\bibinfo {year} {2019})}\BibitemShut {NoStop}%
\bibitem [{\citenamefont {Kawabata}\ and\ \citenamefont
  {Ueda}(2022)}]{kawabata2022nonlinear}%
  \BibitemOpen
  \bibfield  {author} {\bibinfo {author} {\bibfnamefont {K.}~\bibnamefont
  {Kawabata}}\ and\ \bibinfo {author} {\bibfnamefont {M.}~\bibnamefont
  {Ueda}},\ }\href@noop {} {\bibfield  {journal} {\bibinfo  {journal} {Physical
  Review B}\ }\textbf {\bibinfo {volume} {106}},\ \bibinfo {pages} {205104}
  (\bibinfo {year} {2022})}\BibitemShut {NoStop}%
\bibitem [{\citenamefont {Tokura}\ \emph {et~al.}(2019)\citenamefont {Tokura},
  \citenamefont {Yasuda},\ and\ \citenamefont
  {Tsukazaki}}]{tokura2019magnetic}%
  \BibitemOpen
  \bibfield  {author} {\bibinfo {author} {\bibfnamefont {Y.}~\bibnamefont
  {Tokura}}, \bibinfo {author} {\bibfnamefont {K.}~\bibnamefont {Yasuda}}, \
  and\ \bibinfo {author} {\bibfnamefont {A.}~\bibnamefont {Tsukazaki}},\
  }\href@noop {} {\bibfield  {journal} {\bibinfo  {journal} {Nature Reviews
  Physics}\ }\textbf {\bibinfo {volume} {1}},\ \bibinfo {pages} {126} (\bibinfo
  {year} {2019})}\BibitemShut {NoStop}%
\bibitem [{\citenamefont {Gao}\ \emph {et~al.}(2019)\citenamefont {Gao},
  \citenamefont {Venderbos}, \citenamefont {Kim},\ and\ \citenamefont
  {Rappe}}]{gao2019topological}%
  \BibitemOpen
  \bibfield  {author} {\bibinfo {author} {\bibfnamefont {H.}~\bibnamefont
  {Gao}}, \bibinfo {author} {\bibfnamefont {J.~W.}\ \bibnamefont {Venderbos}},
  \bibinfo {author} {\bibfnamefont {Y.}~\bibnamefont {Kim}}, \ and\ \bibinfo
  {author} {\bibfnamefont {A.~M.}\ \bibnamefont {Rappe}},\ }\href@noop {}
  {\bibfield  {journal} {\bibinfo  {journal} {Annual Review of Materials
  Research}\ }\textbf {\bibinfo {volume} {49}},\ \bibinfo {pages} {153}
  (\bibinfo {year} {2019})}\BibitemShut {NoStop}%
\bibitem [{\citenamefont {Hu}\ \emph {et~al.}(2020)\citenamefont {Hu},
  \citenamefont {Ding}, \citenamefont {Gordon}, \citenamefont {Ghosh},
  \citenamefont {Tien}, \citenamefont {Li}, \citenamefont {Linn}, \citenamefont
  {Lien}, \citenamefont {Huang}, \citenamefont {Mackey} \emph
  {et~al.}}]{hu2020realization}%
  \BibitemOpen
  \bibfield  {author} {\bibinfo {author} {\bibfnamefont {C.}~\bibnamefont
  {Hu}}, \bibinfo {author} {\bibfnamefont {L.}~\bibnamefont {Ding}}, \bibinfo
  {author} {\bibfnamefont {K.~N.}\ \bibnamefont {Gordon}}, \bibinfo {author}
  {\bibfnamefont {B.}~\bibnamefont {Ghosh}}, \bibinfo {author} {\bibfnamefont
  {H.-J.}\ \bibnamefont {Tien}}, \bibinfo {author} {\bibfnamefont
  {H.}~\bibnamefont {Li}}, \bibinfo {author} {\bibfnamefont {A.~G.}\
  \bibnamefont {Linn}}, \bibinfo {author} {\bibfnamefont {S.-W.}\ \bibnamefont
  {Lien}}, \bibinfo {author} {\bibfnamefont {C.-Y.}\ \bibnamefont {Huang}},
  \bibinfo {author} {\bibfnamefont {S.}~\bibnamefont {Mackey}},  \emph
  {et~al.},\ }\href@noop {} {\bibfield  {journal} {\bibinfo  {journal} {Science
  advances}\ }\textbf {\bibinfo {volume} {6}},\ \bibinfo {pages} {eaba4275}
  (\bibinfo {year} {2020})}\BibitemShut {NoStop}%
\bibitem [{\citenamefont {Das}\ \emph {et~al.}(2019)\citenamefont {Das},
  \citenamefont {Di~Sante}, \citenamefont {Cilento}, \citenamefont {Bigi},
  \citenamefont {Kopic}, \citenamefont {Soranzio}, \citenamefont {Sterzi},
  \citenamefont {Krieger}, \citenamefont {Vobornik}, \citenamefont {Fujii}
  \emph {et~al.}}]{das2019electronic}%
  \BibitemOpen
  \bibfield  {author} {\bibinfo {author} {\bibfnamefont {P.}~\bibnamefont
  {Das}}, \bibinfo {author} {\bibfnamefont {D.}~\bibnamefont {Di~Sante}},
  \bibinfo {author} {\bibfnamefont {F.}~\bibnamefont {Cilento}}, \bibinfo
  {author} {\bibfnamefont {C.}~\bibnamefont {Bigi}}, \bibinfo {author}
  {\bibfnamefont {D.}~\bibnamefont {Kopic}}, \bibinfo {author} {\bibfnamefont
  {D.}~\bibnamefont {Soranzio}}, \bibinfo {author} {\bibfnamefont
  {A.}~\bibnamefont {Sterzi}}, \bibinfo {author} {\bibfnamefont
  {J.}~\bibnamefont {Krieger}}, \bibinfo {author} {\bibfnamefont
  {I.}~\bibnamefont {Vobornik}}, \bibinfo {author} {\bibfnamefont
  {J.}~\bibnamefont {Fujii}},  \emph {et~al.},\ }\href@noop {} {\bibfield
  {journal} {\bibinfo  {journal} {Electronic Structure}\ }\textbf {\bibinfo
  {volume} {1}},\ \bibinfo {pages} {014003} (\bibinfo {year}
  {2019})}\BibitemShut {NoStop}%
\bibitem [{\citenamefont {Bodo}\ \emph {et~al.}(2022)\citenamefont {Bodo},
  \citenamefont {Desmarais},\ and\ \citenamefont {Erba}}]{bodo2022spin}%
  \BibitemOpen
  \bibfield  {author} {\bibinfo {author} {\bibfnamefont {F.}~\bibnamefont
  {Bodo}}, \bibinfo {author} {\bibfnamefont {J.~K.}\ \bibnamefont {Desmarais}},
  \ and\ \bibinfo {author} {\bibfnamefont {A.}~\bibnamefont {Erba}},\
  }\href@noop {} {\bibfield  {journal} {\bibinfo  {journal} {Physical Review
  B}\ }\textbf {\bibinfo {volume} {105}},\ \bibinfo {pages} {125108} (\bibinfo
  {year} {2022})}\BibitemShut {NoStop}%
\bibitem [{\citenamefont {Armitage}\ \emph {et~al.}(2018)\citenamefont
  {Armitage}, \citenamefont {Mele},\ and\ \citenamefont
  {Vishwanath}}]{armitage2018weyl}%
  \BibitemOpen
  \bibfield  {author} {\bibinfo {author} {\bibfnamefont {N.}~\bibnamefont
  {Armitage}}, \bibinfo {author} {\bibfnamefont {E.}~\bibnamefont {Mele}}, \
  and\ \bibinfo {author} {\bibfnamefont {A.}~\bibnamefont {Vishwanath}},\
  }\href@noop {} {\bibfield  {journal} {\bibinfo  {journal} {Reviews of Modern
  Physics}\ }\textbf {\bibinfo {volume} {90}},\ \bibinfo {pages} {015001}
  (\bibinfo {year} {2018})}\BibitemShut {NoStop}%
\bibitem [{\citenamefont {Shekhar}\ \emph {et~al.}(2015)\citenamefont
  {Shekhar}, \citenamefont {Nayak}, \citenamefont {Sun}, \citenamefont
  {Schmidt}, \citenamefont {Nicklas}, \citenamefont {Leermakers}, \citenamefont
  {Zeitler}, \citenamefont {Skourski}, \citenamefont {Wosnitza}, \citenamefont
  {Liu} \emph {et~al.}}]{shekhar2015extremely}%
  \BibitemOpen
  \bibfield  {author} {\bibinfo {author} {\bibfnamefont {C.}~\bibnamefont
  {Shekhar}}, \bibinfo {author} {\bibfnamefont {A.~K.}\ \bibnamefont {Nayak}},
  \bibinfo {author} {\bibfnamefont {Y.}~\bibnamefont {Sun}}, \bibinfo {author}
  {\bibfnamefont {M.}~\bibnamefont {Schmidt}}, \bibinfo {author} {\bibfnamefont
  {M.}~\bibnamefont {Nicklas}}, \bibinfo {author} {\bibfnamefont
  {I.}~\bibnamefont {Leermakers}}, \bibinfo {author} {\bibfnamefont
  {U.}~\bibnamefont {Zeitler}}, \bibinfo {author} {\bibfnamefont
  {Y.}~\bibnamefont {Skourski}}, \bibinfo {author} {\bibfnamefont
  {J.}~\bibnamefont {Wosnitza}}, \bibinfo {author} {\bibfnamefont
  {Z.}~\bibnamefont {Liu}},  \emph {et~al.},\ }\href@noop {} {\bibfield
  {journal} {\bibinfo  {journal} {Nature Physics}\ }\textbf {\bibinfo {volume}
  {11}},\ \bibinfo {pages} {645} (\bibinfo {year} {2015})}\BibitemShut
  {NoStop}%
\bibitem [{\citenamefont {Chen}\ \emph {et~al.}(2016)\citenamefont {Chen},
  \citenamefont {Luo}, \citenamefont {Xiao}, \citenamefont {Lu}, \citenamefont
  {Zhang}, \citenamefont {Yang}, \citenamefont {Li}, \citenamefont {Pei},
  \citenamefont {Shao}, \citenamefont {Zhang} \emph
  {et~al.}}]{chen2016superconductivity}%
  \BibitemOpen
  \bibfield  {author} {\bibinfo {author} {\bibfnamefont {F.}~\bibnamefont
  {Chen}}, \bibinfo {author} {\bibfnamefont {X.}~\bibnamefont {Luo}}, \bibinfo
  {author} {\bibfnamefont {R.}~\bibnamefont {Xiao}}, \bibinfo {author}
  {\bibfnamefont {W.}~\bibnamefont {Lu}}, \bibinfo {author} {\bibfnamefont
  {B.}~\bibnamefont {Zhang}}, \bibinfo {author} {\bibfnamefont
  {H.}~\bibnamefont {Yang}}, \bibinfo {author} {\bibfnamefont {J.}~\bibnamefont
  {Li}}, \bibinfo {author} {\bibfnamefont {Q.}~\bibnamefont {Pei}}, \bibinfo
  {author} {\bibfnamefont {D.}~\bibnamefont {Shao}}, \bibinfo {author}
  {\bibfnamefont {R.}~\bibnamefont {Zhang}},  \emph {et~al.},\ }\href@noop {}
  {\bibfield  {journal} {\bibinfo  {journal} {Applied Physics Letters}\
  }\textbf {\bibinfo {volume} {108}} (\bibinfo {year} {2016})}\BibitemShut
  {NoStop}%
\bibitem [{\citenamefont {Lv}\ \emph {et~al.}(2015)\citenamefont {Lv},
  \citenamefont {Xu}, \citenamefont {Weng}, \citenamefont {Ma}, \citenamefont
  {Richard}, \citenamefont {Huang}, \citenamefont {Zhao}, \citenamefont {Chen},
  \citenamefont {Matt}, \citenamefont {Bisti} \emph
  {et~al.}}]{lv2015observation}%
  \BibitemOpen
  \bibfield  {author} {\bibinfo {author} {\bibfnamefont {B.}~\bibnamefont
  {Lv}}, \bibinfo {author} {\bibfnamefont {N.}~\bibnamefont {Xu}}, \bibinfo
  {author} {\bibfnamefont {H.}~\bibnamefont {Weng}}, \bibinfo {author}
  {\bibfnamefont {J.}~\bibnamefont {Ma}}, \bibinfo {author} {\bibfnamefont
  {P.}~\bibnamefont {Richard}}, \bibinfo {author} {\bibfnamefont
  {X.}~\bibnamefont {Huang}}, \bibinfo {author} {\bibfnamefont
  {L.}~\bibnamefont {Zhao}}, \bibinfo {author} {\bibfnamefont {G.}~\bibnamefont
  {Chen}}, \bibinfo {author} {\bibfnamefont {C.}~\bibnamefont {Matt}}, \bibinfo
  {author} {\bibfnamefont {F.}~\bibnamefont {Bisti}},  \emph {et~al.},\
  }\href@noop {} {\bibfield  {journal} {\bibinfo  {journal} {Nature Physics}\
  }\textbf {\bibinfo {volume} {11}},\ \bibinfo {pages} {724} (\bibinfo {year}
  {2015})}\BibitemShut {NoStop}%
\bibitem [{\citenamefont {Meng}\ \emph {et~al.}(2020)\citenamefont {Meng},
  \citenamefont {Zhang}, \citenamefont {He}, \citenamefont {Jin}, \citenamefont
  {Dai}, \citenamefont {Liu},\ and\ \citenamefont {Liu}}]{meng2020ternary}%
  \BibitemOpen
  \bibfield  {author} {\bibinfo {author} {\bibfnamefont {W.}~\bibnamefont
  {Meng}}, \bibinfo {author} {\bibfnamefont {X.}~\bibnamefont {Zhang}},
  \bibinfo {author} {\bibfnamefont {T.}~\bibnamefont {He}}, \bibinfo {author}
  {\bibfnamefont {L.}~\bibnamefont {Jin}}, \bibinfo {author} {\bibfnamefont
  {X.}~\bibnamefont {Dai}}, \bibinfo {author} {\bibfnamefont {Y.}~\bibnamefont
  {Liu}}, \ and\ \bibinfo {author} {\bibfnamefont {G.}~\bibnamefont {Liu}},\
  }\href@noop {} {\bibfield  {journal} {\bibinfo  {journal} {Journal of
  Advanced Research}\ }\textbf {\bibinfo {volume} {24}},\ \bibinfo {pages}
  {523} (\bibinfo {year} {2020})}\BibitemShut {NoStop}%
\bibitem [{\citenamefont {Yang}\ and\ \citenamefont
  {Zhang}(2016)}]{yang2016acoustic}%
  \BibitemOpen
  \bibfield  {author} {\bibinfo {author} {\bibfnamefont {Z.}~\bibnamefont
  {Yang}}\ and\ \bibinfo {author} {\bibfnamefont {B.}~\bibnamefont {Zhang}},\
  }\href@noop {} {\bibfield  {journal} {\bibinfo  {journal} {Physical review
  letters}\ }\textbf {\bibinfo {volume} {117}},\ \bibinfo {pages} {224301}
  (\bibinfo {year} {2016})}\BibitemShut {NoStop}%
\bibitem [{\citenamefont {Zeng}\ \emph {et~al.}(2021)\citenamefont {Zeng},
  \citenamefont {Nandy},\ and\ \citenamefont {Tewari}}]{zeng2021nonlinear}%
  \BibitemOpen
  \bibfield  {author} {\bibinfo {author} {\bibfnamefont {C.}~\bibnamefont
  {Zeng}}, \bibinfo {author} {\bibfnamefont {S.}~\bibnamefont {Nandy}}, \ and\
  \bibinfo {author} {\bibfnamefont {S.}~\bibnamefont {Tewari}},\ }\href@noop {}
  {\bibfield  {journal} {\bibinfo  {journal} {Physical Review B}\ }\textbf
  {\bibinfo {volume} {103}},\ \bibinfo {pages} {245119} (\bibinfo {year}
  {2021})}\BibitemShut {NoStop}%
\bibitem [{\citenamefont {Imran}\ and\ \citenamefont
  {Hershfield}(2018)}]{imran2018berry}%
  \BibitemOpen
  \bibfield  {author} {\bibinfo {author} {\bibfnamefont {M.}~\bibnamefont
  {Imran}}\ and\ \bibinfo {author} {\bibfnamefont {S.}~\bibnamefont
  {Hershfield}},\ }\href@noop {} {\bibfield  {journal} {\bibinfo  {journal}
  {Physical Review B}\ }\textbf {\bibinfo {volume} {98}},\ \bibinfo {pages}
  {205139} (\bibinfo {year} {2018})}\BibitemShut {NoStop}%
\bibitem [{\citenamefont {Zyuzin}\ and\ \citenamefont
  {Burkov}(2012)}]{zyuzin2012topological}%
  \BibitemOpen
  \bibfield  {author} {\bibinfo {author} {\bibfnamefont {A.}~\bibnamefont
  {Zyuzin}}\ and\ \bibinfo {author} {\bibfnamefont {A.}~\bibnamefont
  {Burkov}},\ }\href@noop {} {\bibfield  {journal} {\bibinfo  {journal}
  {Physical Review B}\ }\textbf {\bibinfo {volume} {86}},\ \bibinfo {pages}
  {115133} (\bibinfo {year} {2012})}\BibitemShut {NoStop}%
\bibitem [{\citenamefont {Ashby}\ and\ \citenamefont
  {Carbotte}(2014)}]{ashby2014chiral}%
  \BibitemOpen
  \bibfield  {author} {\bibinfo {author} {\bibfnamefont {P.~E.}\ \bibnamefont
  {Ashby}}\ and\ \bibinfo {author} {\bibfnamefont {J.}~\bibnamefont
  {Carbotte}},\ }\href@noop {} {\bibfield  {journal} {\bibinfo  {journal}
  {Physical Review B}\ }\textbf {\bibinfo {volume} {89}},\ \bibinfo {pages}
  {245121} (\bibinfo {year} {2014})}\BibitemShut {NoStop}%
\bibitem [{\citenamefont {Ojanen}(2013)}]{ojanen2013helical}%
  \BibitemOpen
  \bibfield  {author} {\bibinfo {author} {\bibfnamefont {T.}~\bibnamefont
  {Ojanen}},\ }\href@noop {} {\bibfield  {journal} {\bibinfo  {journal}
  {Physical Review B}\ }\textbf {\bibinfo {volume} {87}},\ \bibinfo {pages}
  {245112} (\bibinfo {year} {2013})}\BibitemShut {NoStop}%
\bibitem [{\citenamefont {Sun}\ \emph {et~al.}(2015)\citenamefont {Sun},
  \citenamefont {Wu},\ and\ \citenamefont {Yan}}]{sun2015topological}%
  \BibitemOpen
  \bibfield  {author} {\bibinfo {author} {\bibfnamefont {Y.}~\bibnamefont
  {Sun}}, \bibinfo {author} {\bibfnamefont {S.-C.}\ \bibnamefont {Wu}}, \ and\
  \bibinfo {author} {\bibfnamefont {B.}~\bibnamefont {Yan}},\ }\href@noop {}
  {\bibfield  {journal} {\bibinfo  {journal} {Physical Review B}\ }\textbf
  {\bibinfo {volume} {92}},\ \bibinfo {pages} {115428} (\bibinfo {year}
  {2015})}\BibitemShut {NoStop}%
\bibitem [{\citenamefont {Zhuo}\ \emph {et~al.}(2021)\citenamefont {Zhuo},
  \citenamefont {Lai}, \citenamefont {Yu}, \citenamefont {Yu}, \citenamefont
  {Ma}, \citenamefont {Lu}, \citenamefont {Liu}, \citenamefont {Liu},\ and\
  \citenamefont {Sun}}]{zhuo2021dynamical}%
  \BibitemOpen
  \bibfield  {author} {\bibinfo {author} {\bibfnamefont {X.}~\bibnamefont
  {Zhuo}}, \bibinfo {author} {\bibfnamefont {J.}~\bibnamefont {Lai}}, \bibinfo
  {author} {\bibfnamefont {P.}~\bibnamefont {Yu}}, \bibinfo {author}
  {\bibfnamefont {Z.}~\bibnamefont {Yu}}, \bibinfo {author} {\bibfnamefont
  {J.}~\bibnamefont {Ma}}, \bibinfo {author} {\bibfnamefont {W.}~\bibnamefont
  {Lu}}, \bibinfo {author} {\bibfnamefont {M.}~\bibnamefont {Liu}}, \bibinfo
  {author} {\bibfnamefont {Z.}~\bibnamefont {Liu}}, \ and\ \bibinfo {author}
  {\bibfnamefont {D.}~\bibnamefont {Sun}},\ }\href@noop {} {\bibfield
  {journal} {\bibinfo  {journal} {Light: Science \& Applications}\ }\textbf
  {\bibinfo {volume} {10}},\ \bibinfo {pages} {101} (\bibinfo {year}
  {2021})}\BibitemShut {NoStop}%
\bibitem [{\citenamefont {Gao}\ \emph {et~al.}(2020)\citenamefont {Gao},
  \citenamefont {Kaushik}, \citenamefont {Philip}, \citenamefont {Li},
  \citenamefont {Qin}, \citenamefont {Liu}, \citenamefont {Zhang},
  \citenamefont {Su}, \citenamefont {Chen}, \citenamefont {Weng} \emph
  {et~al.}}]{gao2020chiral}%
  \BibitemOpen
  \bibfield  {author} {\bibinfo {author} {\bibfnamefont {Y.}~\bibnamefont
  {Gao}}, \bibinfo {author} {\bibfnamefont {S.}~\bibnamefont {Kaushik}},
  \bibinfo {author} {\bibfnamefont {E.}~\bibnamefont {Philip}}, \bibinfo
  {author} {\bibfnamefont {Z.}~\bibnamefont {Li}}, \bibinfo {author}
  {\bibfnamefont {Y.}~\bibnamefont {Qin}}, \bibinfo {author} {\bibfnamefont
  {Y.}~\bibnamefont {Liu}}, \bibinfo {author} {\bibfnamefont {W.}~\bibnamefont
  {Zhang}}, \bibinfo {author} {\bibfnamefont {Y.}~\bibnamefont {Su}}, \bibinfo
  {author} {\bibfnamefont {X.}~\bibnamefont {Chen}}, \bibinfo {author}
  {\bibfnamefont {H.}~\bibnamefont {Weng}},  \emph {et~al.},\ }\href@noop {}
  {\bibfield  {journal} {\bibinfo  {journal} {Nature communications}\ }\textbf
  {\bibinfo {volume} {11}},\ \bibinfo {pages} {720} (\bibinfo {year}
  {2020})}\BibitemShut {NoStop}%
\bibitem [{\citenamefont {Sie}\ \emph {et~al.}(2019)\citenamefont {Sie},
  \citenamefont {Nyby}, \citenamefont {Pemmaraju}, \citenamefont {Park},
  \citenamefont {Shen}, \citenamefont {Yang}, \citenamefont {Hoffmann},
  \citenamefont {Ofori-Okai}, \citenamefont {Li}, \citenamefont {Reid} \emph
  {et~al.}}]{sie2019ultrafast}%
  \BibitemOpen
  \bibfield  {author} {\bibinfo {author} {\bibfnamefont {E.~J.}\ \bibnamefont
  {Sie}}, \bibinfo {author} {\bibfnamefont {C.~M.}\ \bibnamefont {Nyby}},
  \bibinfo {author} {\bibfnamefont {C.}~\bibnamefont {Pemmaraju}}, \bibinfo
  {author} {\bibfnamefont {S.~J.}\ \bibnamefont {Park}}, \bibinfo {author}
  {\bibfnamefont {X.}~\bibnamefont {Shen}}, \bibinfo {author} {\bibfnamefont
  {J.}~\bibnamefont {Yang}}, \bibinfo {author} {\bibfnamefont {M.~C.}\
  \bibnamefont {Hoffmann}}, \bibinfo {author} {\bibfnamefont {B.}~\bibnamefont
  {Ofori-Okai}}, \bibinfo {author} {\bibfnamefont {R.}~\bibnamefont {Li}},
  \bibinfo {author} {\bibfnamefont {A.~H.}\ \bibnamefont {Reid}},  \emph
  {et~al.},\ }\href@noop {} {\bibfield  {journal} {\bibinfo  {journal}
  {Nature}\ }\textbf {\bibinfo {volume} {565}},\ \bibinfo {pages} {61}
  (\bibinfo {year} {2019})}\BibitemShut {NoStop}%
\bibitem [{\citenamefont {Caglieris}\ \emph {et~al.}(2018)\citenamefont
  {Caglieris}, \citenamefont {Wuttke}, \citenamefont {Sykora}, \citenamefont
  {S{\"u}ss}, \citenamefont {Shekhar}, \citenamefont {Felser}, \citenamefont
  {B{\"u}chner},\ and\ \citenamefont {Hess}}]{caglieris2018anomalous}%
  \BibitemOpen
  \bibfield  {author} {\bibinfo {author} {\bibfnamefont {F.}~\bibnamefont
  {Caglieris}}, \bibinfo {author} {\bibfnamefont {C.}~\bibnamefont {Wuttke}},
  \bibinfo {author} {\bibfnamefont {S.}~\bibnamefont {Sykora}}, \bibinfo
  {author} {\bibfnamefont {V.}~\bibnamefont {S{\"u}ss}}, \bibinfo {author}
  {\bibfnamefont {C.}~\bibnamefont {Shekhar}}, \bibinfo {author} {\bibfnamefont
  {C.}~\bibnamefont {Felser}}, \bibinfo {author} {\bibfnamefont
  {B.}~\bibnamefont {B{\"u}chner}}, \ and\ \bibinfo {author} {\bibfnamefont
  {C.}~\bibnamefont {Hess}},\ }\href@noop {} {\bibfield  {journal} {\bibinfo
  {journal} {Physical Review B}\ }\textbf {\bibinfo {volume} {98}},\ \bibinfo
  {pages} {201107} (\bibinfo {year} {2018})}\BibitemShut {NoStop}%
\bibitem [{\citenamefont {Sun}\ \emph {et~al.}(2016)\citenamefont {Sun},
  \citenamefont {Zhang}, \citenamefont {Felser},\ and\ \citenamefont
  {Yan}}]{sun2016strong}%
  \BibitemOpen
  \bibfield  {author} {\bibinfo {author} {\bibfnamefont {Y.}~\bibnamefont
  {Sun}}, \bibinfo {author} {\bibfnamefont {Y.}~\bibnamefont {Zhang}}, \bibinfo
  {author} {\bibfnamefont {C.}~\bibnamefont {Felser}}, \ and\ \bibinfo {author}
  {\bibfnamefont {B.}~\bibnamefont {Yan}},\ }\href@noop {} {\bibfield
  {journal} {\bibinfo  {journal} {Physical Review Letters}\ }\textbf {\bibinfo
  {volume} {117}},\ \bibinfo {pages} {146403} (\bibinfo {year}
  {2016})}\BibitemShut {NoStop}%
\bibitem [{\citenamefont {Grassano}\ \emph {et~al.}(2020)\citenamefont
  {Grassano}, \citenamefont {Pulci}, \citenamefont {Cannuccia},\ and\
  \citenamefont {Bechstedt}}]{grassano2020influence}%
  \BibitemOpen
  \bibfield  {author} {\bibinfo {author} {\bibfnamefont {D.}~\bibnamefont
  {Grassano}}, \bibinfo {author} {\bibfnamefont {O.}~\bibnamefont {Pulci}},
  \bibinfo {author} {\bibfnamefont {E.}~\bibnamefont {Cannuccia}}, \ and\
  \bibinfo {author} {\bibfnamefont {F.}~\bibnamefont {Bechstedt}},\ }\href@noop
  {} {\bibfield  {journal} {\bibinfo  {journal} {The European Physical Journal
  B}\ }\textbf {\bibinfo {volume} {93}},\ \bibinfo {pages} {1} (\bibinfo {year}
  {2020})}\BibitemShut {NoStop}%
\bibitem [{\citenamefont {Du}\ \emph {et~al.}(2016)\citenamefont {Du},
  \citenamefont {Wang}, \citenamefont {Chen}, \citenamefont {Mao},
  \citenamefont {Khan}, \citenamefont {Xu}, \citenamefont {Zhou}, \citenamefont
  {Zhang}, \citenamefont {Yang}, \citenamefont {Chen} \emph
  {et~al.}}]{du2016large}%
  \BibitemOpen
  \bibfield  {author} {\bibinfo {author} {\bibfnamefont {J.}~\bibnamefont
  {Du}}, \bibinfo {author} {\bibfnamefont {H.}~\bibnamefont {Wang}}, \bibinfo
  {author} {\bibfnamefont {Q.}~\bibnamefont {Chen}}, \bibinfo {author}
  {\bibfnamefont {Q.}~\bibnamefont {Mao}}, \bibinfo {author} {\bibfnamefont
  {R.}~\bibnamefont {Khan}}, \bibinfo {author} {\bibfnamefont {B.}~\bibnamefont
  {Xu}}, \bibinfo {author} {\bibfnamefont {Y.}~\bibnamefont {Zhou}}, \bibinfo
  {author} {\bibfnamefont {Y.}~\bibnamefont {Zhang}}, \bibinfo {author}
  {\bibfnamefont {J.}~\bibnamefont {Yang}}, \bibinfo {author} {\bibfnamefont
  {B.}~\bibnamefont {Chen}},  \emph {et~al.},\ }\href@noop {} {\bibfield
  {journal} {\bibinfo  {journal} {Science China Physics, Mechanics \&
  Astronomy}\ }\textbf {\bibinfo {volume} {59}},\ \bibinfo {pages} {1}
  (\bibinfo {year} {2016})}\BibitemShut {NoStop}%
\bibitem [{\citenamefont {Huang}\ \emph {et~al.}(2015)\citenamefont {Huang},
  \citenamefont {Xu}, \citenamefont {Belopolski}, \citenamefont {Lee},
  \citenamefont {Chang}, \citenamefont {Wang}, \citenamefont {Alidoust},
  \citenamefont {Bian}, \citenamefont {Neupane}, \citenamefont {Bansil} \emph
  {et~al.}}]{huang2015inversion}%
  \BibitemOpen
  \bibfield  {author} {\bibinfo {author} {\bibfnamefont {S.-M.}\ \bibnamefont
  {Huang}}, \bibinfo {author} {\bibfnamefont {S.-Y.}\ \bibnamefont {Xu}},
  \bibinfo {author} {\bibfnamefont {I.}~\bibnamefont {Belopolski}}, \bibinfo
  {author} {\bibfnamefont {C.-C.}\ \bibnamefont {Lee}}, \bibinfo {author}
  {\bibfnamefont {G.}~\bibnamefont {Chang}}, \bibinfo {author} {\bibfnamefont
  {B.}~\bibnamefont {Wang}}, \bibinfo {author} {\bibfnamefont {N.}~\bibnamefont
  {Alidoust}}, \bibinfo {author} {\bibfnamefont {G.}~\bibnamefont {Bian}},
  \bibinfo {author} {\bibfnamefont {M.}~\bibnamefont {Neupane}}, \bibinfo
  {author} {\bibfnamefont {A.}~\bibnamefont {Bansil}},  \emph {et~al.},\
  }\href@noop {} {\bibfield  {journal} {\bibinfo  {journal} {arXiv preprint
  arXiv:1501.00755}\ } (\bibinfo {year} {2015})}\BibitemShut {NoStop}%
\bibitem [{\citenamefont {Liu}\ \emph {et~al.}(2016)\citenamefont {Liu},
  \citenamefont {Yang}, \citenamefont {Sun}, \citenamefont {Zhang},
  \citenamefont {Peng}, \citenamefont {Yang}, \citenamefont {Chen},
  \citenamefont {Zhang}, \citenamefont {Guo}, \citenamefont {Prabhakaran} \emph
  {et~al.}}]{liu2016evolution}%
  \BibitemOpen
  \bibfield  {author} {\bibinfo {author} {\bibfnamefont {Z.}~\bibnamefont
  {Liu}}, \bibinfo {author} {\bibfnamefont {L.}~\bibnamefont {Yang}}, \bibinfo
  {author} {\bibfnamefont {Y.}~\bibnamefont {Sun}}, \bibinfo {author}
  {\bibfnamefont {T.}~\bibnamefont {Zhang}}, \bibinfo {author} {\bibfnamefont
  {H.}~\bibnamefont {Peng}}, \bibinfo {author} {\bibfnamefont {H.}~\bibnamefont
  {Yang}}, \bibinfo {author} {\bibfnamefont {C.}~\bibnamefont {Chen}}, \bibinfo
  {author} {\bibfnamefont {Y.~f.}\ \bibnamefont {Zhang}}, \bibinfo {author}
  {\bibfnamefont {Y.}~\bibnamefont {Guo}}, \bibinfo {author} {\bibfnamefont
  {D.}~\bibnamefont {Prabhakaran}},  \emph {et~al.},\ }\href@noop {} {\bibfield
   {journal} {\bibinfo  {journal} {Nature materials}\ }\textbf {\bibinfo
  {volume} {15}},\ \bibinfo {pages} {27} (\bibinfo {year} {2016})}\BibitemShut
  {NoStop}%
\bibitem [{\citenamefont {Weng}\ \emph {et~al.}(2015)\citenamefont {Weng},
  \citenamefont {Fang}, \citenamefont {Fang}, \citenamefont {Bernevig},\ and\
  \citenamefont {Dai}}]{weng2015weyl}%
  \BibitemOpen
  \bibfield  {author} {\bibinfo {author} {\bibfnamefont {H.}~\bibnamefont
  {Weng}}, \bibinfo {author} {\bibfnamefont {C.}~\bibnamefont {Fang}}, \bibinfo
  {author} {\bibfnamefont {Z.}~\bibnamefont {Fang}}, \bibinfo {author}
  {\bibfnamefont {B.~A.}\ \bibnamefont {Bernevig}}, \ and\ \bibinfo {author}
  {\bibfnamefont {X.}~\bibnamefont {Dai}},\ }\href@noop {} {\bibfield
  {journal} {\bibinfo  {journal} {Physical Review X}\ }\textbf {\bibinfo
  {volume} {5}},\ \bibinfo {pages} {011029} (\bibinfo {year}
  {2015})}\BibitemShut {NoStop}%
\bibitem [{\citenamefont {Du}\ \emph {et~al.}(2017)\citenamefont {Du},
  \citenamefont {Bo}, \citenamefont {Wang}, \citenamefont {Kan}, \citenamefont
  {Duan}, \citenamefont {Savrasov},\ and\ \citenamefont
  {Wan}}]{du2017emergence}%
  \BibitemOpen
  \bibfield  {author} {\bibinfo {author} {\bibfnamefont {Y.}~\bibnamefont
  {Du}}, \bibinfo {author} {\bibfnamefont {X.}~\bibnamefont {Bo}}, \bibinfo
  {author} {\bibfnamefont {D.}~\bibnamefont {Wang}}, \bibinfo {author}
  {\bibfnamefont {E.-j.}\ \bibnamefont {Kan}}, \bibinfo {author} {\bibfnamefont
  {C.-G.}\ \bibnamefont {Duan}}, \bibinfo {author} {\bibfnamefont {S.~Y.}\
  \bibnamefont {Savrasov}}, \ and\ \bibinfo {author} {\bibfnamefont
  {X.}~\bibnamefont {Wan}},\ }\href@noop {} {\bibfield  {journal} {\bibinfo
  {journal} {Physical Review B}\ }\textbf {\bibinfo {volume} {96}},\ \bibinfo
  {pages} {235152} (\bibinfo {year} {2017})}\BibitemShut {NoStop}%
\bibitem [{\citenamefont {Markovi{\'c}}\ \emph {et~al.}(2019)\citenamefont
  {Markovi{\'c}}, \citenamefont {Hooley}, \citenamefont {Clark}, \citenamefont
  {Mazzola}, \citenamefont {Watson}, \citenamefont {Riley}, \citenamefont
  {Volckaert}, \citenamefont {Underwood}, \citenamefont {Dyer}, \citenamefont
  {Murgatroyd} \emph {et~al.}}]{markovic2019weyl}%
  \BibitemOpen
  \bibfield  {author} {\bibinfo {author} {\bibfnamefont {I.}~\bibnamefont
  {Markovi{\'c}}}, \bibinfo {author} {\bibfnamefont {C.}~\bibnamefont
  {Hooley}}, \bibinfo {author} {\bibfnamefont {O.~J.}\ \bibnamefont {Clark}},
  \bibinfo {author} {\bibfnamefont {F.}~\bibnamefont {Mazzola}}, \bibinfo
  {author} {\bibfnamefont {M.~D.}\ \bibnamefont {Watson}}, \bibinfo {author}
  {\bibfnamefont {J.~M.}\ \bibnamefont {Riley}}, \bibinfo {author}
  {\bibfnamefont {K.}~\bibnamefont {Volckaert}}, \bibinfo {author}
  {\bibfnamefont {K.}~\bibnamefont {Underwood}}, \bibinfo {author}
  {\bibfnamefont {M.}~\bibnamefont {Dyer}}, \bibinfo {author} {\bibfnamefont
  {P.}~\bibnamefont {Murgatroyd}},  \emph {et~al.},\ }\href@noop {} {\bibfield
  {journal} {\bibinfo  {journal} {Nature Communications}\ }\textbf {\bibinfo
  {volume} {10}},\ \bibinfo {pages} {5485} (\bibinfo {year}
  {2019})}\BibitemShut {NoStop}%
\bibitem [{\citenamefont {Ruan}\ \emph {et~al.}(2016)\citenamefont {Ruan},
  \citenamefont {Jian}, \citenamefont {Yao}, \citenamefont {Zhang},
  \citenamefont {Zhang},\ and\ \citenamefont {Xing}}]{ruan2016symmetry}%
  \BibitemOpen
  \bibfield  {author} {\bibinfo {author} {\bibfnamefont {J.}~\bibnamefont
  {Ruan}}, \bibinfo {author} {\bibfnamefont {S.-K.}\ \bibnamefont {Jian}},
  \bibinfo {author} {\bibfnamefont {H.}~\bibnamefont {Yao}}, \bibinfo {author}
  {\bibfnamefont {H.}~\bibnamefont {Zhang}}, \bibinfo {author} {\bibfnamefont
  {S.-C.}\ \bibnamefont {Zhang}}, \ and\ \bibinfo {author} {\bibfnamefont
  {D.}~\bibnamefont {Xing}},\ }\href@noop {} {\bibfield  {journal} {\bibinfo
  {journal} {Nature communications}\ }\textbf {\bibinfo {volume} {7}},\
  \bibinfo {pages} {11136} (\bibinfo {year} {2016})}\BibitemShut {NoStop}%
\bibitem [{\citenamefont {Grassano}\ \emph {et~al.}(2018)\citenamefont
  {Grassano}, \citenamefont {Pulci}, \citenamefont {Mosca~Conte},\ and\
  \citenamefont {Bechstedt}}]{grassano2018validity}%
  \BibitemOpen
  \bibfield  {author} {\bibinfo {author} {\bibfnamefont {D.}~\bibnamefont
  {Grassano}}, \bibinfo {author} {\bibfnamefont {O.}~\bibnamefont {Pulci}},
  \bibinfo {author} {\bibfnamefont {A.}~\bibnamefont {Mosca~Conte}}, \ and\
  \bibinfo {author} {\bibfnamefont {F.}~\bibnamefont {Bechstedt}},\ }\href@noop
  {} {\bibfield  {journal} {\bibinfo  {journal} {Scientific reports}\ }\textbf
  {\bibinfo {volume} {8}},\ \bibinfo {pages} {1} (\bibinfo {year}
  {2018})}\BibitemShut {NoStop}%
\bibitem [{\citenamefont {Lee}\ \emph {et~al.}(2015)\citenamefont {Lee},
  \citenamefont {Xu}, \citenamefont {Huang}, \citenamefont {Sanchez},
  \citenamefont {Belopolski}, \citenamefont {Chang}, \citenamefont {Bian},
  \citenamefont {Alidoust}, \citenamefont {Zheng}, \citenamefont {Neupane}
  \emph {et~al.}}]{lee2015fermi}%
  \BibitemOpen
  \bibfield  {author} {\bibinfo {author} {\bibfnamefont {C.-C.}\ \bibnamefont
  {Lee}}, \bibinfo {author} {\bibfnamefont {S.-Y.}\ \bibnamefont {Xu}},
  \bibinfo {author} {\bibfnamefont {S.-M.}\ \bibnamefont {Huang}}, \bibinfo
  {author} {\bibfnamefont {D.~S.}\ \bibnamefont {Sanchez}}, \bibinfo {author}
  {\bibfnamefont {I.}~\bibnamefont {Belopolski}}, \bibinfo {author}
  {\bibfnamefont {G.}~\bibnamefont {Chang}}, \bibinfo {author} {\bibfnamefont
  {G.}~\bibnamefont {Bian}}, \bibinfo {author} {\bibfnamefont {N.}~\bibnamefont
  {Alidoust}}, \bibinfo {author} {\bibfnamefont {H.}~\bibnamefont {Zheng}},
  \bibinfo {author} {\bibfnamefont {M.}~\bibnamefont {Neupane}},  \emph
  {et~al.},\ }\href@noop {} {\bibfield  {journal} {\bibinfo  {journal}
  {Physical Review B}\ }\textbf {\bibinfo {volume} {92}},\ \bibinfo {pages}
  {235104} (\bibinfo {year} {2015})}\BibitemShut {NoStop}%
\bibitem [{\citenamefont {Pandey}\ and\ \citenamefont
  {Pandey}(2023{\natexlab{b}})}]{pandey2023py}%
  \BibitemOpen
  \bibfield  {author} {\bibinfo {author} {\bibfnamefont {V.}~\bibnamefont
  {Pandey}}\ and\ \bibinfo {author} {\bibfnamefont {S.~K.}\ \bibnamefont
  {Pandey}},\ }\href@noop {} {\bibfield  {journal} {\bibinfo  {journal}
  {Computer Physics Communications}\ }\textbf {\bibinfo {volume} {283}},\
  \bibinfo {pages} {108570} (\bibinfo {year} {2023}{\natexlab{b}})}\BibitemShut
  {NoStop}%
\bibitem [{\citenamefont {Sihi}\ and\ \citenamefont
  {Pandey}(2023)}]{sihi2023track}%
  \BibitemOpen
  \bibfield  {author} {\bibinfo {author} {\bibfnamefont {A.}~\bibnamefont
  {Sihi}}\ and\ \bibinfo {author} {\bibfnamefont {S.~K.}\ \bibnamefont
  {Pandey}},\ }\href@noop {} {\bibfield  {journal} {\bibinfo  {journal}
  {Computer Physics Communications}\ }\textbf {\bibinfo {volume} {285}},\
  \bibinfo {pages} {108640} (\bibinfo {year} {2023})}\BibitemShut {NoStop}%
\bibitem [{\citenamefont {Giannozzi}\ \emph {et~al.}(2009)\citenamefont
  {Giannozzi}, \citenamefont {Baroni}, \citenamefont {Bonini}, \citenamefont
  {Calandra}, \citenamefont {Car}, \citenamefont {Cavazzoni}, \citenamefont
  {Ceresoli}, \citenamefont {Chiarotti}, \citenamefont {Cococcioni},
  \citenamefont {Dabo} \emph {et~al.}}]{giannozzi2009quantum}%
  \BibitemOpen
  \bibfield  {author} {\bibinfo {author} {\bibfnamefont {P.}~\bibnamefont
  {Giannozzi}}, \bibinfo {author} {\bibfnamefont {S.}~\bibnamefont {Baroni}},
  \bibinfo {author} {\bibfnamefont {N.}~\bibnamefont {Bonini}}, \bibinfo
  {author} {\bibfnamefont {M.}~\bibnamefont {Calandra}}, \bibinfo {author}
  {\bibfnamefont {R.}~\bibnamefont {Car}}, \bibinfo {author} {\bibfnamefont
  {C.}~\bibnamefont {Cavazzoni}}, \bibinfo {author} {\bibfnamefont
  {D.}~\bibnamefont {Ceresoli}}, \bibinfo {author} {\bibfnamefont {G.~L.}\
  \bibnamefont {Chiarotti}}, \bibinfo {author} {\bibfnamefont {M.}~\bibnamefont
  {Cococcioni}}, \bibinfo {author} {\bibfnamefont {I.}~\bibnamefont {Dabo}},
  \emph {et~al.},\ }\href@noop {} {\bibfield  {journal} {\bibinfo  {journal}
  {Journal of physics: Condensed matter}\ }\textbf {\bibinfo {volume} {21}},\
  \bibinfo {pages} {395502} (\bibinfo {year} {2009})}\BibitemShut {NoStop}%
\bibitem [{\citenamefont {Blaha}\ \emph {et~al.}(2001)\citenamefont {Blaha},
  \citenamefont {Schwarz}, \citenamefont {Madsen}, \citenamefont {Kvasnicka},
  \citenamefont {Luitz} \emph {et~al.}}]{blaha2001wien2k}%
  \BibitemOpen
  \bibfield  {author} {\bibinfo {author} {\bibfnamefont {P.}~\bibnamefont
  {Blaha}}, \bibinfo {author} {\bibfnamefont {K.}~\bibnamefont {Schwarz}},
  \bibinfo {author} {\bibfnamefont {G.~K.}\ \bibnamefont {Madsen}}, \bibinfo
  {author} {\bibfnamefont {D.}~\bibnamefont {Kvasnicka}}, \bibinfo {author}
  {\bibfnamefont {J.}~\bibnamefont {Luitz}},  \emph {et~al.},\ }\href@noop {}
  {\bibfield  {journal} {\bibinfo  {journal} {An augmented plane wave+ local
  orbitals program for calculating crystal properties}\ }\textbf {\bibinfo
  {volume} {60}} (\bibinfo {year} {2001})}\BibitemShut {NoStop}%
\bibitem [{\citenamefont {Tomczak}\ \emph {et~al.}(2010)\citenamefont
  {Tomczak}, \citenamefont {Haule}, \citenamefont {Miyake}, \citenamefont
  {Georges},\ and\ \citenamefont {Kotliar}}]{tomczak2010thermopower}%
  \BibitemOpen
  \bibfield  {author} {\bibinfo {author} {\bibfnamefont {J.~M.}\ \bibnamefont
  {Tomczak}}, \bibinfo {author} {\bibfnamefont {K.}~\bibnamefont {Haule}},
  \bibinfo {author} {\bibfnamefont {T.}~\bibnamefont {Miyake}}, \bibinfo
  {author} {\bibfnamefont {A.}~\bibnamefont {Georges}}, \ and\ \bibinfo
  {author} {\bibfnamefont {G.}~\bibnamefont {Kotliar}},\ }\href@noop {}
  {\bibfield  {journal} {\bibinfo  {journal} {Physical Review B}\ }\textbf
  {\bibinfo {volume} {82}},\ \bibinfo {pages} {085104} (\bibinfo {year}
  {2010})}\BibitemShut {NoStop}%
\bibitem [{\citenamefont {Oudovenko}\ \emph {et~al.}(2006)\citenamefont
  {Oudovenko}, \citenamefont {P{\'a}lsson}, \citenamefont {Haule},
  \citenamefont {Kotliar},\ and\ \citenamefont
  {Savrasov}}]{oudovenko2006electronic}%
  \BibitemOpen
  \bibfield  {author} {\bibinfo {author} {\bibfnamefont {V.}~\bibnamefont
  {Oudovenko}}, \bibinfo {author} {\bibfnamefont {G.}~\bibnamefont
  {P{\'a}lsson}}, \bibinfo {author} {\bibfnamefont {K.}~\bibnamefont {Haule}},
  \bibinfo {author} {\bibfnamefont {G.}~\bibnamefont {Kotliar}}, \ and\
  \bibinfo {author} {\bibfnamefont {S.}~\bibnamefont {Savrasov}},\ }\href@noop
  {} {\bibfield  {journal} {\bibinfo  {journal} {Physical Review B}\ }\textbf
  {\bibinfo {volume} {73}},\ \bibinfo {pages} {035120} (\bibinfo {year}
  {2006})}\BibitemShut {NoStop}%
\bibitem [{\citenamefont {Imada}\ \emph {et~al.}(1998)\citenamefont {Imada},
  \citenamefont {Fujimori},\ and\ \citenamefont {Tokura}}]{imada1998metal}%
  \BibitemOpen
  \bibfield  {author} {\bibinfo {author} {\bibfnamefont {M.}~\bibnamefont
  {Imada}}, \bibinfo {author} {\bibfnamefont {A.}~\bibnamefont {Fujimori}}, \
  and\ \bibinfo {author} {\bibfnamefont {Y.}~\bibnamefont {Tokura}},\
  }\href@noop {} {\bibfield  {journal} {\bibinfo  {journal} {Reviews of modern
  physics}\ }\textbf {\bibinfo {volume} {70}},\ \bibinfo {pages} {1039}
  (\bibinfo {year} {1998})}\BibitemShut {NoStop}%
\bibitem [{\citenamefont {Madsen}\ and\ \citenamefont
  {Singh}(2006)}]{madsen2006boltztrap}%
  \BibitemOpen
  \bibfield  {author} {\bibinfo {author} {\bibfnamefont {G.~K.}\ \bibnamefont
  {Madsen}}\ and\ \bibinfo {author} {\bibfnamefont {D.~J.}\ \bibnamefont
  {Singh}},\ }\href@noop {} {\bibfield  {journal} {\bibinfo  {journal}
  {Computer Physics Communications}\ }\textbf {\bibinfo {volume} {175}},\
  \bibinfo {pages} {67} (\bibinfo {year} {2006})}\BibitemShut {NoStop}%
\end{thebibliography}%
\bibliographystyle{apsrev4-1}

\beginsupplement

\section{Supplementary Materials}

\begin{figure*}
    \centering
    {
        \includegraphics[width=0.60\linewidth, height=6.45cm]{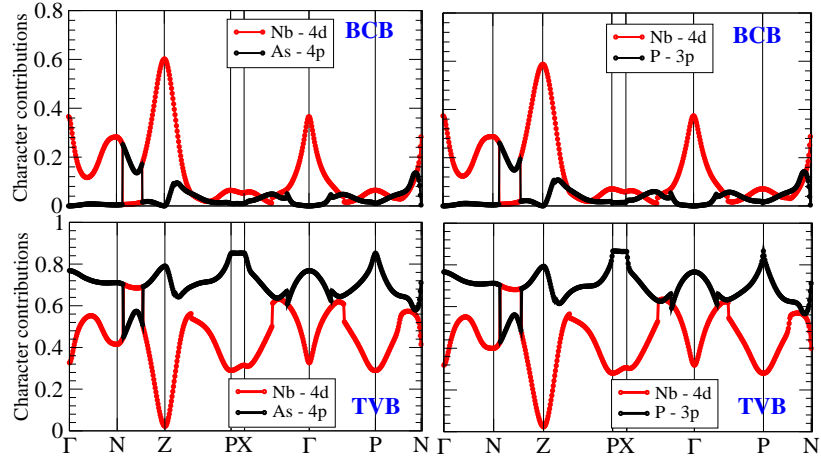}
    }
    \caption
    {{ {\footnotesize (Color online) \textit{IDPI} in \textit{N-Z} direction for TVB \& BCB corresponding to NbAs \& NbP.} }
    }
    \label{fig:foobar}
\end{figure*}

\begin{figure*}
    \centering
    {
        \includegraphics[width=0.95\linewidth, height=5.5cm]{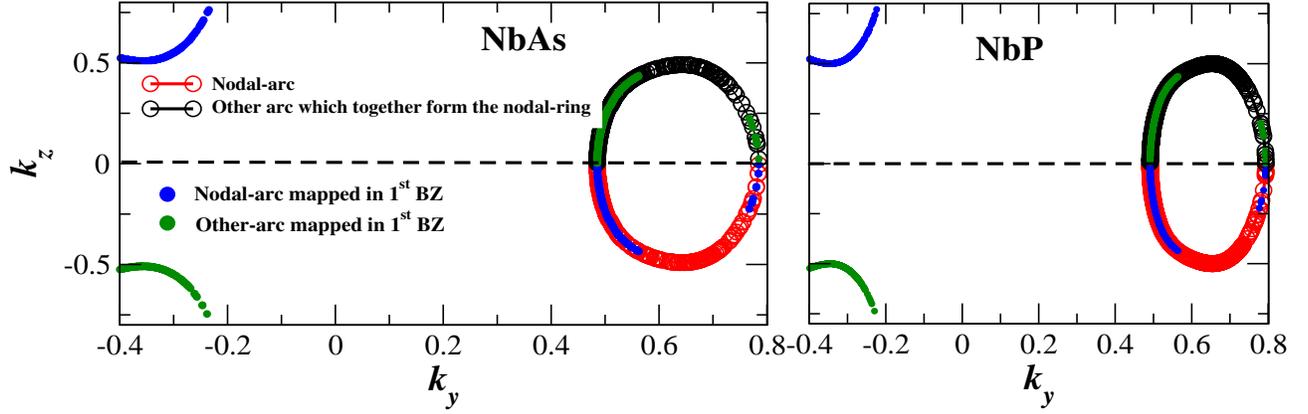}
    }
    \caption
    { {{\footnotesize(Color online) Plot showing the portion of \textit{nodal-arcs} associated with the first Brillouin zone of NbAs \& NbP.}}
    }
    \label{fig:foobar}
\end{figure*}

\begin{figure*}
    \centering
    {
        \includegraphics[width=0.60\linewidth, height=5.5cm]{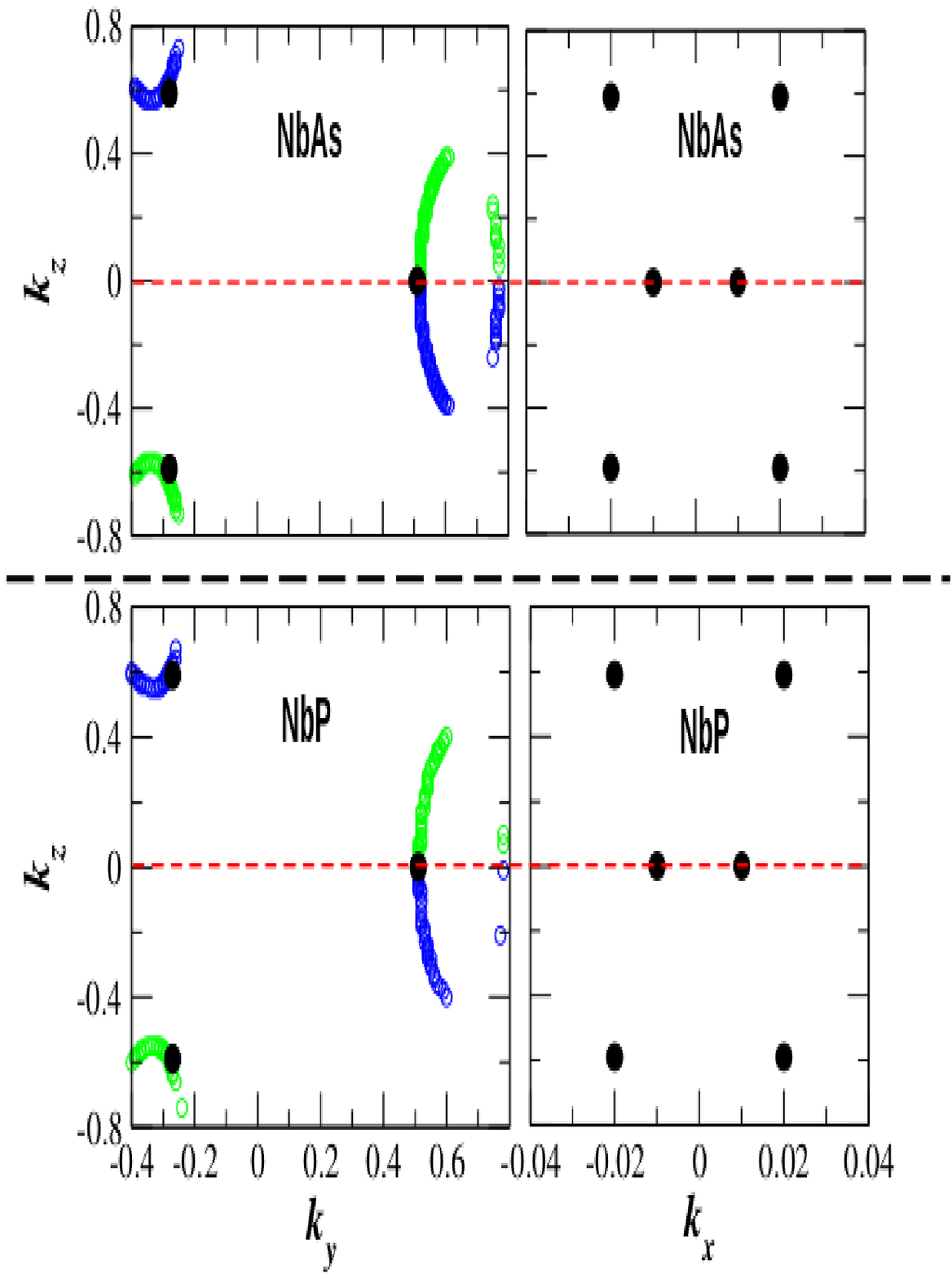}
    }  
    \caption
    { {{\footnotesize(Color online) The position of Weyl-nodes in NbAs \& NbP with respect to the coordinates of \textit{nodal-arcs} from which they are evolved due to SOC.}}
    }
    \label{fig:foobar}
\end{figure*}

\begin{figure*}
    \centering
    {
        \includegraphics[width=0.60\linewidth, height=6.0cm]{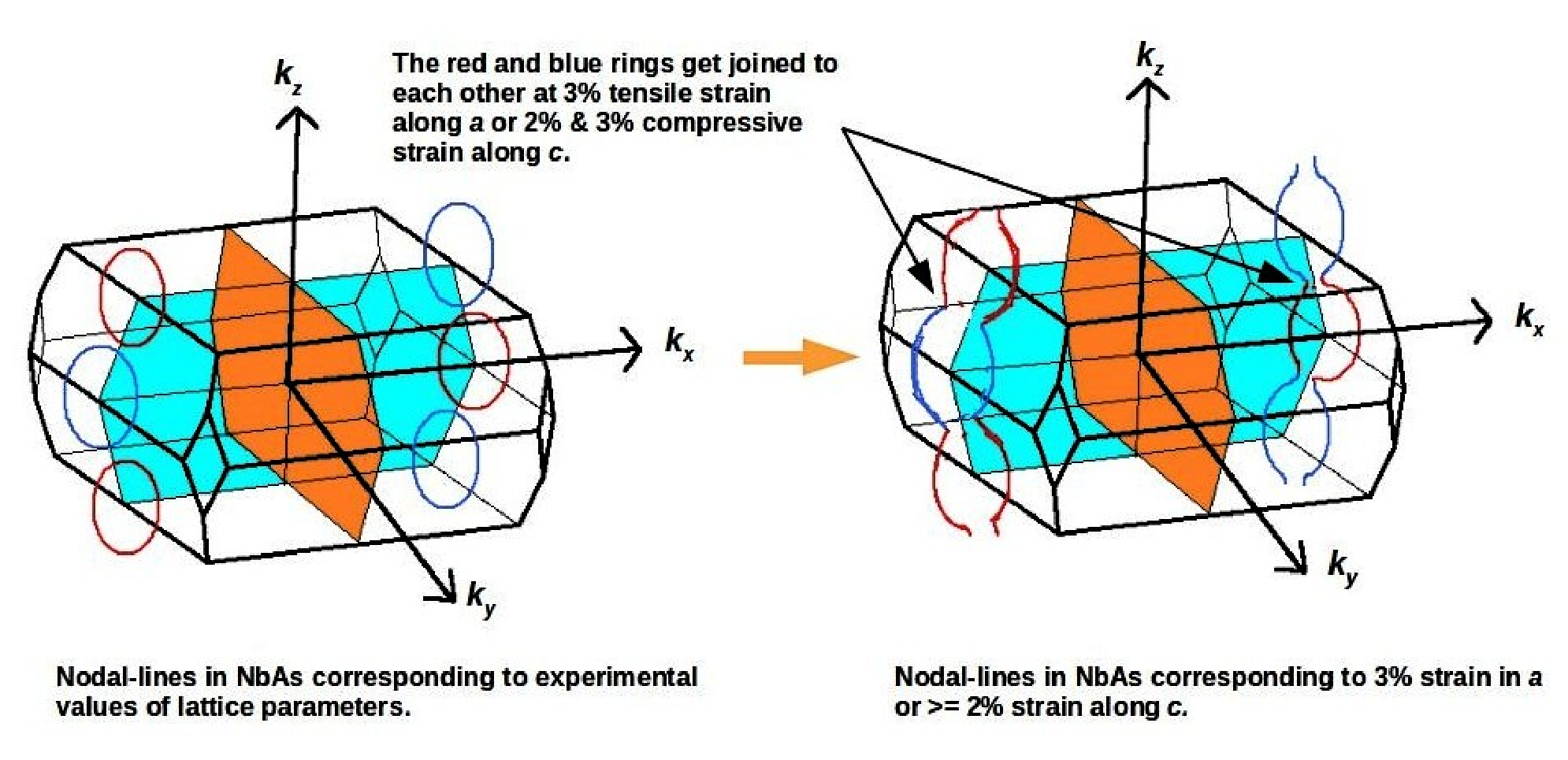}
    }  
    \caption
    { {{(Color online) Schematic showing the partial merging of nodal-lines in extended BZ upon the application of strain along $a$ or $c$ direction of NbAs \& NbP.}}
    }
    \label{straineff}
\end{figure*}

\begin{figure*}
    \centering
    \subfigure[]
    {
        \includegraphics[width=0.45\linewidth, height=5.5cm]{a_strainNbPNL.eps}
    }
    \subfigure[]
    {
        \includegraphics[width=0.45\linewidth, height=5.5cm]{c_strainNbPNL.eps}
    }\vspace*{0.1in}
    
    \caption
    { {{ (Color online) Plot (a) ((b)) shows the effect of axial strain, applied along the $a$ ($c$) direction, on the nodal-rings situated in $k_x$=0 plane of NbP.}}
    }
    \label{fig_cone1_taas}
\end{figure*}

\begin{figure*}
    \centering
    \subfigure[]
    {
        \includegraphics[width=0.45\linewidth, height=5.5cm]{a_strainTaAsNL.eps}
    }
    \subfigure[]
    {
        \includegraphics[width=0.45\linewidth, height=5.5cm]{c_strainTaAsNL.eps}
    }\vspace*{0.1in}
    
    \caption
    { {{ (Color online) Plot (a) ((b)) shows the effect of axial strain, applied along the $a$ ($c$) direction, on the nodal-rings situated in $k_x$=0 plane of TaAs.}}
    }
    \label{fig_cone1_taas}
\end{figure*}

\begin{figure*}
    \centering
    \subfigure[]
    {
        \includegraphics[width=0.45\linewidth, height=5.5cm]{a_strainTaPNL.eps}
    }
    \subfigure[]
    {
        \includegraphics[width=0.45\linewidth, height=5.5cm]{c_strainTaPNL.eps}
    }\vspace*{0.1in}
    
    \caption
    { {{ (Color online) Plot (a) ((b)) shows the effect of axial strain, applied along the $a$ ($c$) direction, on the nodal-rings situated in $k_x$=0 plane of TaP.}}
    }
    \label{fig_cone1_taas}
\end{figure*}

\begin{figure*}
    \centering
    \subfigure[]
    {
        \includegraphics[width=0.45\linewidth, height=5.5cm]{a_change_n_energyNbP.eps}\hspace*{0.2in}
    }
    \subfigure[]
    {
        \includegraphics[width=0.45\linewidth, height=5.5cm]{a_change_p_energyNbP.eps}
    }
    {
        \includegraphics[width=0.45\linewidth, height=5.5cm]{c_change_n_energyNbP.eps}\hspace*{0.2in}
    }
    \subfigure[]
    {
        \includegraphics[width=0.45\linewidth, height=5.5cm]{c_change_p_energyNbP.eps}
    }
    \caption
    { {{ (Color online) Plots (a) \& (b) ((c) \& (d)) shows the position of node points and their energy with respect to Fermi energy (scalled to 0 meV shown in dashed blue line) when the strain is applied along $a$ ($c$) direction in NbP.}}
    }
    \label{fig_cone3_taas}
\end{figure*}

\begin{figure*}
    \centering
    \subfigure[]
    {
        \includegraphics[width=0.45\linewidth, height=5.5cm]{a_change_n_energyTaAs.eps}\hspace*{0.2in}
    }
    \subfigure[]
    {
        \includegraphics[width=0.45\linewidth, height=5.5cm]{a_change_p_energyTaAs.eps}
    }
    {
        \includegraphics[width=0.45\linewidth, height=5.5cm]{c_change_n_energyTaAs.eps}\hspace*{0.2in}
    }
    \subfigure[]
    {
        \includegraphics[width=0.45\linewidth, height=5.5cm]{c_change_p_energyTaAs.eps}
    }
    \caption
    { {{ (Color online) Plots (a) \& (b) ((c) \& (d)) shows the position of node points and their energy with respect to Fermi energy (scalled to 0 meV shown in dashed blue line) when the strain is applied along $a$ ($c$) direction in TaAs.}}
    }
    \label{fig_cone3_taas}
\end{figure*}

\begin{figure*}
    \centering
    \subfigure[]
    {
        \includegraphics[width=0.45\linewidth, height=5.5cm]{a_change_n_energyTaP.eps}\hspace*{0.2in}
    }
    \subfigure[]
    {
        \includegraphics[width=0.45\linewidth, height=5.5cm]{a_change_p_energyTaP.eps}
    }
    {
        \includegraphics[width=0.45\linewidth, height=5.5cm]{c_change_n_energyTaP.eps}\hspace*{0.2in}
    }
    \subfigure[]
    {
        \includegraphics[width=0.45\linewidth, height=5.5cm]{c_change_p_energyTaP.eps}
    }
    \caption
    { {{ (Color online) Plots (a) \& (b) ((c) \& (d)) shows the position of node points and their energy with respect to Fermi energy (scalled to 0 meV shown in dashed blue line) when the strain is applied along $a$ ($c$) direction in TaP.}}
    }
    \label{fig_cone3_taas}
\end{figure*}

\begin{figure*}
    \centering
    \subfigure[]
    {
        \includegraphics[width=0.30\linewidth, height=4.0cm]{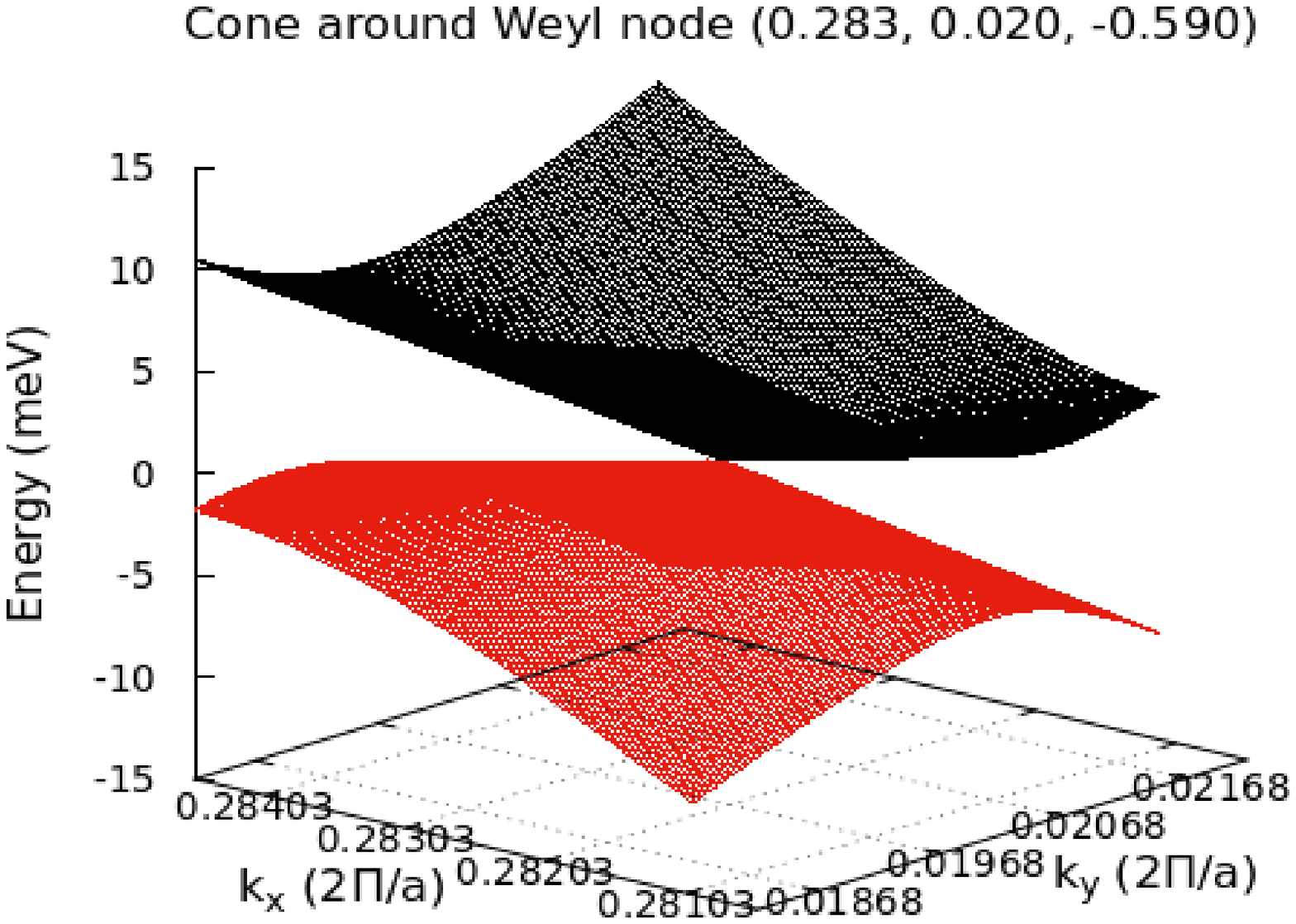}
    }
    \subfigure[]
    {
        \includegraphics[width=0.30\linewidth, height=4.0cm]{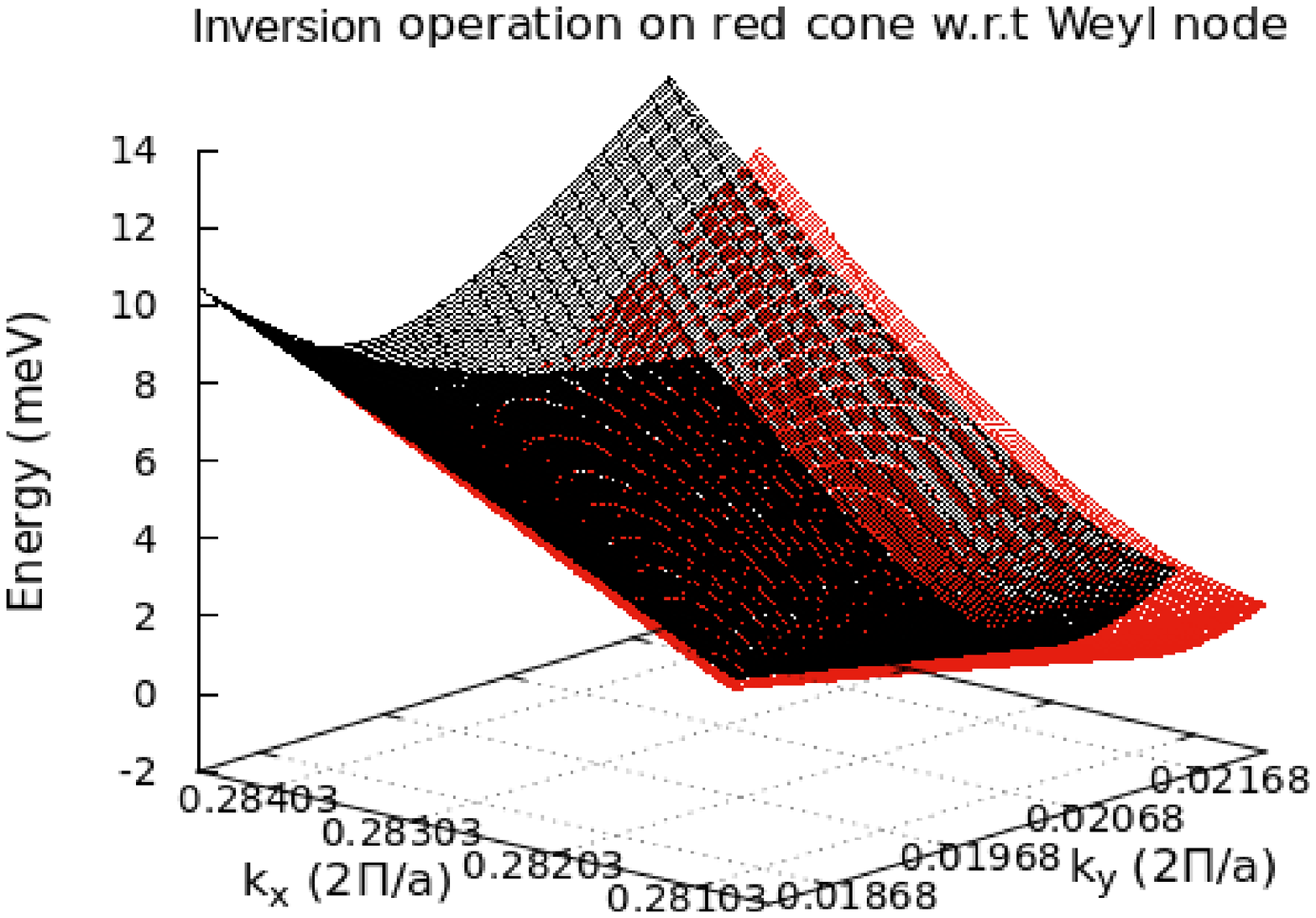}
    }\\
    \subfigure[]
    {
        \includegraphics[width=0.30\linewidth, height=4.0cm]{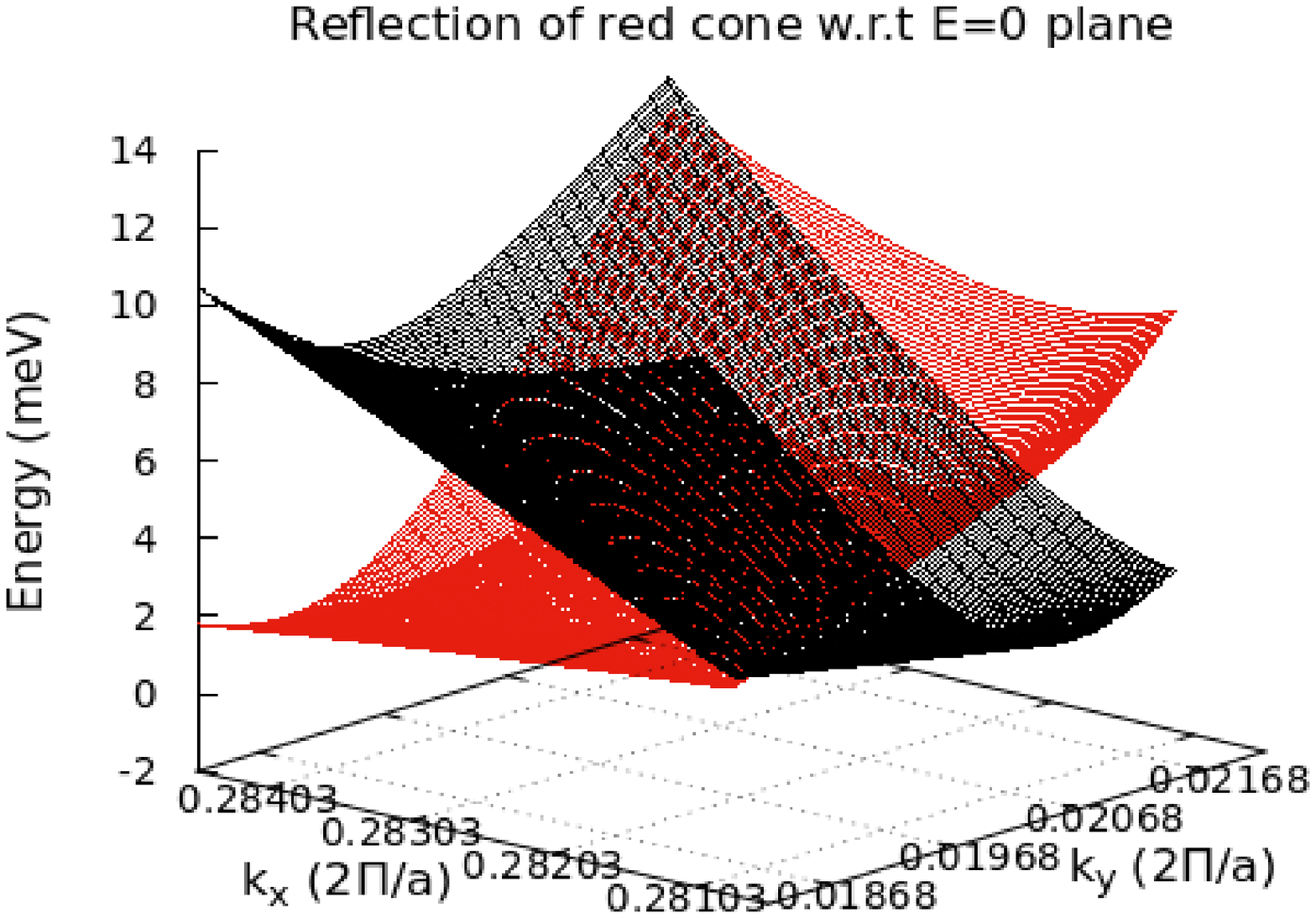}
    }
    \subfigure[]
    {
        \includegraphics[width=0.30\linewidth, height=4.0cm]{seebeck/TaAs/2/TaAs_2.eps}
    }
    \caption
    { {{ \footnotesize (Color online) Plot (a) shows the Weyl cone around W2 point of TaAs. Plot (b) shows the extent to which red cone overlaps the balck one when inversion operation is applied on it with respect to the Weyl node. Plot (c) shows the extent to which red cone overlaps the balck one when mirror-reflection operation is applied on it with respect to the plane parallel to $k_x$-$k_y$ plane and passing through Weyl energy. Plot (d) shows the components of $S^{xx}$, $S^{yy}$ \& $S^{zz}$ contributed from this cone.}}
    }
    \label{fig_cone_taas}
\end{figure*}

\begin{figure*}
    \centering
    \subfigure[]
    {
        \includegraphics[width=0.30\linewidth, height=4.0cm]{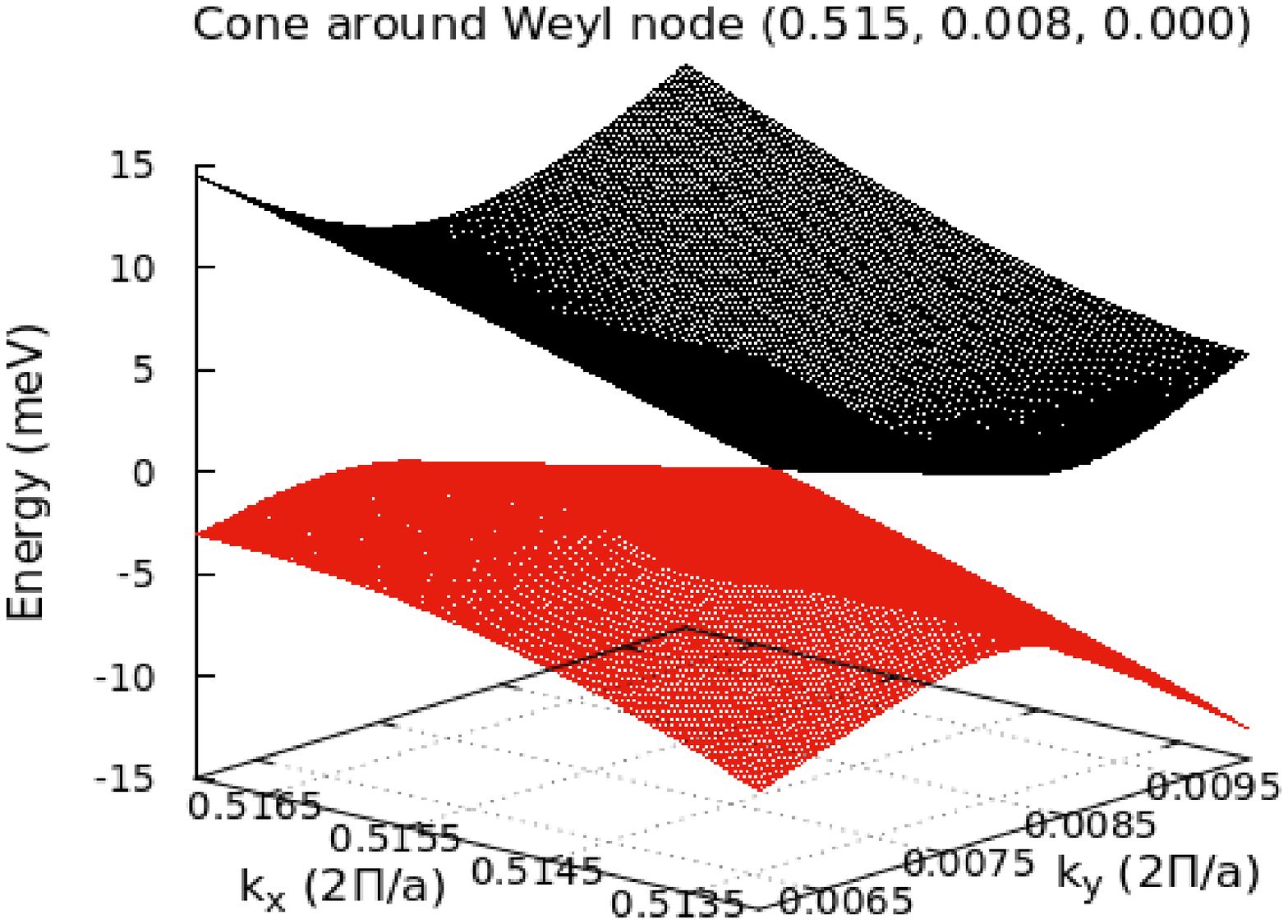}
    }
    \subfigure[]
    {
        \includegraphics[width=0.30\linewidth, height=4.0cm]{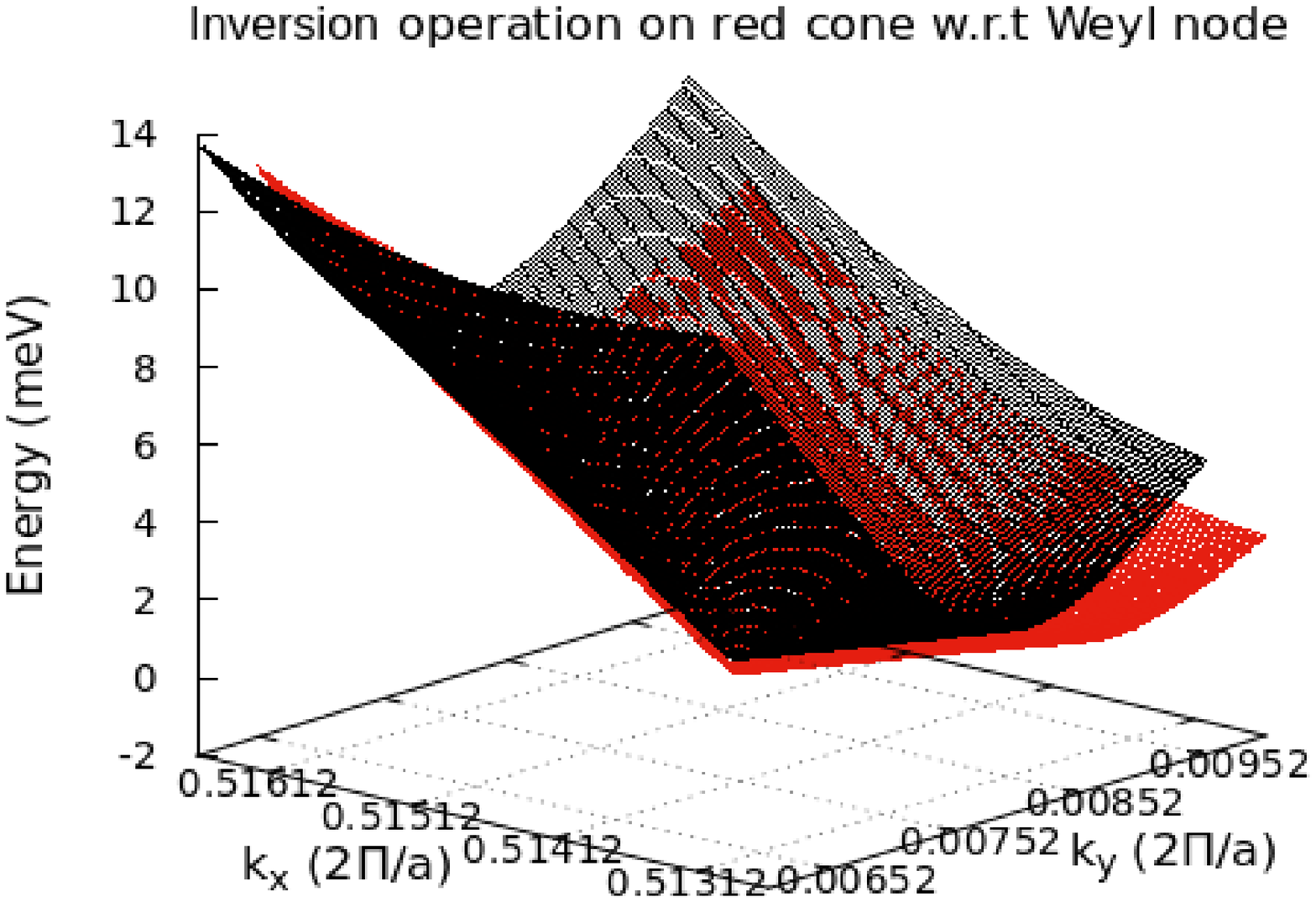}
    }\\
    \subfigure[]
    {
        \includegraphics[width=0.30\linewidth, height=4.0cm]{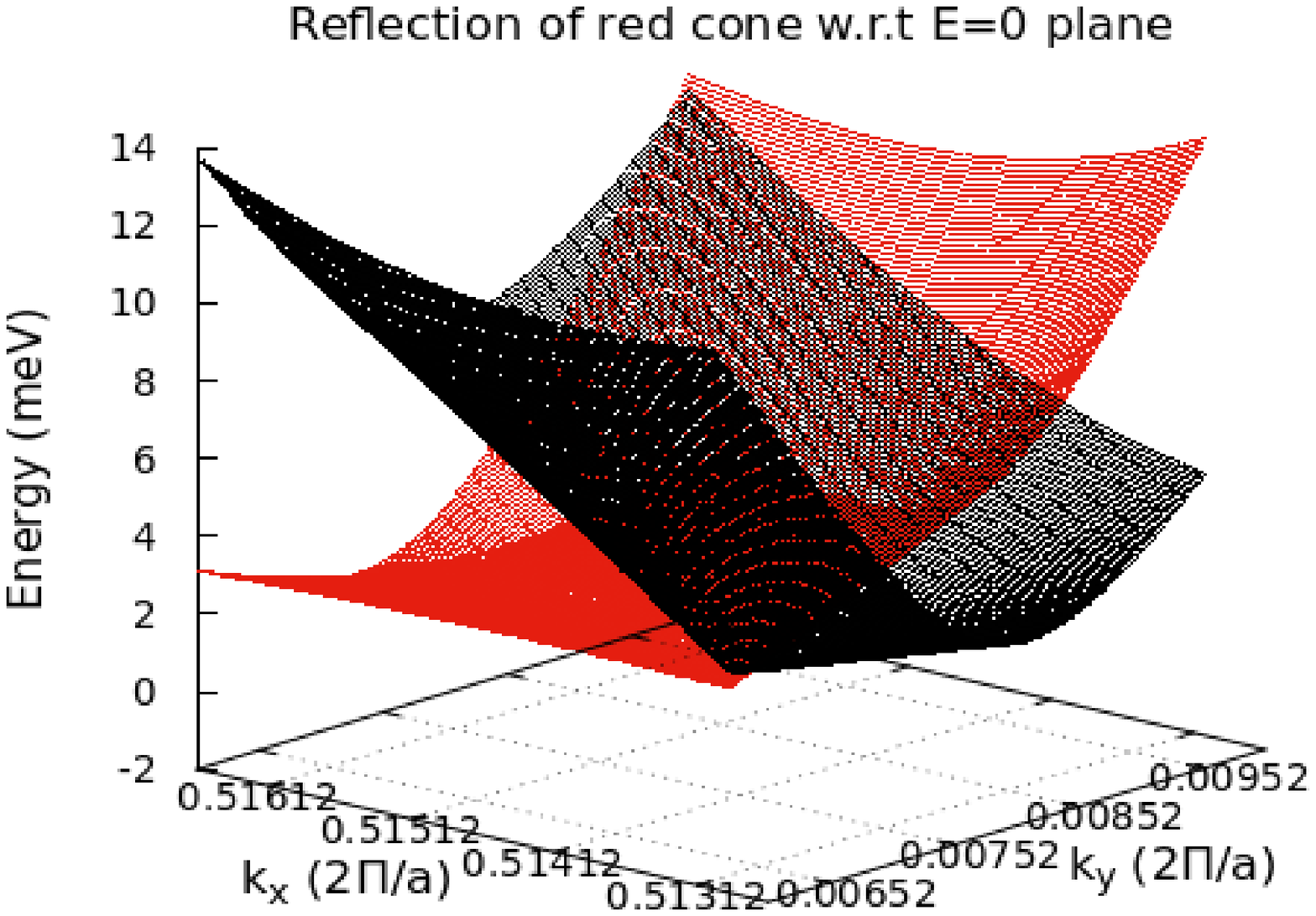}
    }
    \subfigure[]
    {
        \includegraphics[width=0.30\linewidth, height=4.0cm]{seebeck/TaAs/3/TaAs_3.eps}
    }
    \caption
    { {{ \footnotesize (Color online) Plot (a) shows the Weyl cone around W1 point of TaAs. Plot (b) shows the extent to which red cone overlaps the balck one when inversion operation is applied on it with respect to the Weyl node. Plot (c) shows the extent to which red cone overlaps the balck one when mirror-reflection operation is applied on it with respect to the plane parallel to $k_x$-$k_y$ plane and passing through Weyl energy. Plot (d) shows the components of $S^{xx}$, $S^{yy}$ \& $S^{zz}$ contributed from this cone.}}
    }
    \label{fig_cone_taas}
\end{figure*}

\begin{figure*}
    \centering
    \subfigure[]
    {
        \includegraphics[width=0.30\linewidth, height=4.0cm]{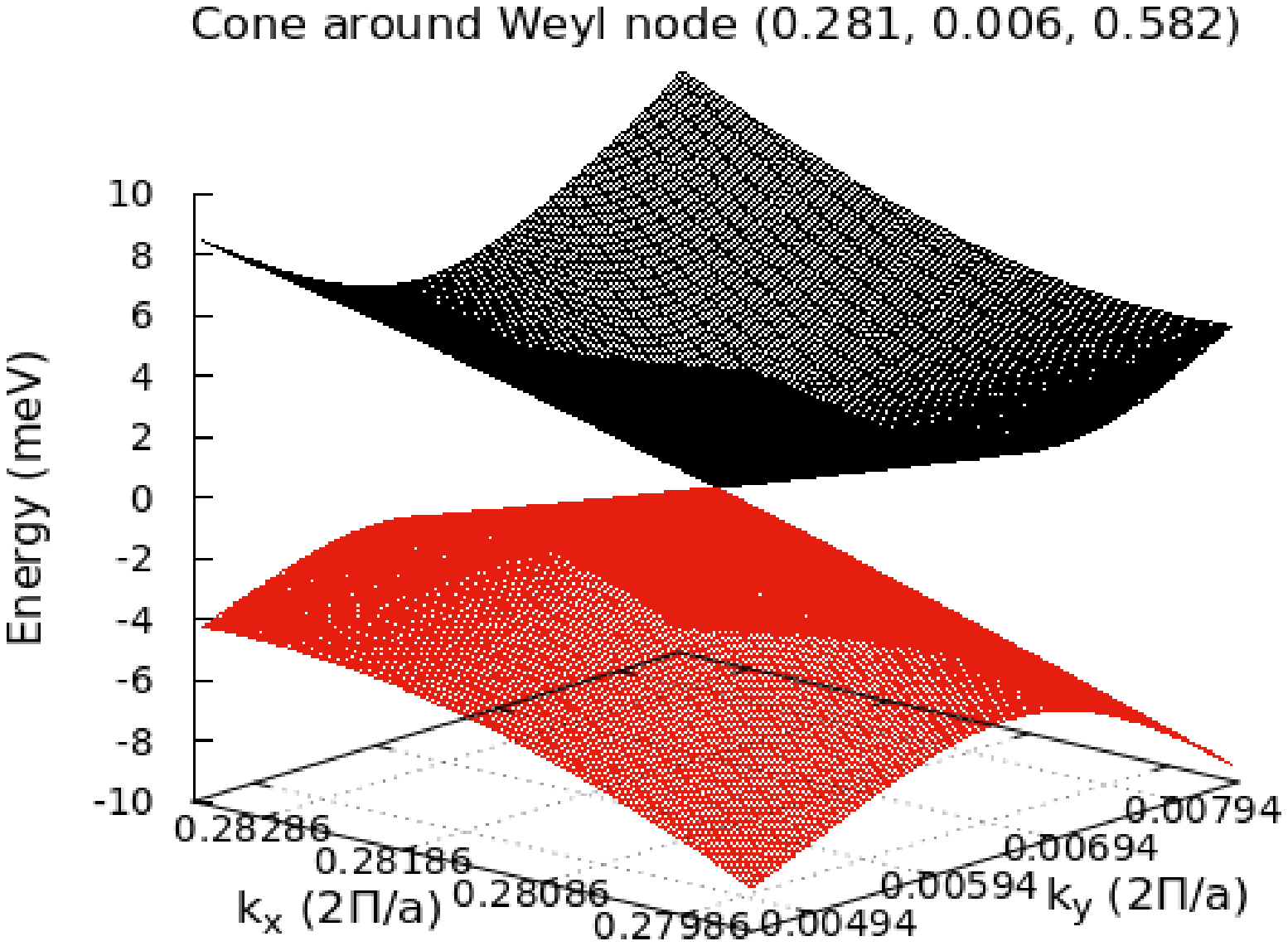}
    }
    \subfigure[]
    {
        \includegraphics[width=0.30\linewidth, height=4.0cm]{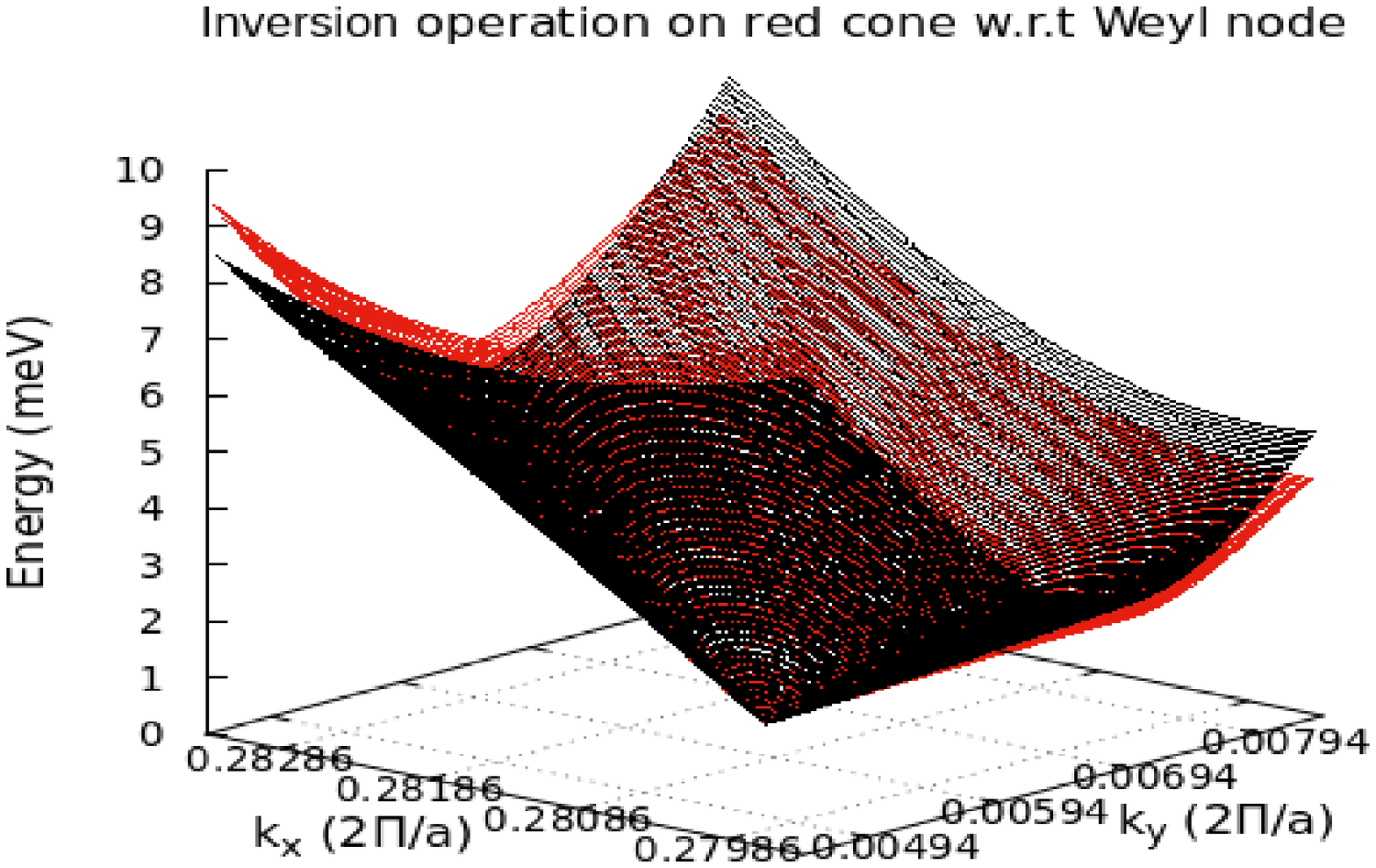}
    }\\
    \subfigure[]
    {
        \includegraphics[width=0.30\linewidth, height=4.0cm]{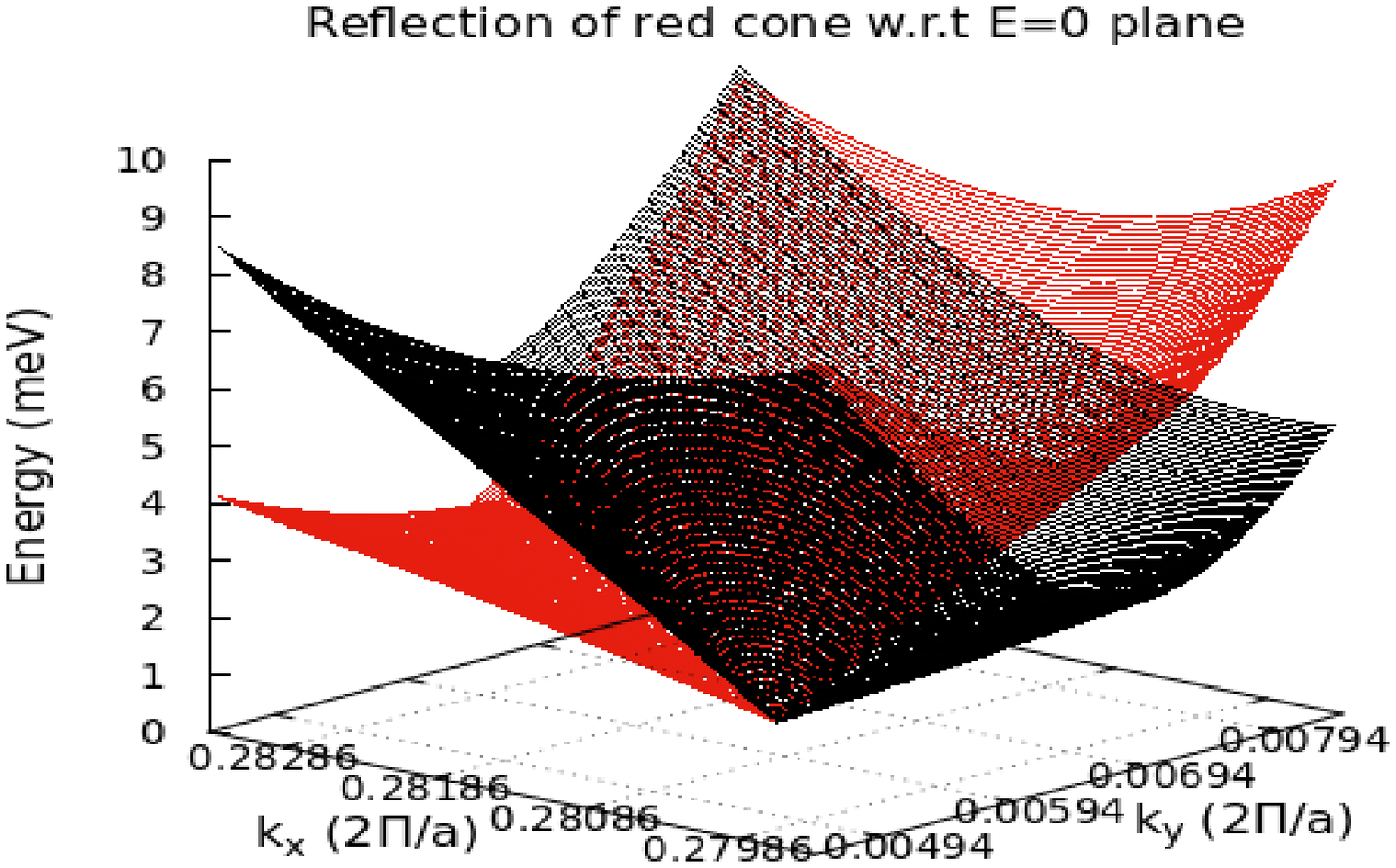}
    }
    \subfigure[]
    {
        \includegraphics[width=0.30\linewidth, height=4.0cm]{seebeck/NbAs/1/NbAs_1.eps}
    }
    \caption
    { {{ \footnotesize (Color online) Plot (a) shows the Weyl cone around W2 point of NbAs. Plot (b) shows the extent to which red cone overlaps the balck one when inversion operation is applied on it with respect to the Weyl node. Plot (c) shows the extent to which red cone overlaps the balck one when mirror-reflection operation is applied on it with respect to the plane parallel to $k_x$-$k_y$ plane and passing through Weyl energy. Plot (d) shows the components of $S^{xx}$, $S^{yy}$ \& $S^{zz}$ contributed from this cone.}}
    }
    \label{fig_cone_taas}
\end{figure*}

\begin{figure*}
    \centering
    \subfigure[]
    {
        \includegraphics[width=0.30\linewidth, height=4.0cm]{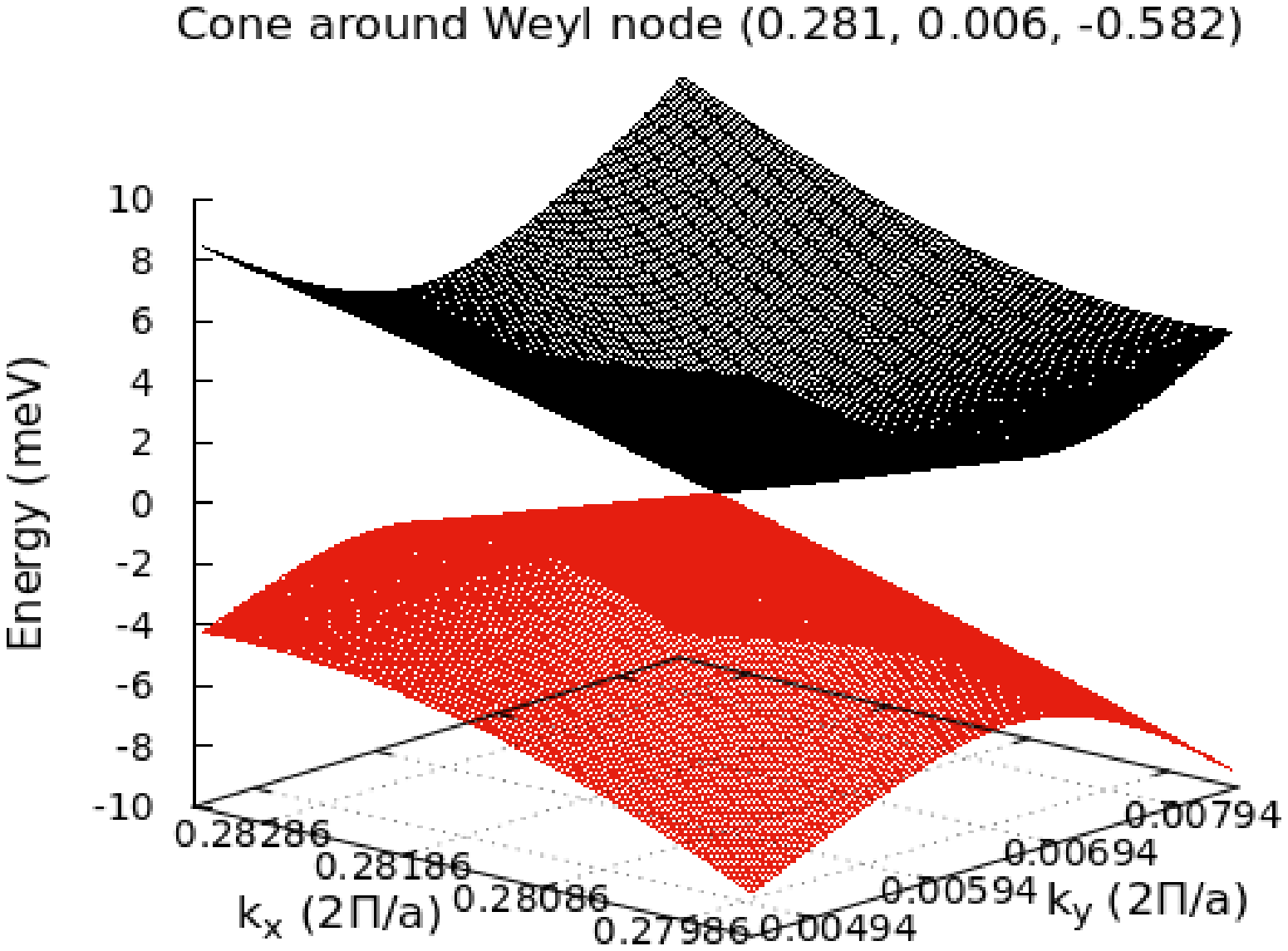}
    }
    \subfigure[]
    {
        \includegraphics[width=0.30\linewidth, height=4.0cm]{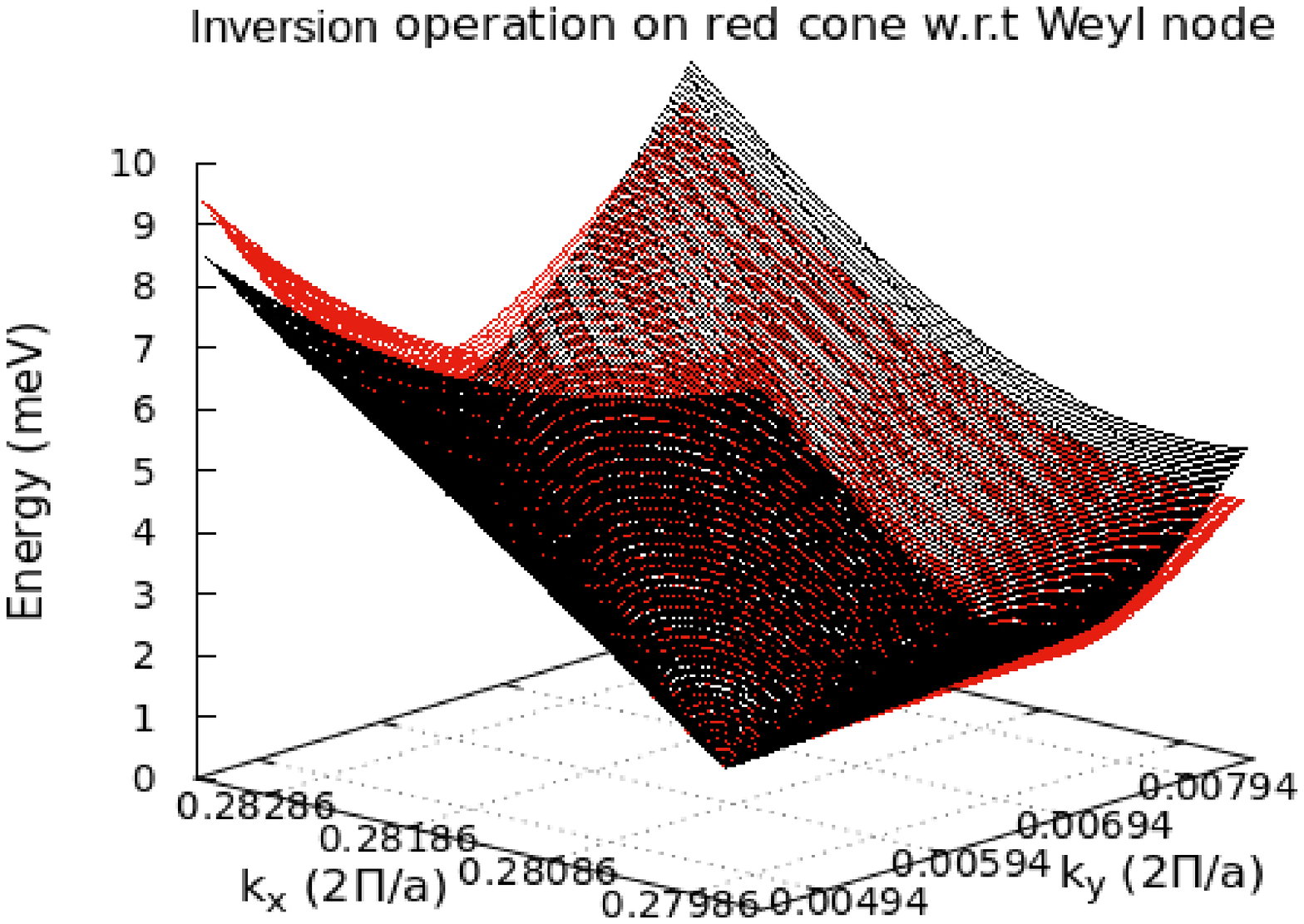}
    }\\
    \subfigure[]
    {
        \includegraphics[width=0.30\linewidth, height=4.0cm]{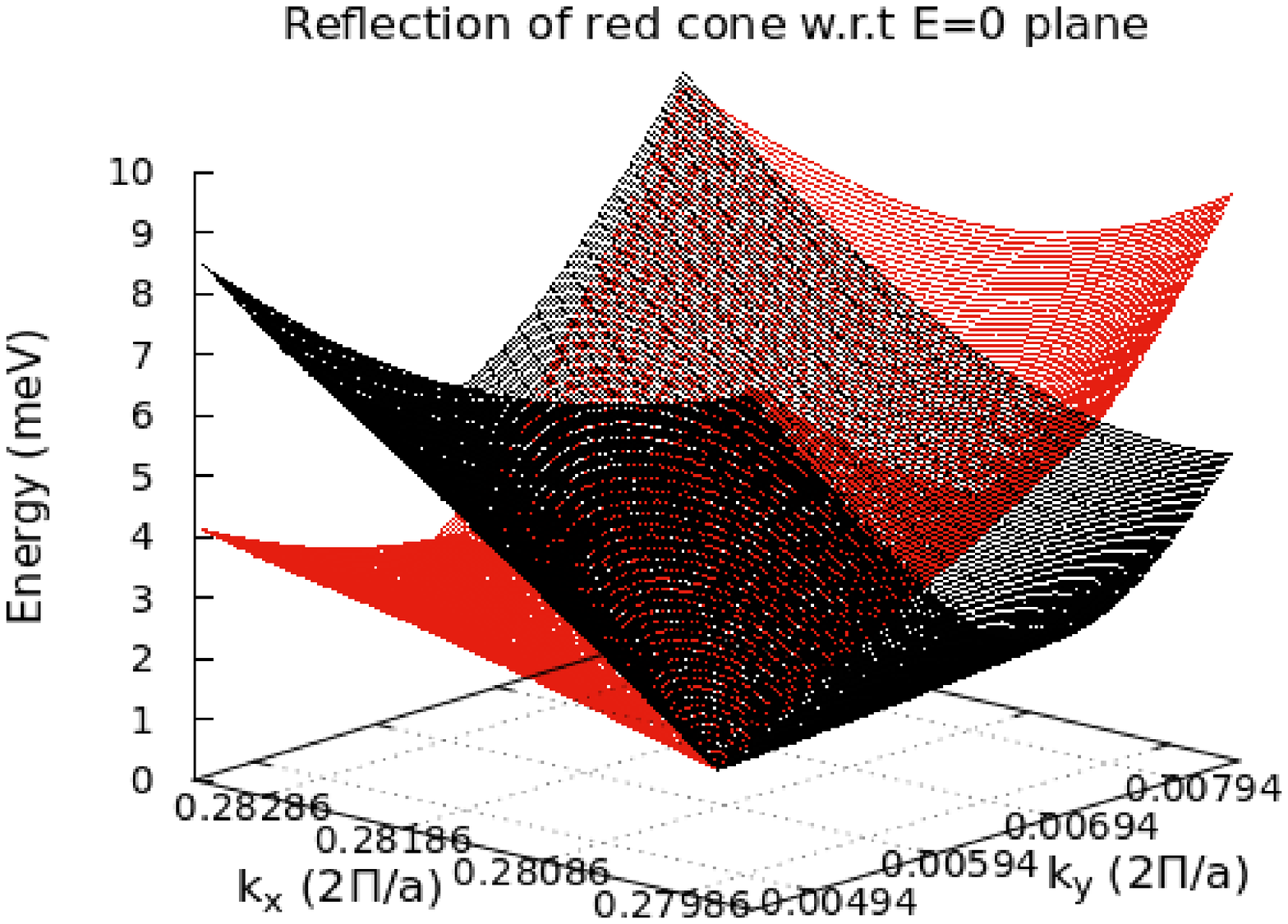}
    }
    \subfigure[]
    {
        \includegraphics[width=0.30\linewidth, height=4.0cm]{seebeck/NbAs/2/NbAs_2.eps}
    }
    \caption
    { {{ \footnotesize (Color online) Plot (a) shows the Weyl cone around W2 point of NbAs. Plot (b) shows the extent to which red cone overlaps the balck one when inversion operation is applied on it with respect to the Weyl node. Plot (c) shows the extent to which red cone overlaps the balck one when mirror-reflection operation is applied on it with respect to the plane parallel to $k_x$-$k_y$ plane and passing through Weyl energy. Plot (d) shows the components of $S^{xx}$, $S^{yy}$ \& $S^{zz}$ contributed from this cone.}}
    }
    \label{fig_cone_taas}
\end{figure*}

\begin{figure*}
    \centering
    \subfigure[]
    {
        \includegraphics[width=0.30\linewidth, height=4.0cm]{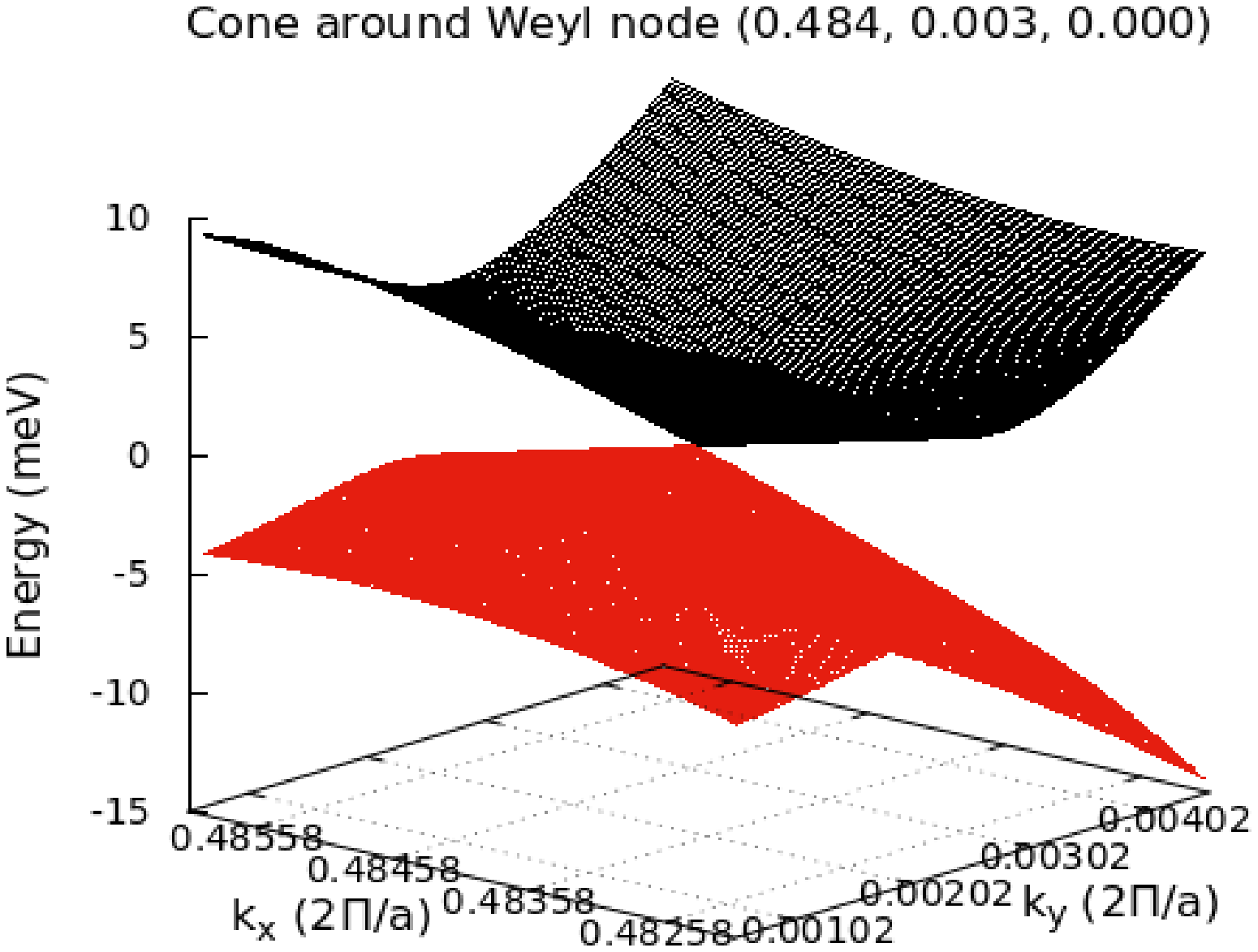}
    }
    \subfigure[]
    {
        \includegraphics[width=0.30\linewidth, height=4.0cm]{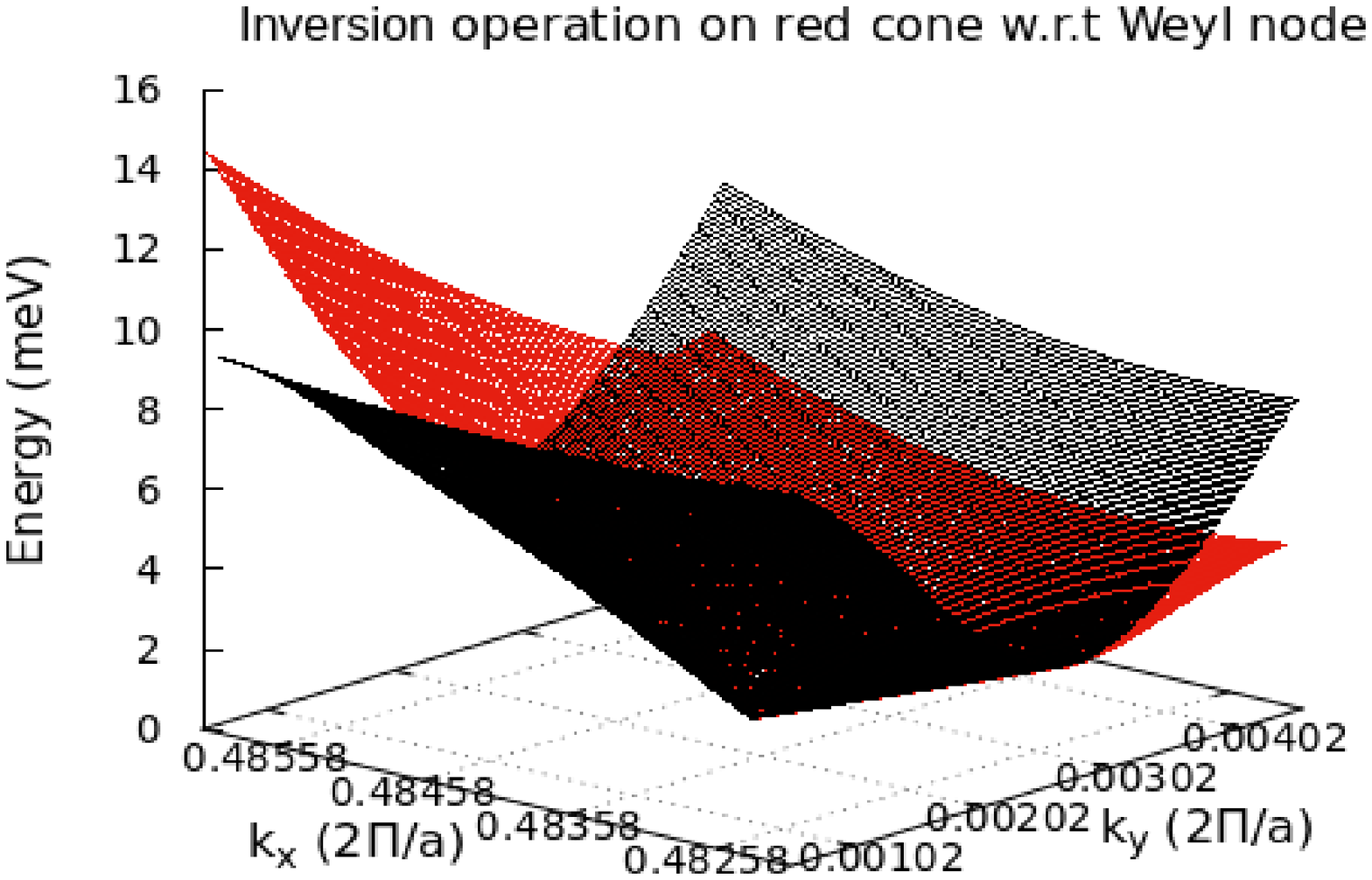}
    }\\
    \subfigure[]
    {
        \includegraphics[width=0.30\linewidth, height=4.0cm]{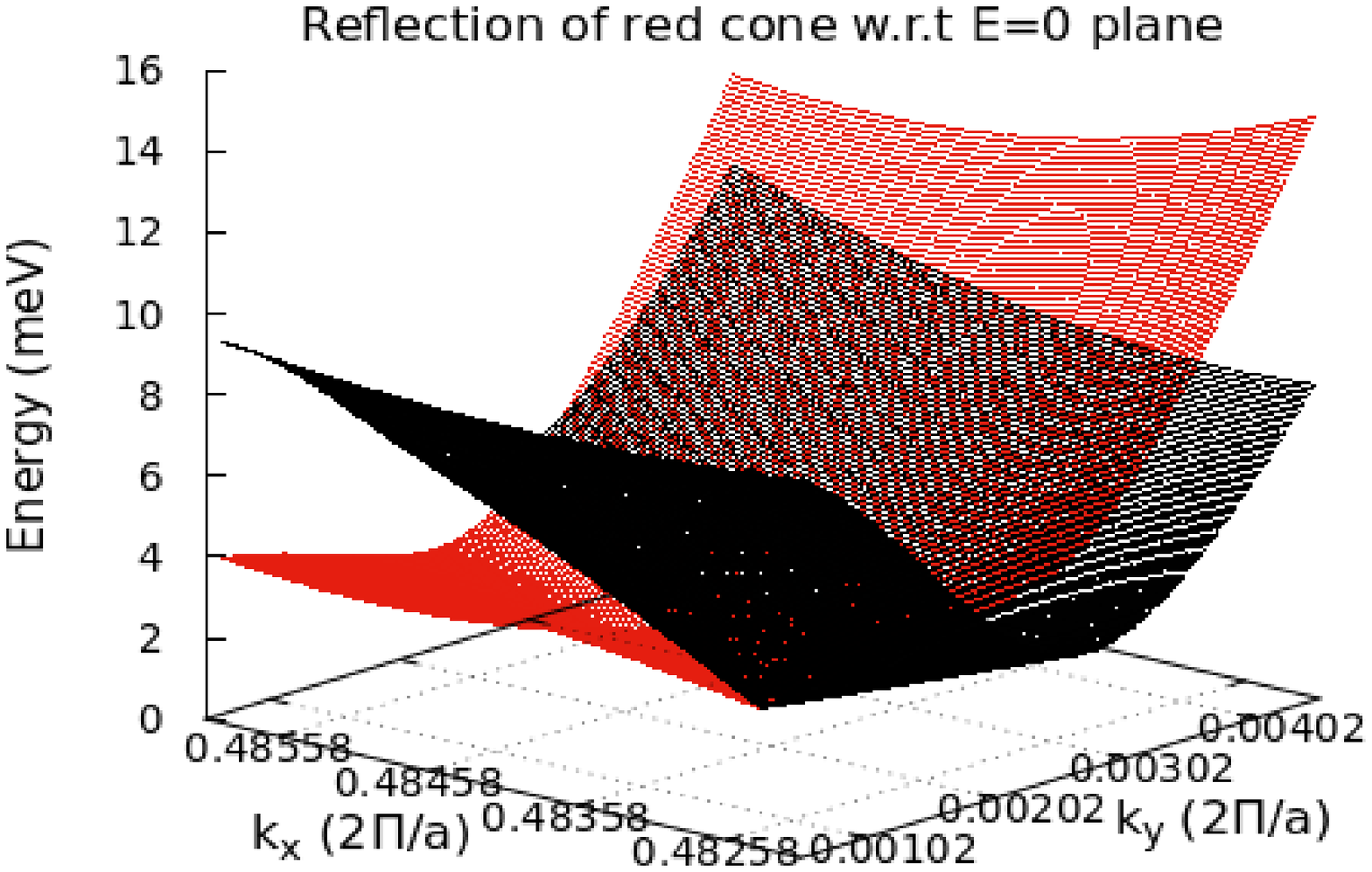}
    }
    \subfigure[]
    {
        \includegraphics[width=0.30\linewidth, height=4.0cm]{seebeck/NbAs/3/NbAs_3.eps}
    }
    \caption
    { {{ \footnotesize (Color online) Plot (a) shows the Weyl cone around W1 point of NbAs. Plot (b) shows the extent to which red cone overlaps the balck one when inversion operation is applied on it with respect to the Weyl node. Plot (c) shows the extent to which red cone overlaps the balck one when mirror-reflection operation is applied on it with respect to the plane parallel to $k_x$-$k_y$ plane and passing through Weyl energy. Plot (d) shows the components of $S^{xx}$, $S^{yy}$ \& $S^{zz}$ contributed from this cone.}}
    }
    \label{fig_cone_taas}
\end{figure*}

\begin{figure*}
    \centering
    \subfigure[]
    {
        \includegraphics[width=0.30\linewidth, height=4.0cm]{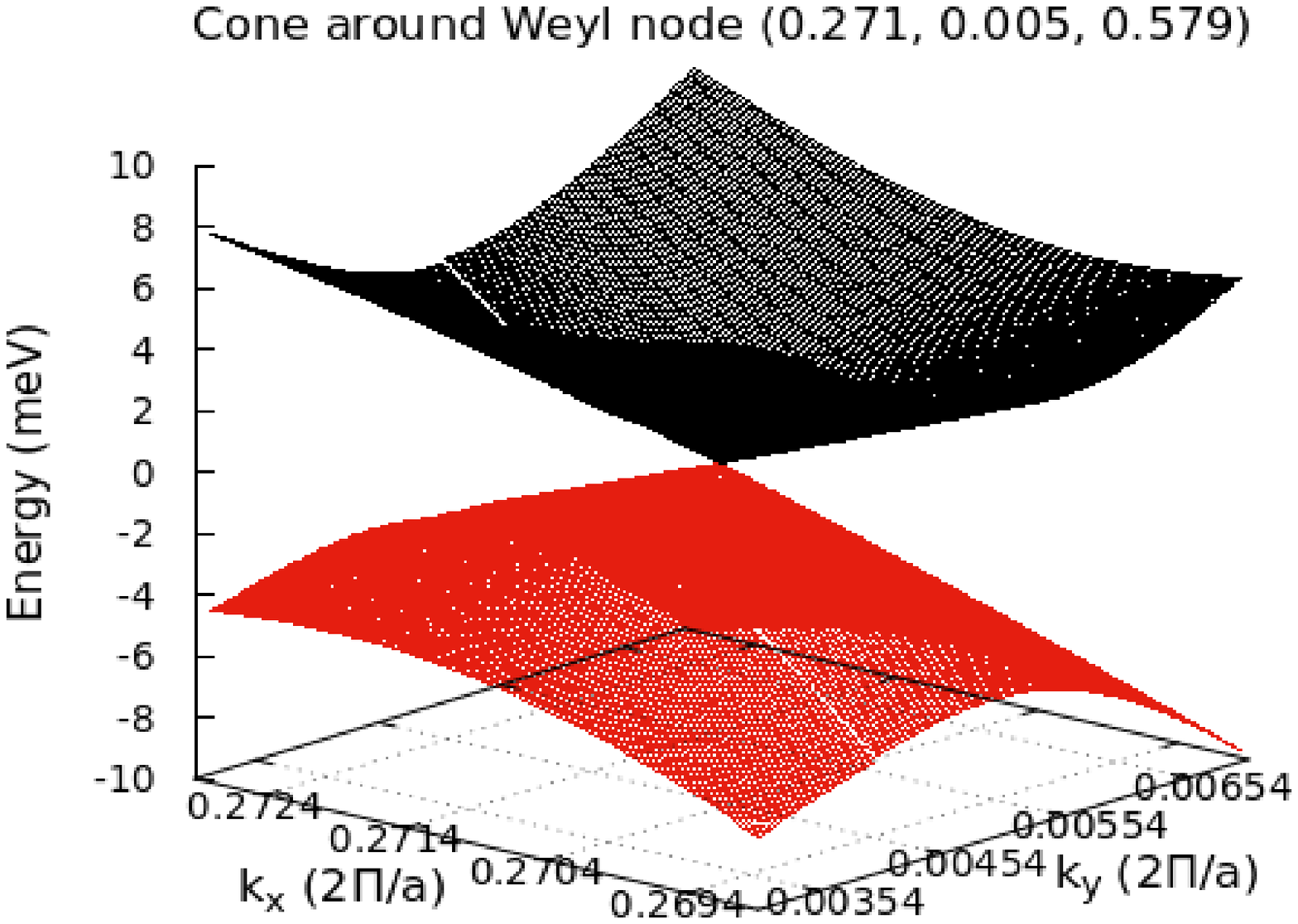}
    }
    \subfigure[]
    {
        \includegraphics[width=0.30\linewidth, height=4.0cm]{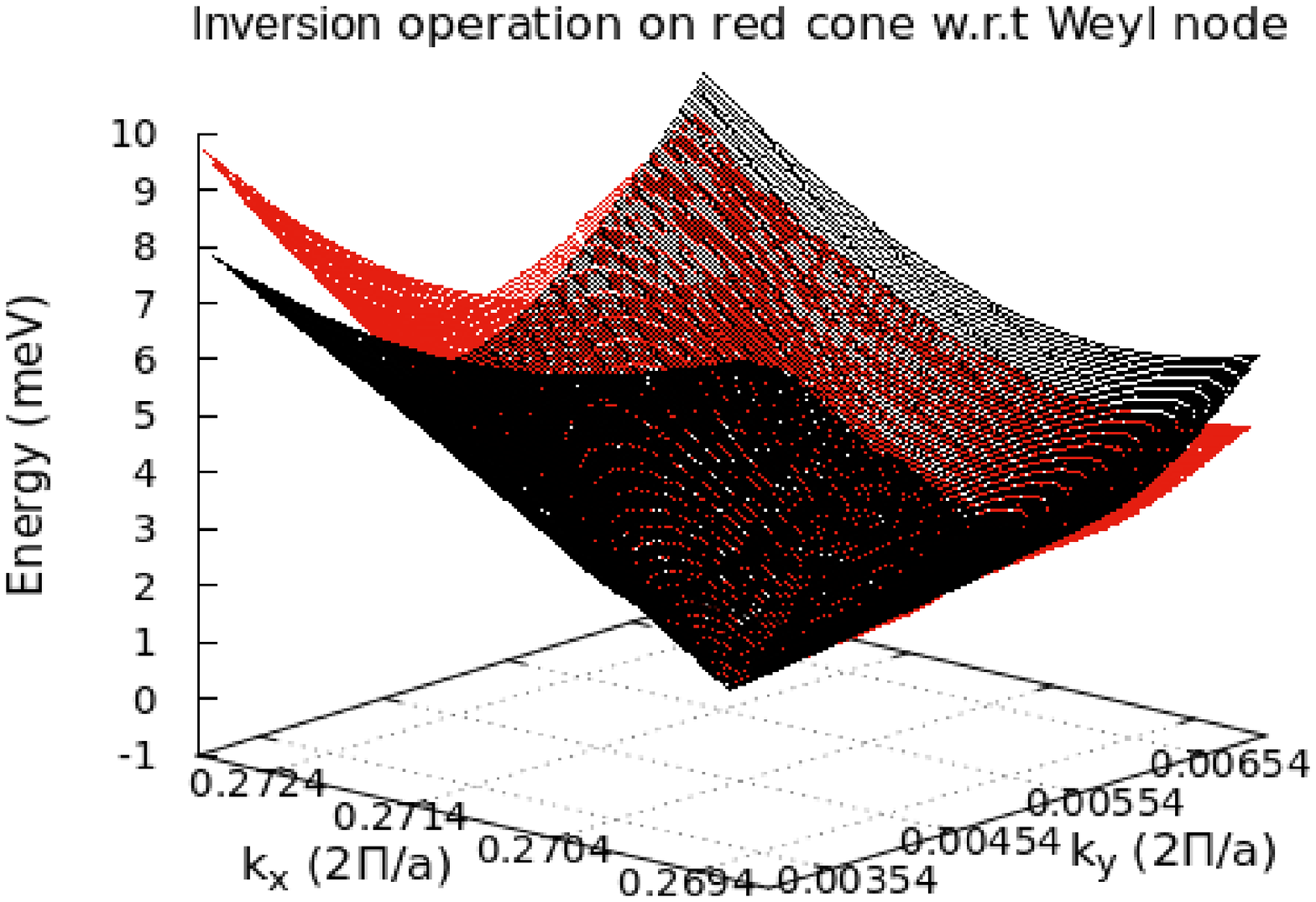}
    }\\
    \subfigure[]
    {
        \includegraphics[width=0.30\linewidth, height=4.0cm]{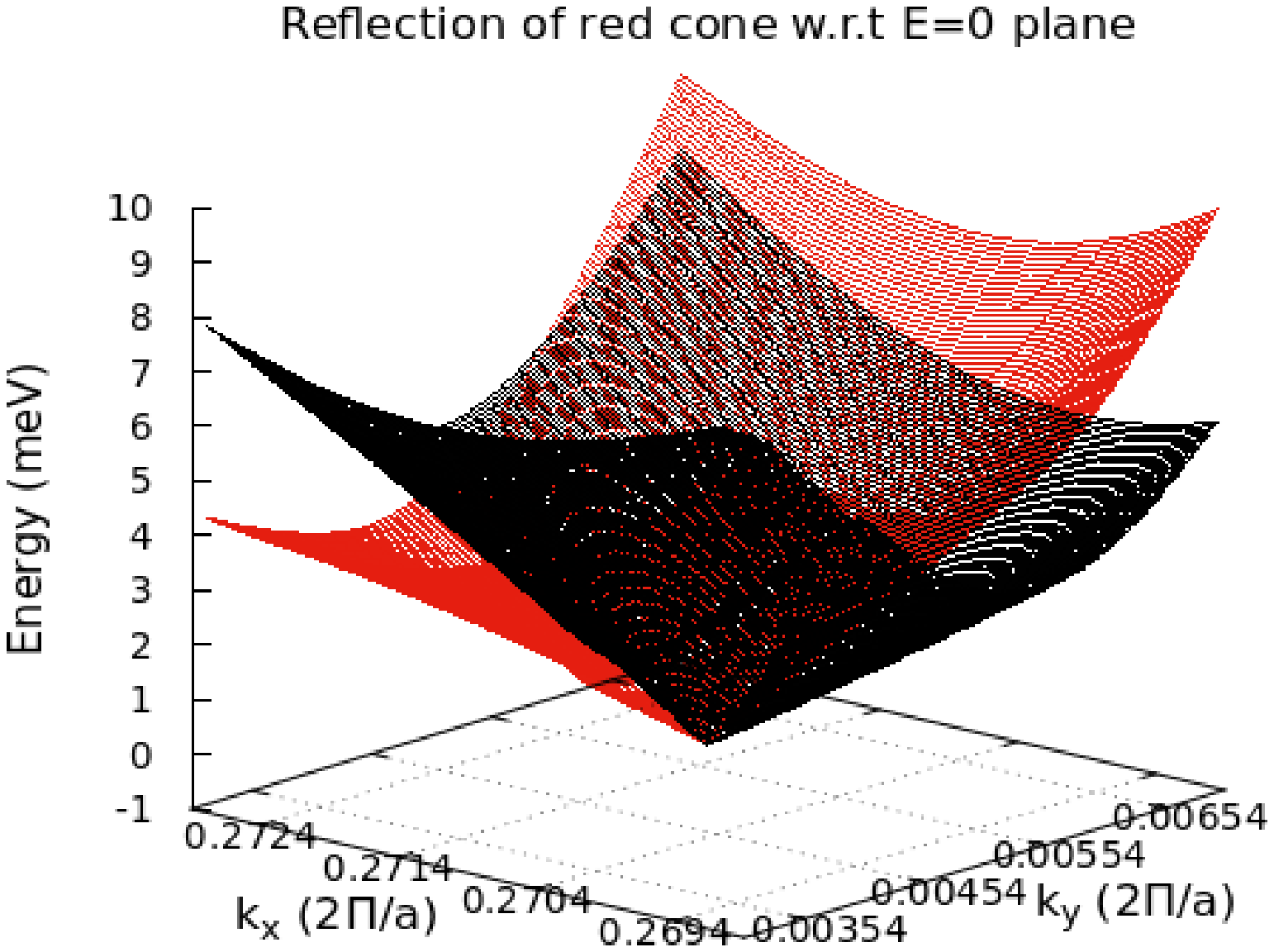}
    }
    \subfigure[]
    {
        \includegraphics[width=0.30\linewidth, height=4.0cm]{seebeck/NbP/1/NbP_1.eps}
    }
    \caption
    { {{ \footnotesize (Color online) Plot (a) shows the Weyl cone around W2 point of NbP. Plot (b) shows the extent to which red cone overlaps the balck one when inversion operation is applied on it with respect to the Weyl node. Plot (c) shows the extent to which red cone overlaps the balck one when mirror-reflection operation is applied on it with respect to the plane parallel to $k_x$-$k_y$ plane and passing through Weyl energy. Plot (d) shows the components of $S^{xx}$, $S^{yy}$ \& $S^{zz}$ contributed from this cone.}}
    }
    \label{fig_cone_taas}
\end{figure*}

\begin{figure*}
    \centering
    \subfigure[]
    {
        \includegraphics[width=0.30\linewidth, height=4.0cm]{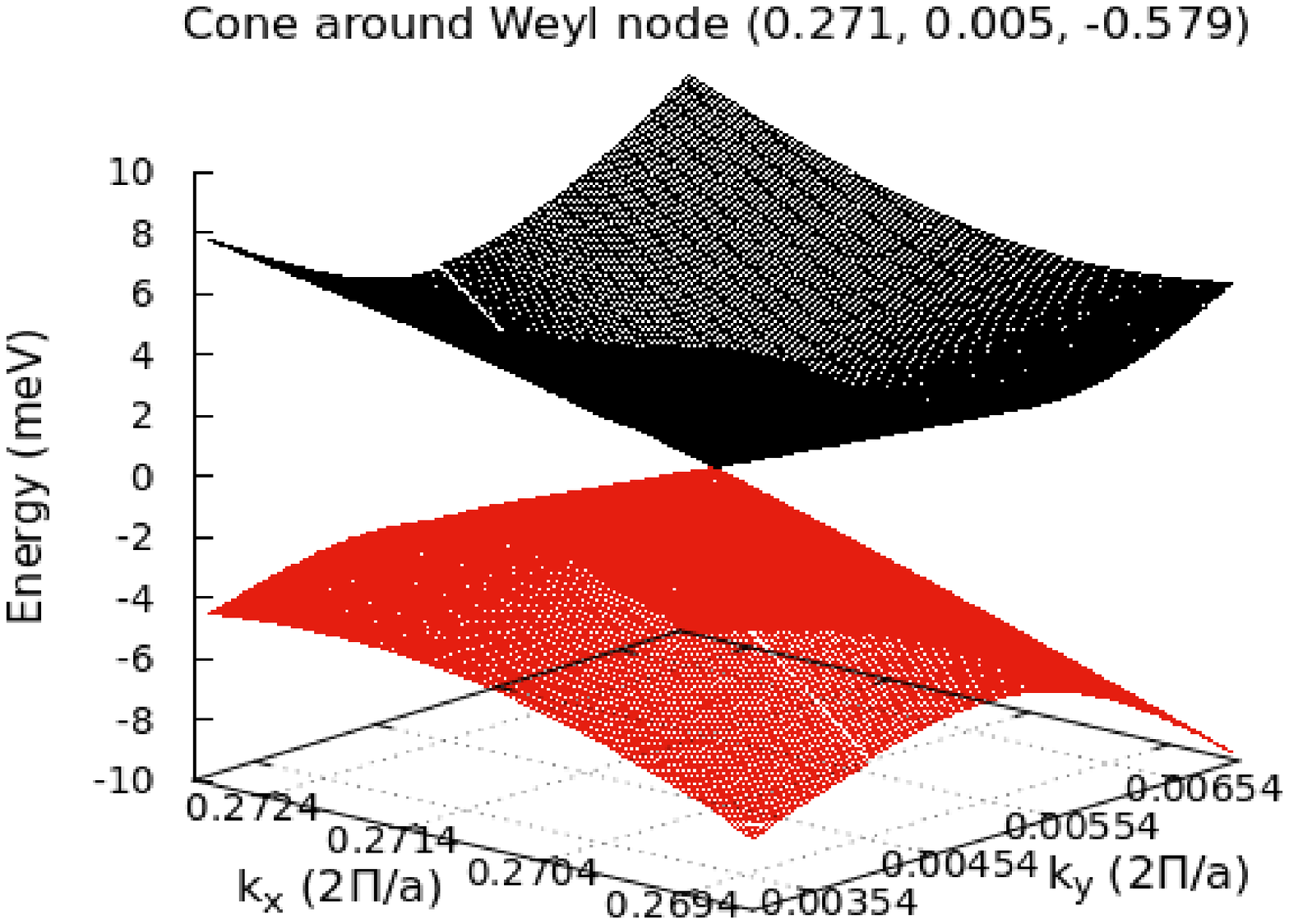}
    }
    \subfigure[]
    {
        \includegraphics[width=0.30\linewidth, height=4.0cm]{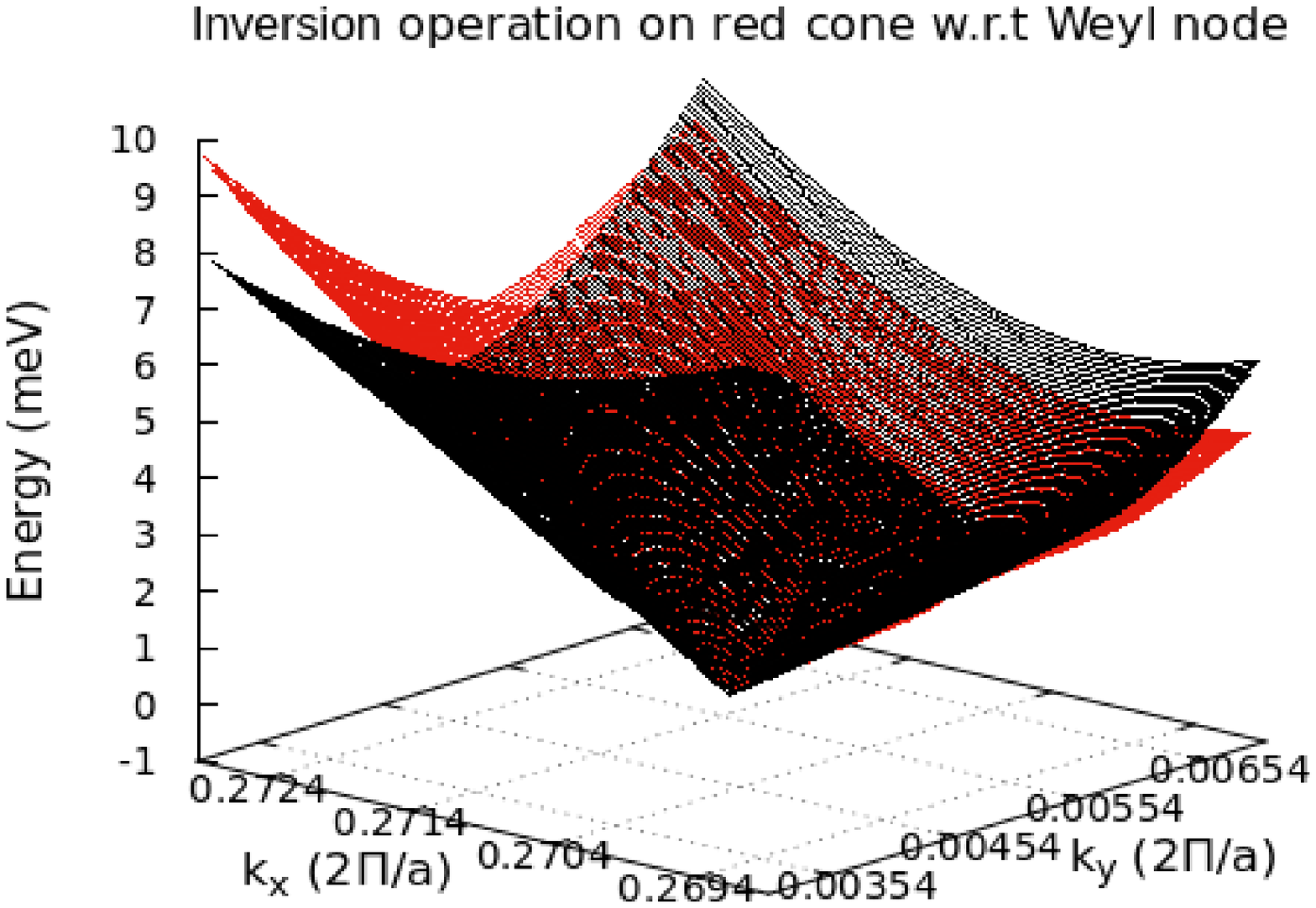}
    }\\
    \subfigure[]
    {
        \includegraphics[width=0.30\linewidth, height=4.0cm]{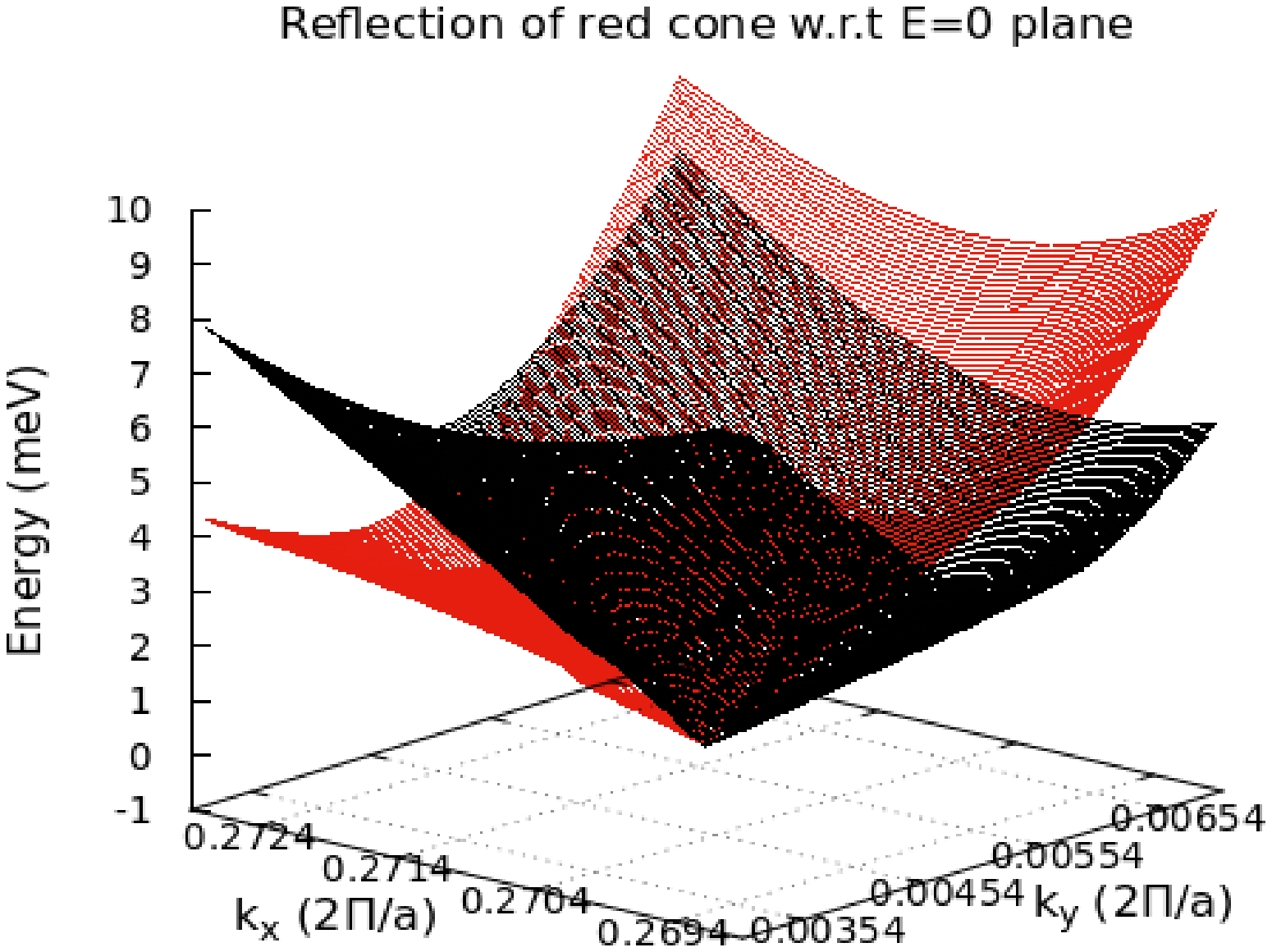}
    }
    \subfigure[]
    {
        \includegraphics[width=0.30\linewidth, height=4.0cm]{seebeck/NbP/2/NbP_2.eps}
    }
    \caption
    { {{ \footnotesize (Color online) Plot (a) shows the Weyl cone around W2 point of NbP. Plot (b) shows the extent to which red cone overlaps the balck one when inversion operation is applied on it with respect to the Weyl node. Plot (c) shows the extent to which red cone overlaps the balck one when mirror-reflection operation is applied on it with respect to the plane parallel to $k_x$-$k_y$ plane and passing through Weyl energy. Plot (d) shows the components of $S^{xx}$, $S^{yy}$ \& $S^{zz}$ contributed from this cone.}}
    }
    \label{fig_cone_taas}
\end{figure*}

\begin{figure*}
    \centering
    \subfigure[]
    {
        \includegraphics[width=0.30\linewidth, height=4.0cm]{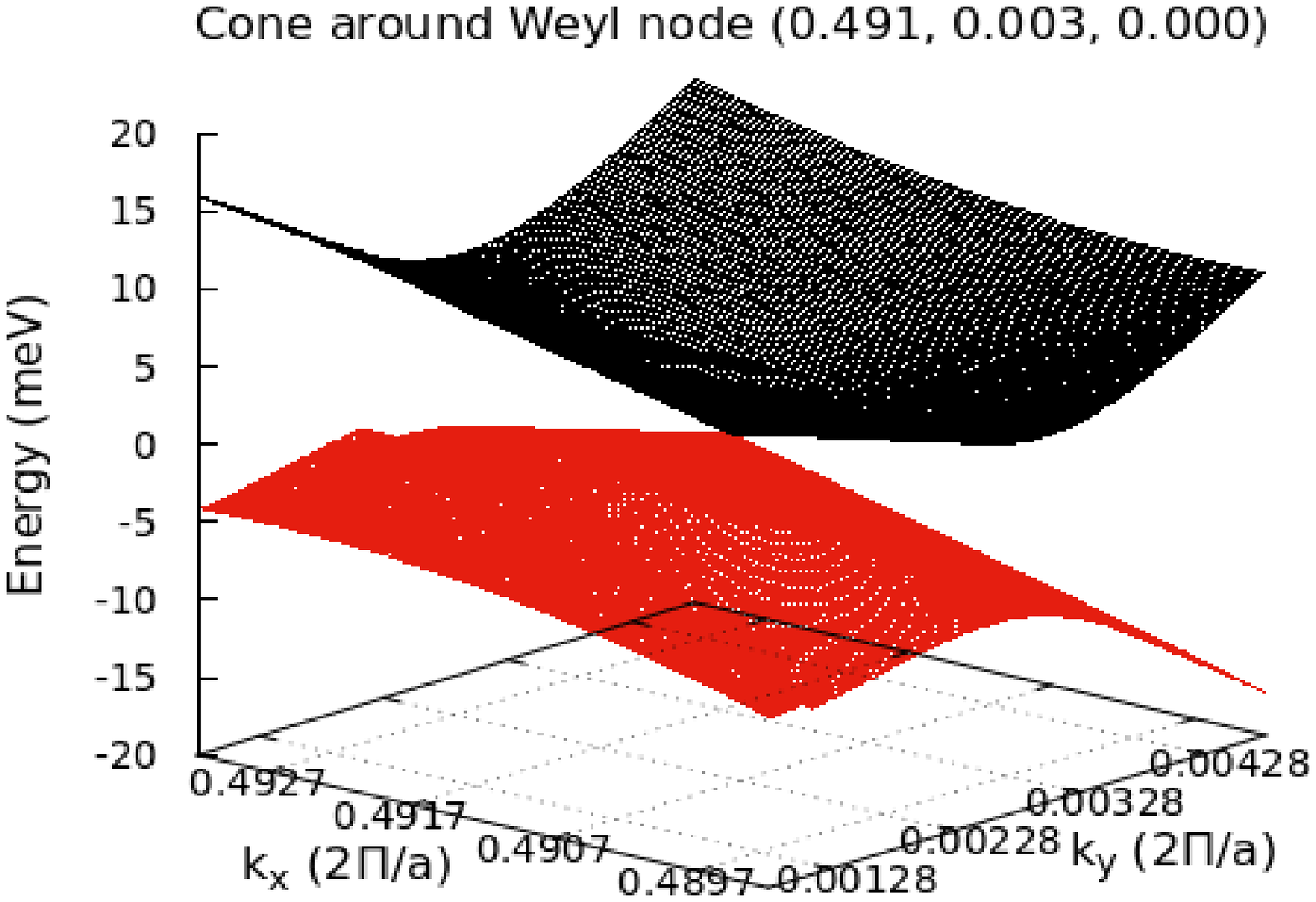}
    }
    \subfigure[]
    {
        \includegraphics[width=0.30\linewidth, height=4.0cm]{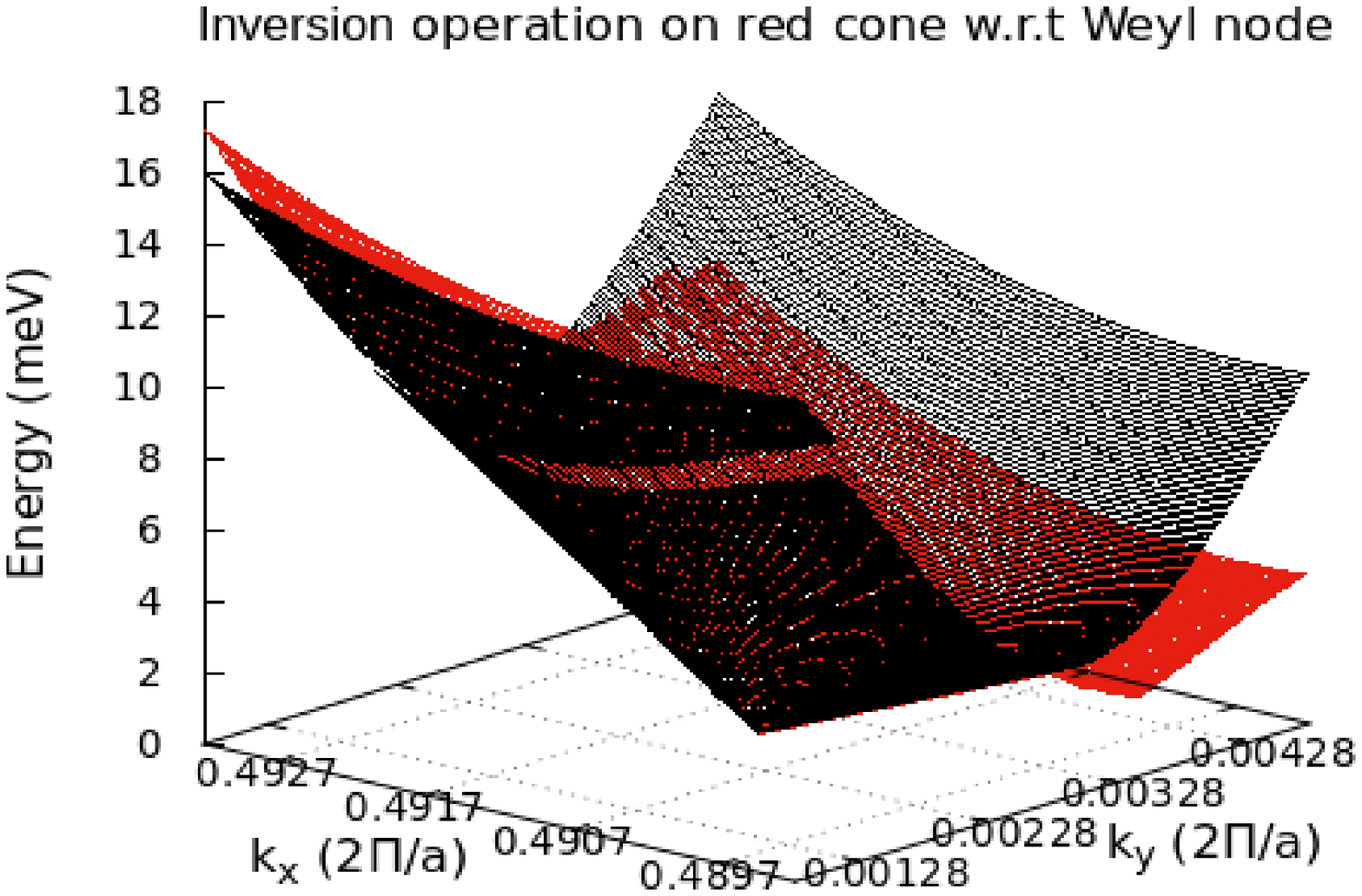}
    }\\
    \subfigure[]
    {
        \includegraphics[width=0.30\linewidth, height=4.0cm]{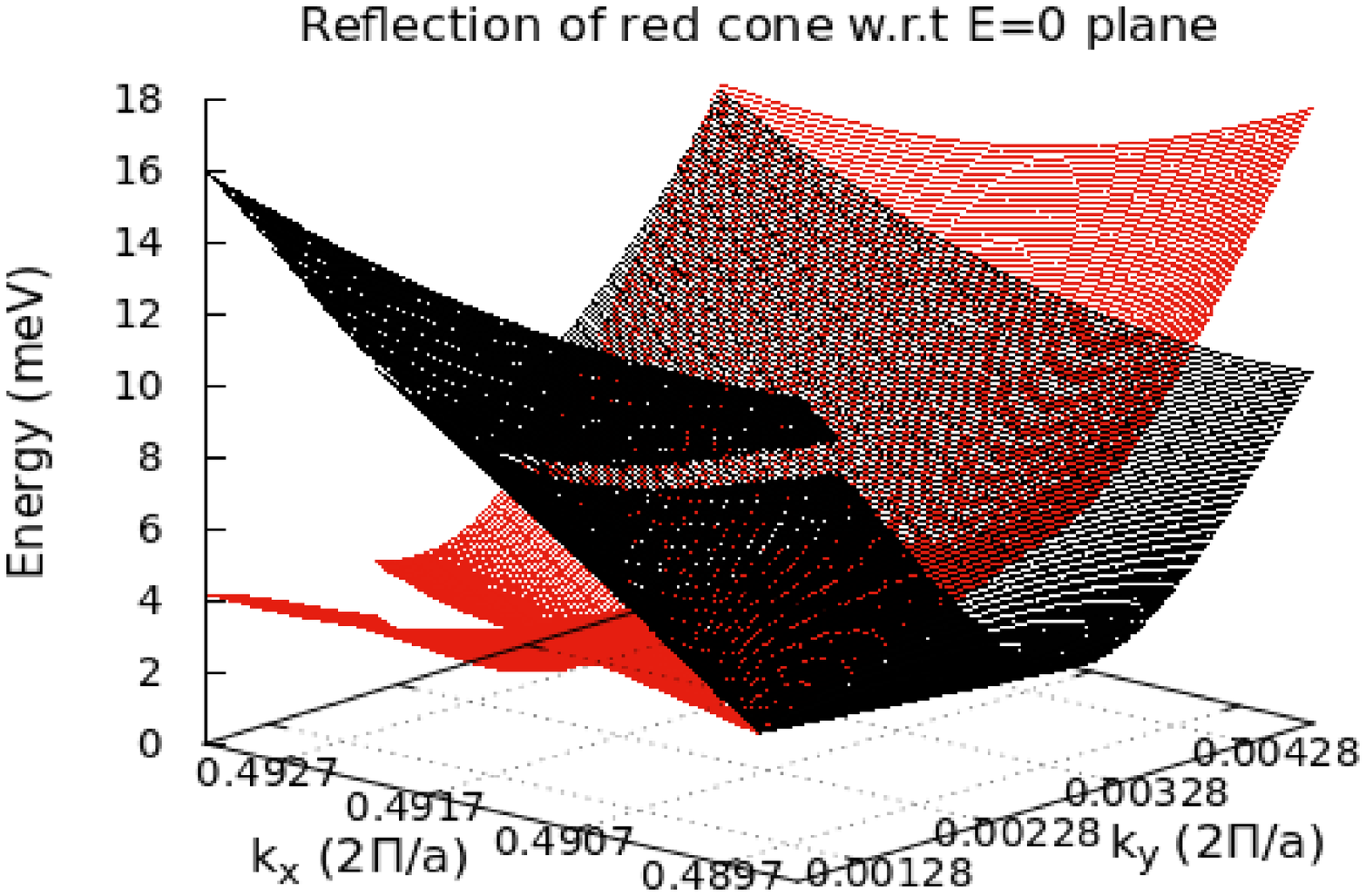}
    }
    \subfigure[]
    {
        \includegraphics[width=0.30\linewidth, height=4.0cm]{seebeck/NbP/3/NbP_3.eps}
    }
    \caption
    { {{ \footnotesize (Color online) Plot (a) shows the Weyl cone around W1 point of NbP. Plot (b) shows the extent to which red cone overlaps the balck one when inversion operation is applied on it with respect to the Weyl node. Plot (c) shows the extent to which red cone overlaps the balck one when mirror-reflection operation is applied on it with respect to the plane parallel to $k_x$-$k_y$ plane and passing through Weyl energy. Plot (d) shows the components of $S^{xx}$, $S^{yy}$ \& $S^{zz}$ contributed from this cone.}}
    }
    \label{fig_cone_taas}
\end{figure*}

\begin{figure*}
    \centering
    \subfigure[]
    {
        \includegraphics[width=0.30\linewidth, height=4.0cm]{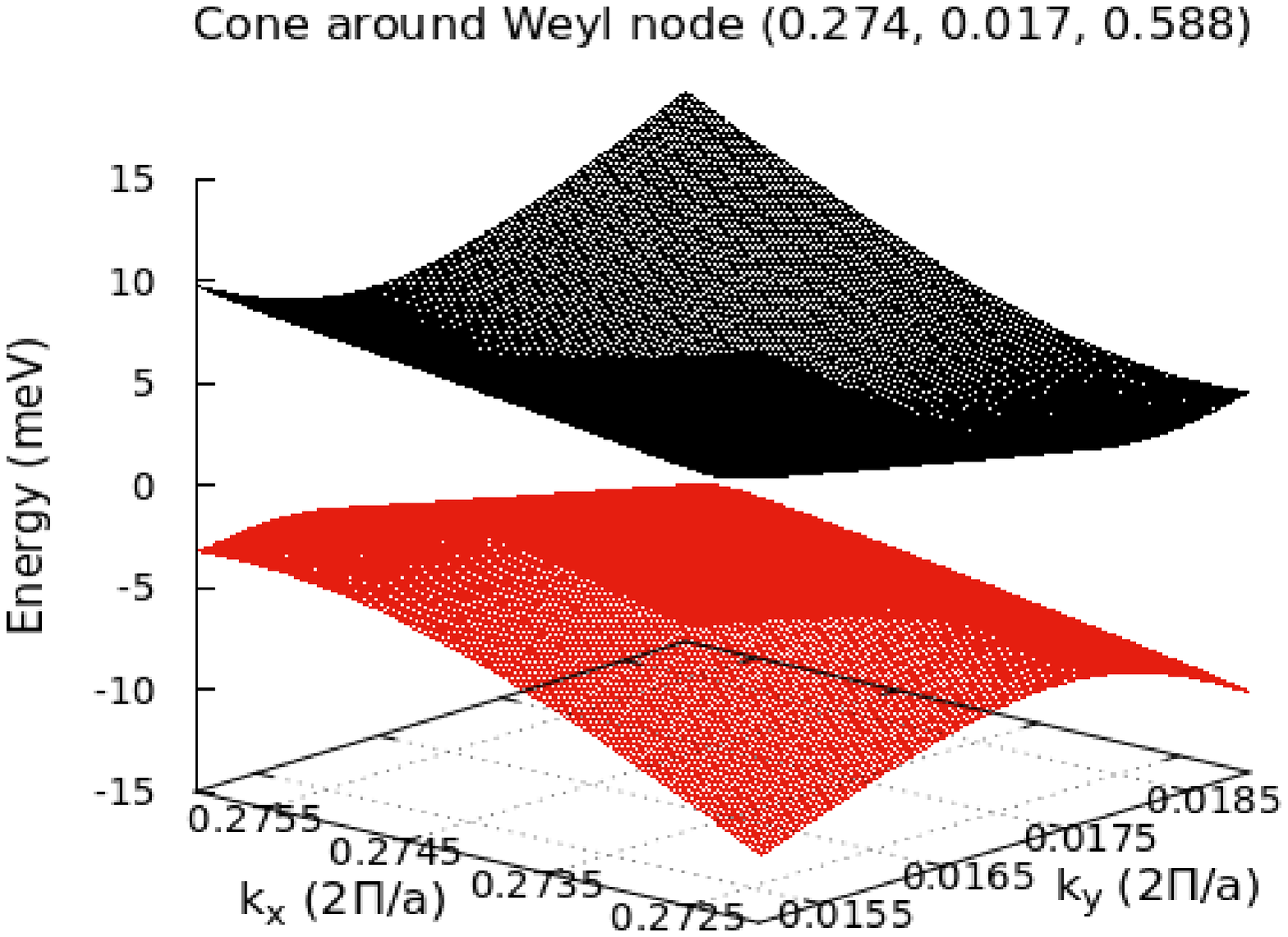}
    }
    \subfigure[]
    {
        \includegraphics[width=0.30\linewidth, height=4.0cm]{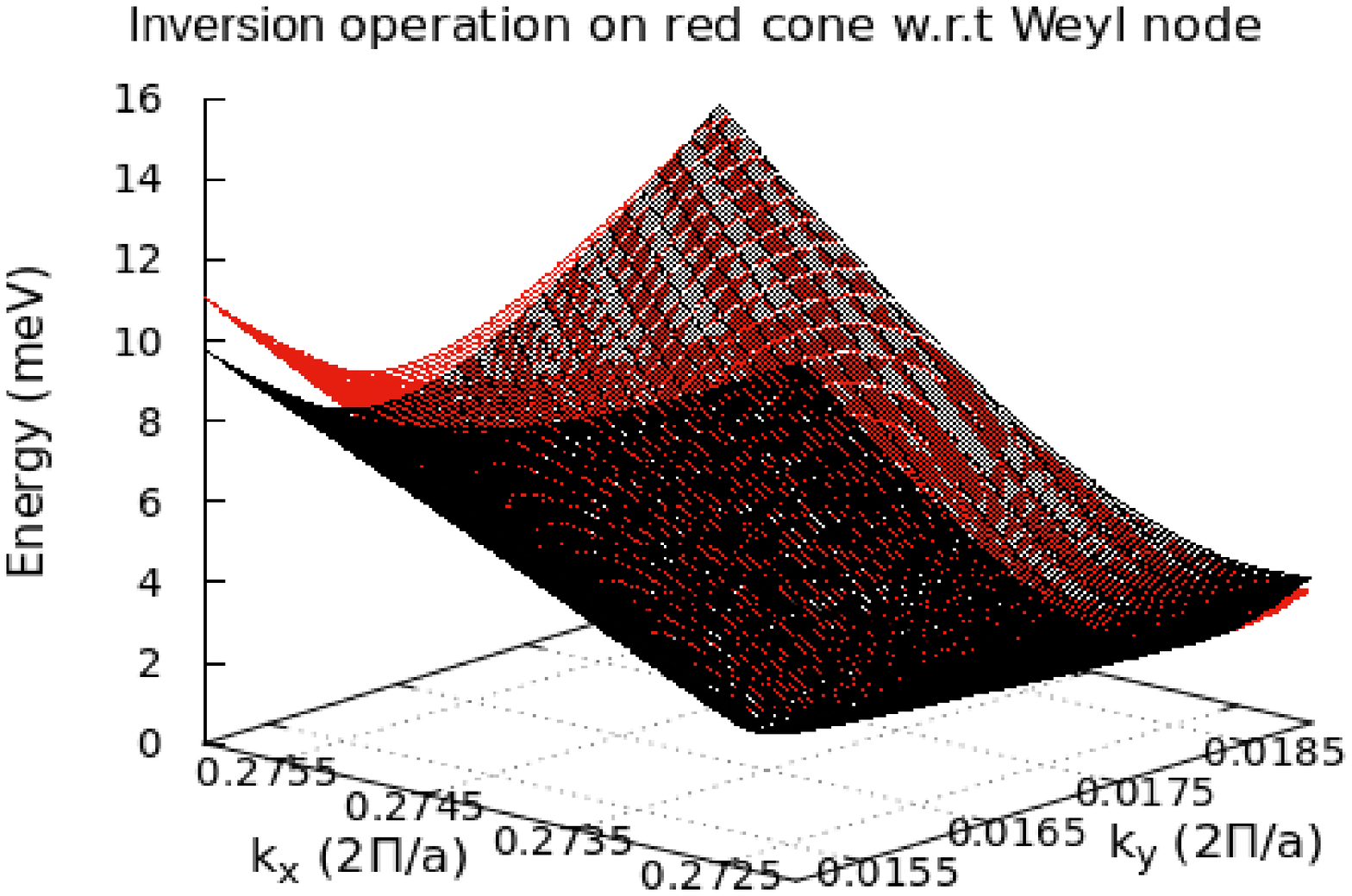}
    }\\
    \subfigure[]
    {
        \includegraphics[width=0.30\linewidth, height=4.0cm]{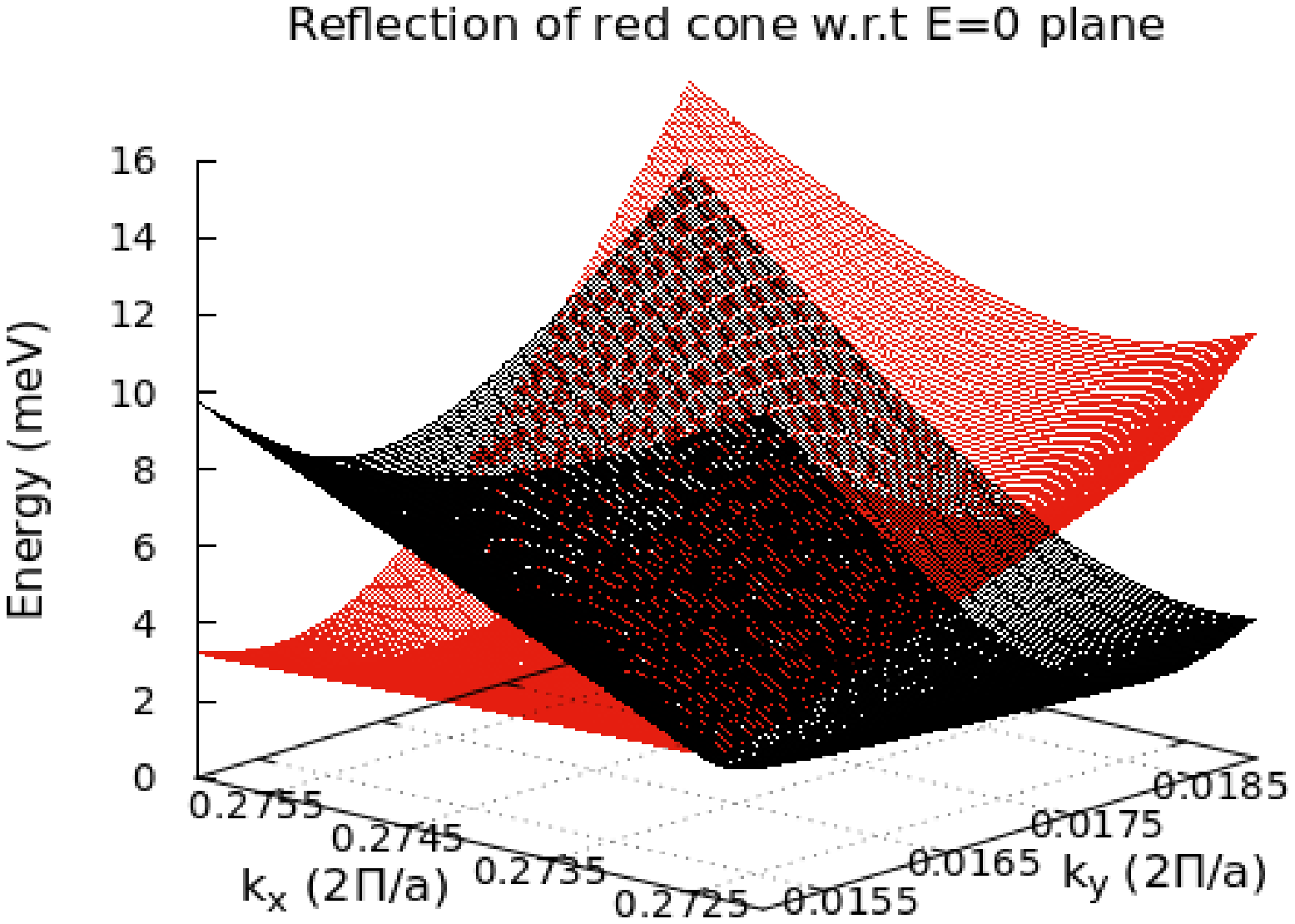}
    }
    \subfigure[]
    {
        \includegraphics[width=0.30\linewidth, height=4.0cm]{seebeck/TaP/1/TaP_1.eps}
    }
    \caption
    { {{ \footnotesize (Color online) Plot (a) shows the Weyl cone around W2 point of TaP. Plot (b) shows the extent to which red cone overlaps the balck one when inversion operation is applied on it with respect to the Weyl node. Plot (c) shows the extent to which red cone overlaps the balck one when mirror-reflection operation is applied on it with respect to the plane parallel to $k_x$-$k_y$ plane and passing through Weyl energy. Plot (d) shows the components of $S^{xx}$, $S^{yy}$ \& $S^{zz}$ contributed from this cone.}}
    }
    \label{fig_cone_taas}
\end{figure*}

\begin{figure*}
    \centering
    \subfigure[]
    {
        \includegraphics[width=0.30\linewidth, height=4.0cm]{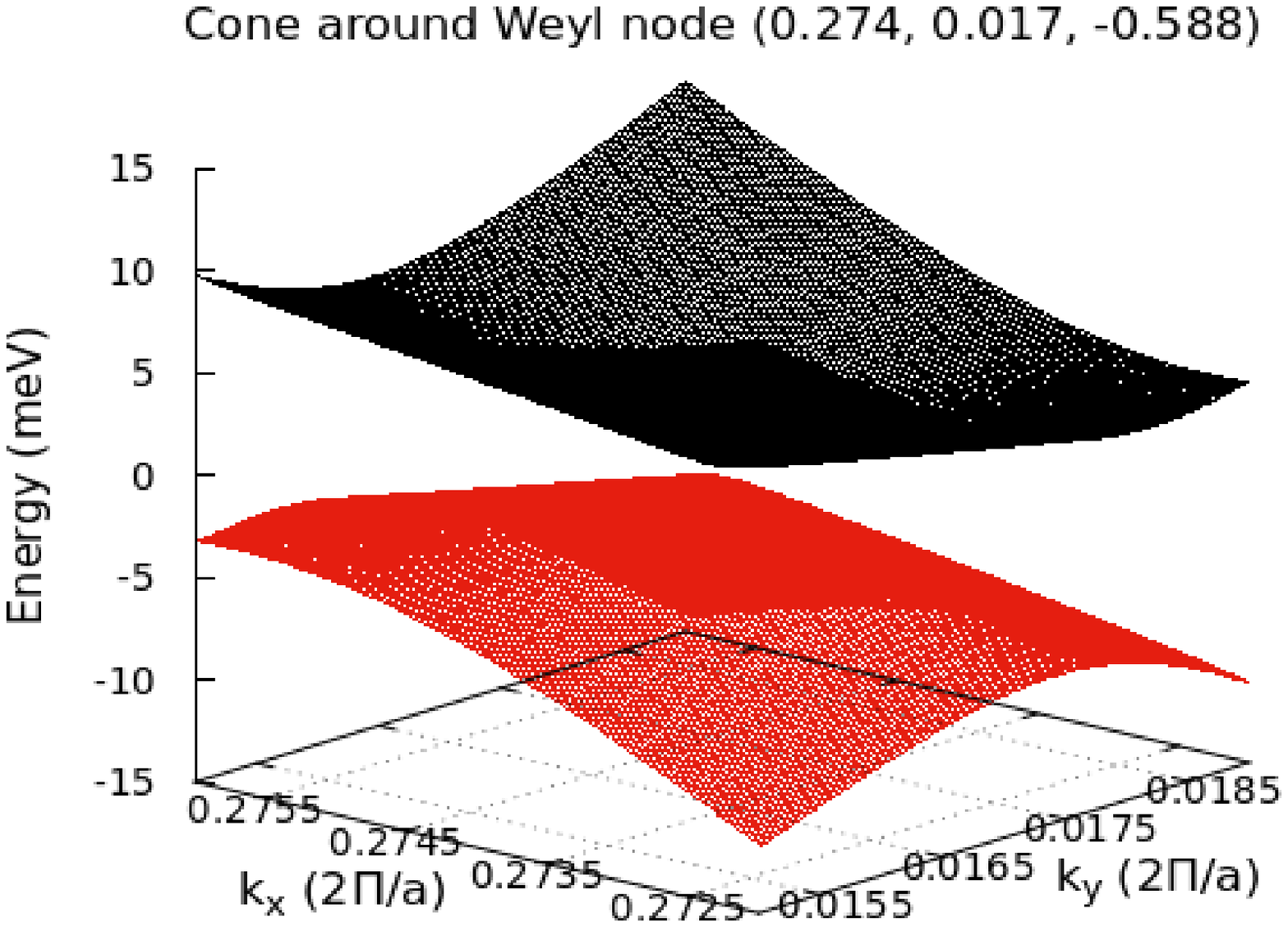}
    }
    \subfigure[]
    {
        \includegraphics[width=0.30\linewidth, height=4.0cm]{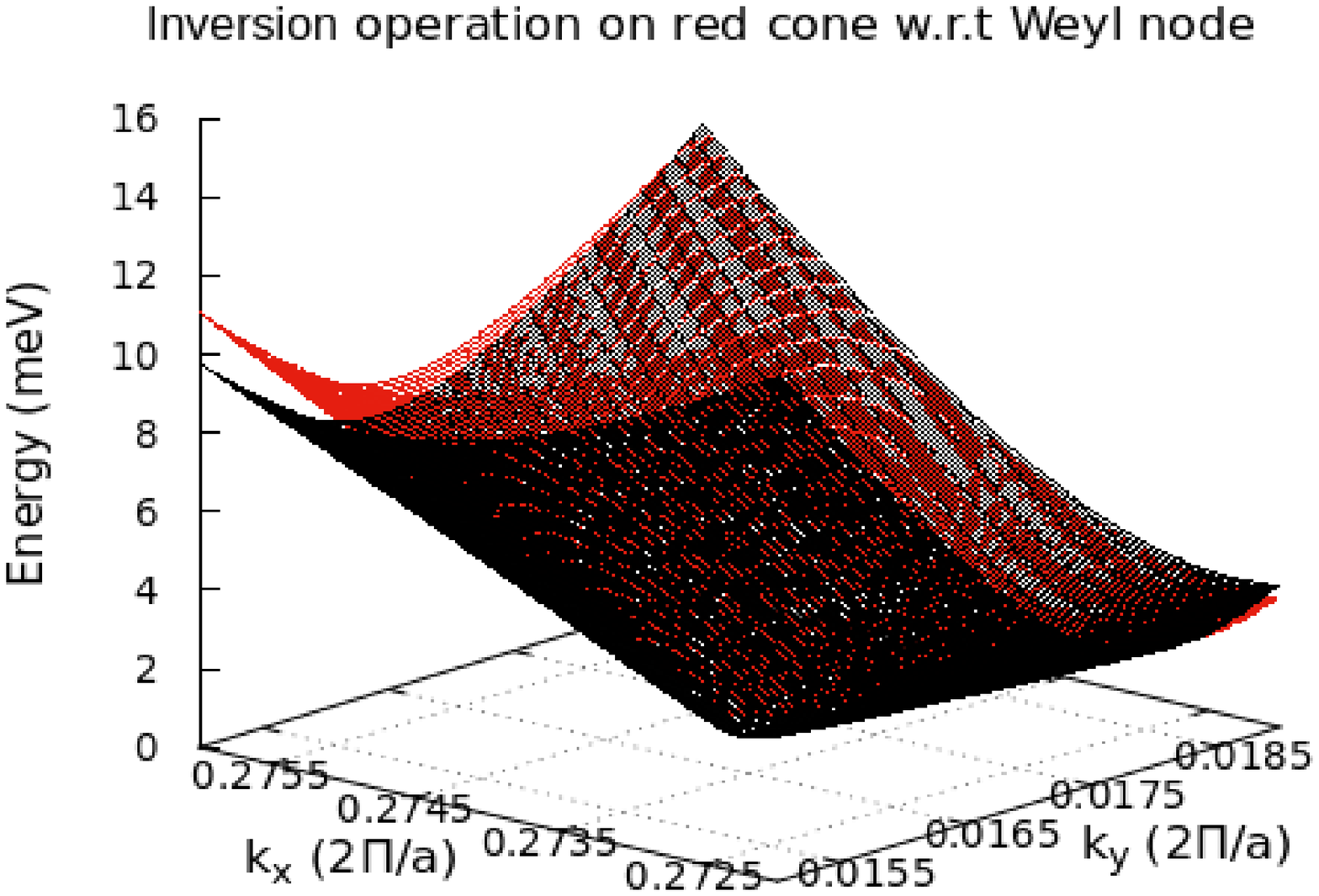}
    }\\
    \subfigure[]
    {
        \includegraphics[width=0.30\linewidth, height=4.0cm]{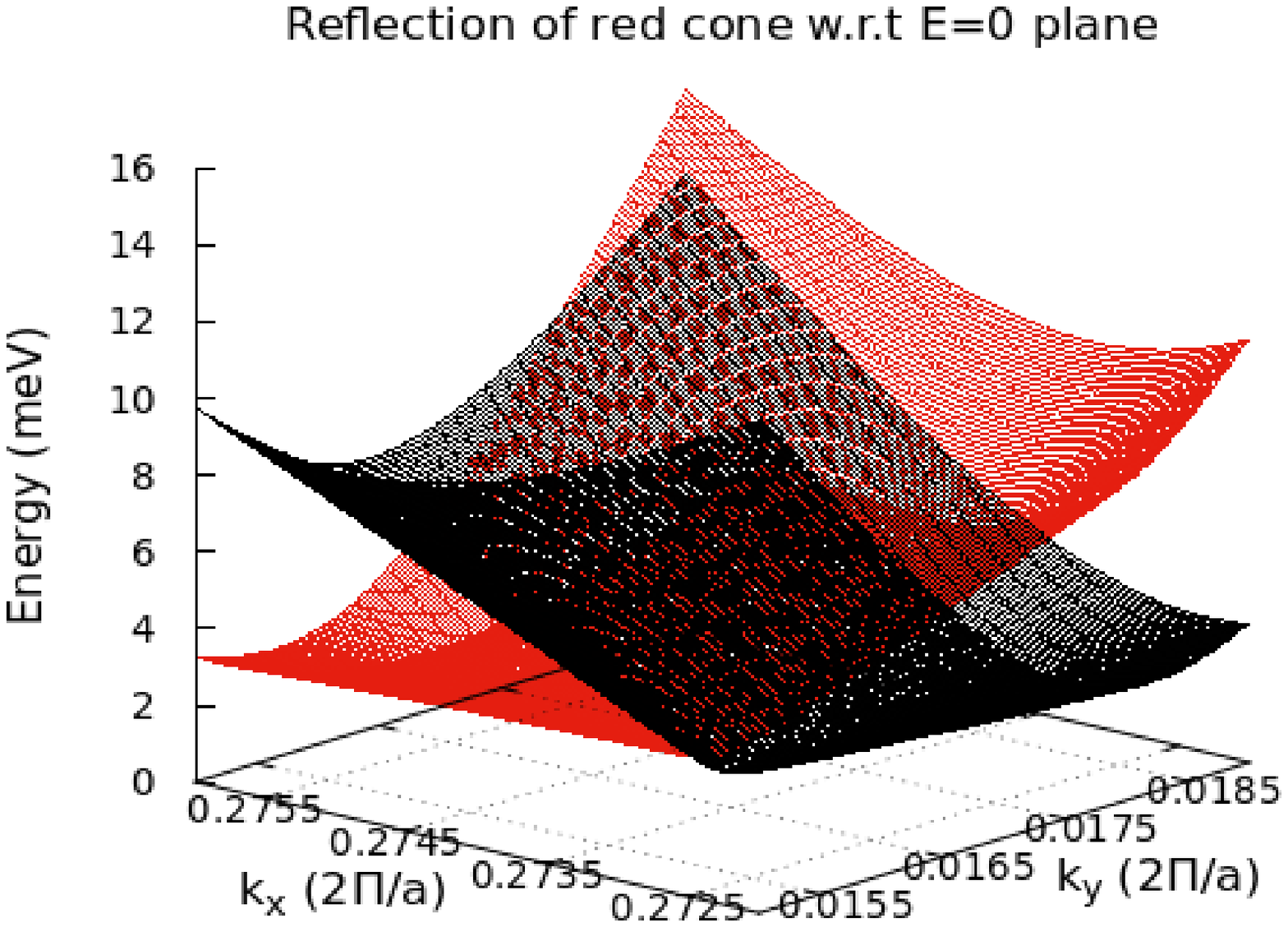}
    }
    \subfigure[]
    {
        \includegraphics[width=0.30\linewidth, height=4.0cm]{seebeck/TaP/2/TaP_2.eps}
    }
    \caption
    { {{ \footnotesize (Color online) Plot (a) shows the Weyl cone around W2 point of TaP. Plot (b) shows the extent to which red cone overlaps the balck one when inversion operation is applied on it with respect to the Weyl node. Plot (c) shows the extent to which red cone overlaps the balck one when mirror-reflection operation is applied on it with respect to the plane parallel to $k_x$-$k_y$ plane and passing through Weyl energy. Plot (d) shows the components of $S^{xx}$, $S^{yy}$ \& $S^{zz}$ contributed from this cone.}}
    }
    \label{fig_cone_taas}
\end{figure*}

\begin{figure*}
    \centering
    \subfigure[]
    {
        \includegraphics[width=0.30\linewidth, height=4.0cm]{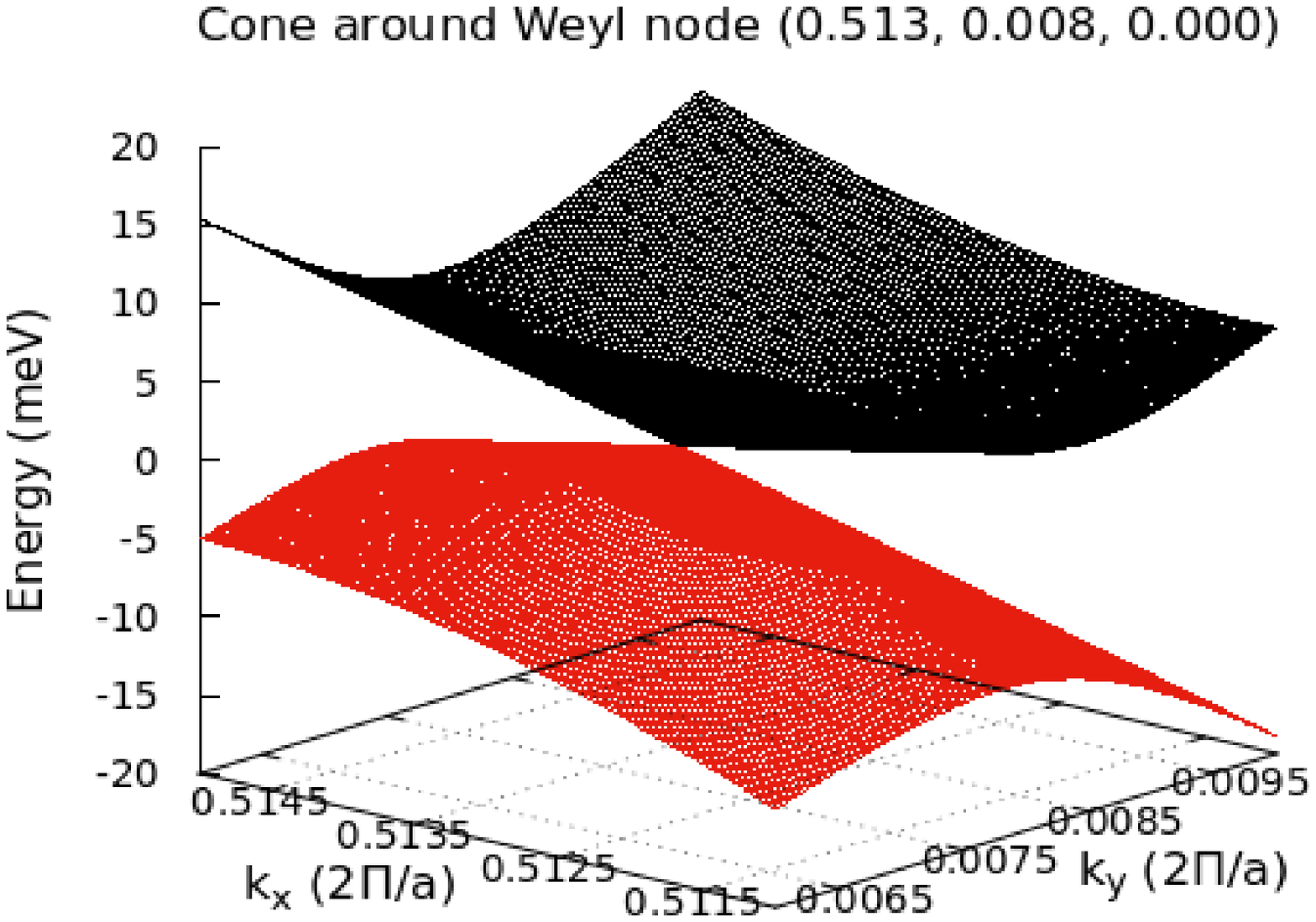}
    }
    \subfigure[]
    {
        \includegraphics[width=0.30\linewidth, height=4.0cm]{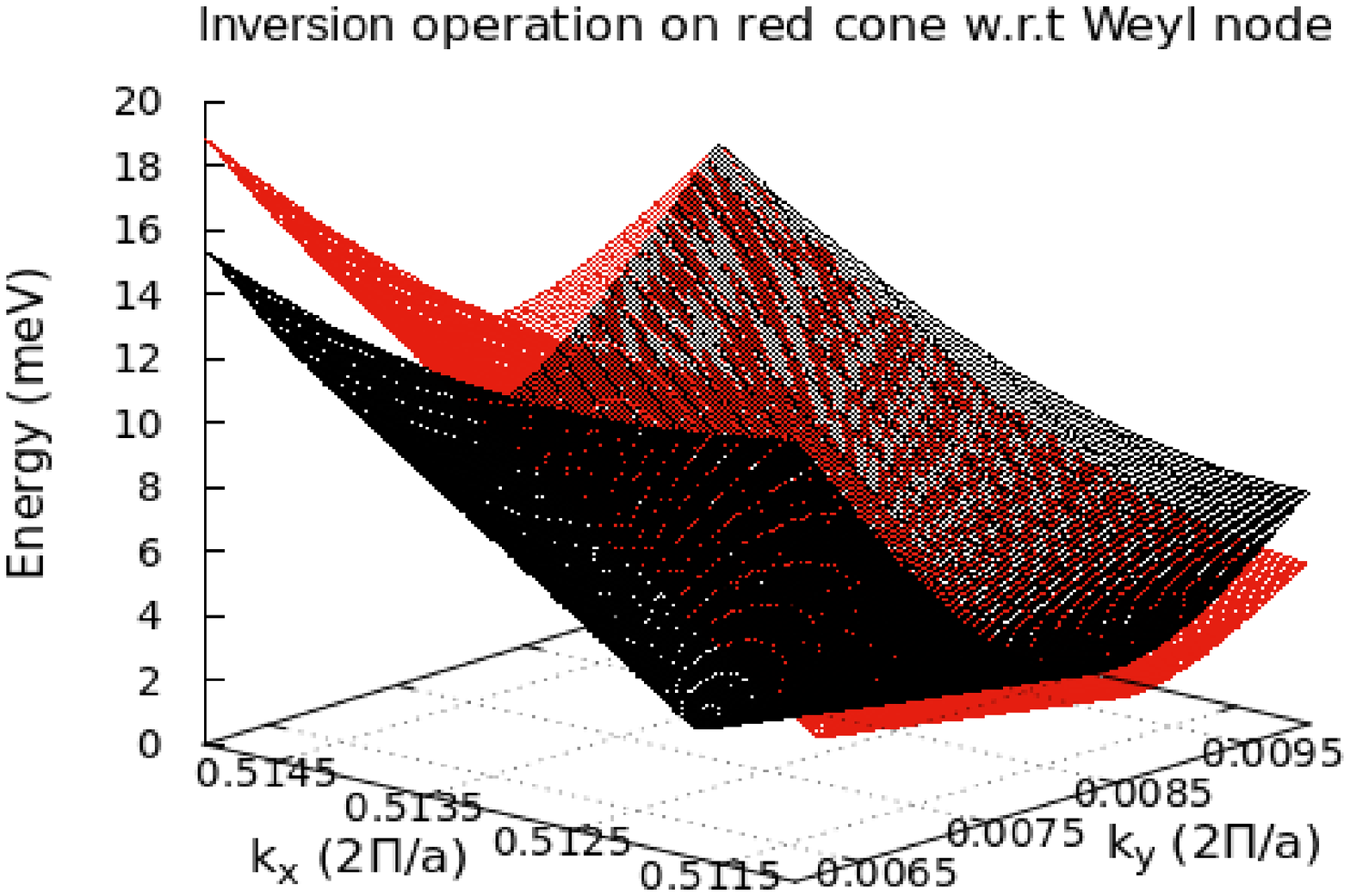}
    }\\
    \subfigure[]
    {
        \includegraphics[width=0.30\linewidth, height=4.0cm]{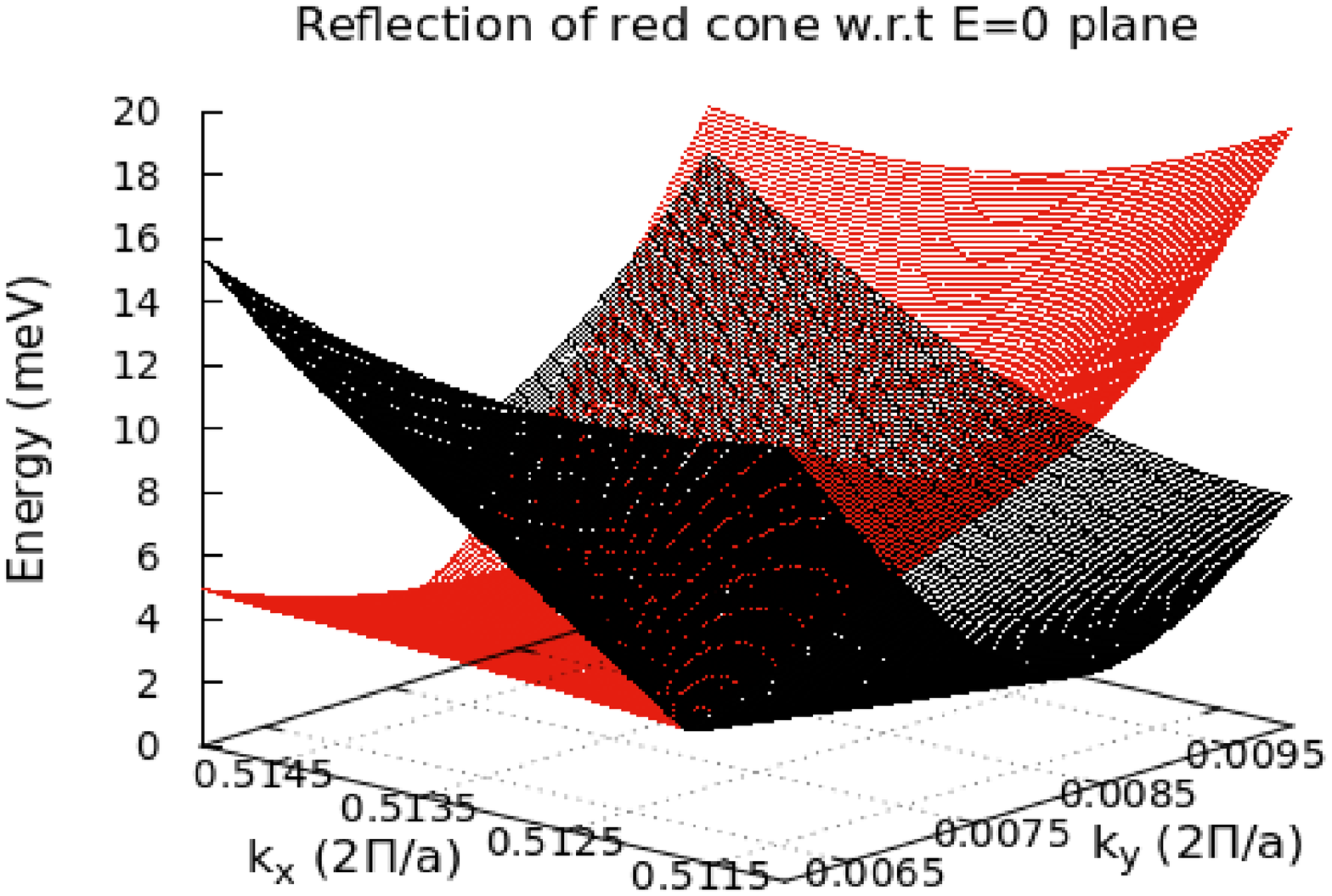}
    }
    \subfigure[]
    {
        \includegraphics[width=0.30\linewidth, height=4.0cm]{seebeck/TaP/3/TaP_3.eps}
    }
    \caption
    { {{ \footnotesize (Color online) Plot (a) shows the Weyl cone around W1 point of TaP. Plot (b) shows the extent to which red cone overlaps the balck one when inversion operation is applied on it with respect to the Weyl node. Plot (c) shows the extent to which red cone overlaps the balck one when mirror-reflection operation is applied on it with respect to the plane parallel to $k_x$-$k_y$ plane and passing through Weyl energy. Plot (d) shows the components of $S^{xx}$, $S^{yy}$ \& $S^{zz}$ contributed from this cone.}}
    }
    \label{fig_cone_taas}
\end{figure*}

\end{document}